\documentclass[aps,rmp,twocolumn,10pt]{revtex4-1}

\usepackage{wrapfig}

\usepackage{amsmath,amssymb,amsthm}
\usepackage{tikz}
\usetikzlibrary{arrows}
\usepackage{verbatim}
\usepackage{palatino}
\usepackage{mathpazo}
\usepackage{stmaryrd}
\usepackage{mathtools}
\mathtoolsset{centercolon}
\usepackage{thm-restate}

\usepackage{enumitem}

\usepackage{dsfont}

\usepackage{hyperref}
\definecolor{darkred}  {rgb}{0.5,0,0}
\definecolor{darkblue} {rgb}{0,0,0.5}
\definecolor{darkgreen}{rgb}{0,0.5,0}
\hypersetup{
  pdftitle = {QRT Proposal},
  pdfauthor = {},
  colorlinks = true,
  urlcolor  = blue,         
  linkcolor = darkblue,     
  citecolor = darkgreen,    
  filecolor = darkred       
}

\theoremstyle{definition}
\newtheorem{definition}{Definition}
\newtheorem*{remark}{Remark}

\newcommand{\bra}[1]{\langle #1|}
\newcommand{\ket}[1]{|#1\rangle}

\newcommand{\op}[2]{|#1\rangle \langle #2|}

\DeclareMathOperator{\tr}{Tr}

\newcommand{\wt}[1]{\widetilde{#1}}

\newcommand{\mc}[1]{\mathcal{#1}}
\newcommand{\mbf}[1]{\mathbf{#1}}
\newcommand{\mbb}[1]{\mathbb{#1}}

\newcommand{\supp}{\text{supp}}

\newcommand{\lr}{\rangle\langle}
\newcommand{\la}{\langle}
\newcommand{\ra}{\rangle}
\def\id{\mathsf{id}}

\renewcommand{\geq}{\geqslant}
\renewcommand{\leq}{\leqslant}

\newcommand{\mB}{\mathcal{B}}
\newcommand{\mQ}{\mathcal{Q}}
\newcommand{\mH}{\mathcal{H}}
\newcommand{\mF}{\mathcal{F}}
\newcommand{\mS}{\mathcal{S}}
\newcommand{\mO}{\mathcal{O}}
\newcommand{\mR}{\mathcal{R}}
\newcommand{\mU}{\mathcal{U}}

\newcommand{\mC}{\mathcal{C}}
\newcommand{\mG}{\mathcal{G}}

\newcommand{\toO}{\xrightarrow{\mc{O}}}

\newcommand{\toOmax}{\xrightarrow{\mc{O}_{\max}}}

\newcommand{\eqO}{\overset{\mc{O}}{\approx}}

\newcommand{\toOn}{\xrightarrow{\mc{O}_n}}

\newcommand{\toSO}{\xrightarrow{S\mc{O}}}

\newcommand{\tocO}{\xrightarrow{c\mc{O}}}

\newcommand{\eqdef}{\coloneqq}

\newcommand{\be}{\begin{equation}}
\newcommand{\ee}{\end{equation}}

\definecolor{indiagreen}{rgb}{0.07, 0.53, 0.03}

\newcommand{\p}{\mathbf{p}}
\newcommand{\q}{\mathbf{q}}
\newcommand{\e}{\mathbf{e}}
\newcommand{\rr}{\mathbf{r}}
\newcommand{\bt}{\mathbf{t}}
\newcommand{\g}{\mathbf{g}}

\newcommand{\1}{\mathds{1}}

\newcommand{\da}{\downarrow}

\begin{document}

\title{Quantum Resource Theories}

\author{Eric Chitambar}\email{echitamb@illinois.edu}
\affiliation{Department of Electrical and Computer Engineering, Coordinated Science Laboratory,\\ University of Illinois at Urbana-Champaign, Urbana, IL 61801}
\author{Gilad Gour}\email{gour@ucalgary.ca}
\affiliation{
Department of Mathematics and Statistics,
University of Calgary, AB, Canada T2N 1N4} 
\affiliation{
Institute for Quantum Science and Technology,
University of Calgary, AB, Canada T2N 1N4}

\date{\today}

\begin{abstract}
Quantum resource theories (QRTs) offer a highly versatile and powerful framework for studying different phenomena in quantum physics.  From quantum entanglement to quantum computation, resource theories can be used to quantify a desirable quantum effect, develop new protocols for its detection, and identify processes that optimize its use for a given application.  Particularly, QRTs have revolutionized the way we think about familiar properties of physical systems like entanglement, elevating them from being just interesting fundamental phenomena to being useful in performing practical tasks. The basic methodology of a general QRT involves partitioning all quantum states into two groups, one consisting of free states and the other consisting of resource states.  Accompanying the set of free states is a collection of free quantum operations arising from natural restrictions placed on the physical system, restrictions that force the free operations to act invariantly on the set of free states.  The QRT then studies what information processing tasks become possible using the restricted operations.  Despite the large degree of freedom in how one defines the free states and free operations, unexpected similarities emerge among different QRTs in terms of resource measures and resource convertibility.  As a result, objects that appear quite distinct on the surface, such as entanglement and quantum reference frames, appear to have great similarity on a deeper structural level.  This article reviews the general framework of a quantum resource theory, focusing on common structural features, operational tasks, and resource measures.  To illustrate these concepts, an overview is provided on some of the more commonly studied QRTs in the literature.

\end{abstract}

\maketitle

\tableofcontents

\section{Introduction and Motivation}

Basic economic principles dictate that objects acquire value when they cannot be easily obtained.  From this perspective, value is a property that emerges relative to physical capabilities.  A resource theory for a given scenario extends this principle by categorizing actions in terms of being either free or prohibited, and then analyzing what can be accomplished using the allowable operations.  Certain objects cannot be generated in this setting and they are considered to be a resource.  For example, a camper is forced to consider what types of food can be prepared using a camping stove and non-perishable ingredients.  The ability to bake and refrigerate is prohibited, and any ingredient requiring, say, a refrigerator is a resource for the camper.

\begin{figure}[b]
    \includegraphics[width=0.3\textwidth]{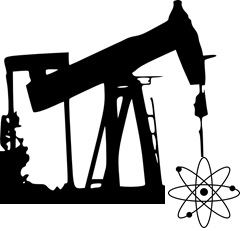}
  \caption{\linespread{1}\selectfont{\small In a quantum resource theory, the precious commodity is some physical property or phenomenon that emerges according to the principles of quantum mechanics.  The paradigmatic example is quantum entanglement.}}
\end{figure}
In recent years, the resource theory perspective has flourished within the quantum information community.  Instead of the resources being cooking ingredients for a camper or fuel for an automobile driver, the resources considered within quantum physics involve objects and phenomena at the atomic and sub-atomic levels.  Resource theories of this sort are called \textit{quantum resource theories}.  It is quite natural to apply a resource-theoretic outlook to the study of quantum systems since processes like decoherence rapidly eliminate most quantum behavior of a system.  Like an oil digger, one must exert considerable experimental effort to witness and control the subtle effects of quantum mechanics.

While the technical details will be covered in this review article, the basic idea of a quantum resource theory is to study quantum information processing under a restricted set of physical operations.  The permissible operations are called ``free,'' and because they do not encompass all physical processes that quantum mechanics allows, only certain physically realizable states of a quantum system can be prepared.  These accessible states are likewise called ``free,'' and any state that is not free is called a resource state.  Thus a quantum resource theory identifies every physical process as being either free or prohibited, and similarly it classifies every quantum state as being either free or a resource.  

\begin{figure}[b]
    \includegraphics[width=0.3\textwidth]{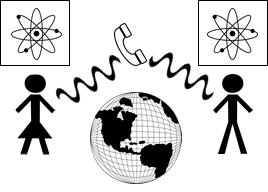}
  \caption{\linespread{1}\selectfont{\small Quantum entanglement is a quantum resource in the ``distant-lab'' scenario \cite{Plenio-2007a} where the free operations are LOCC.}}
  \label{Fig:LOCC}
\end{figure}

The most celebrated example of a quantum resource theory is the theory of entanglement.  For two or more quantum systems, entanglement can be characterized as a resource when the allowed dynamics are local quantum operations and classical communication (LOCC).  For example, as depicted in Fig. \ref{Fig:LOCC}, Alice and Bob may be working in their own quantum laboratory while being separated from each other by some large distance.  Due to current technological limitations, the only communication channel connecting their laboratories is classical, such as a telephone.  Hence Alice cannot directly send quantum states to Bob and vice versa, and the free operations in this resource theory consists of LOCC.  While the classical communication channel allows for the preparation of classically correlated states between the two laboratories, not every type of joint quantum state can be realized for Alice and Bob's systems using LOCC.  A state is said to be entangled, and therefore a resource, precisely when it \textit{cannot} be generated using the free operations of LOCC. For instance, if Alice and Bob each control a single spin-1/2 quantum system, the singlet state $\sqrt{1/2}(\ket{01}-\ket{10})$ cannot be created by LOCC and it is therefore called an entangled state.

Inspired by its success researchers have adopted the resource theory framework within many other areas of quantum information and physics.  For example, asymmetry and quantum reference frames, quantum thermodynamics,  quantum coherence and superposition, secret correlations in quantum and classical systems, non-Gaussianity in bosonic systems, ``magic states'' in stabilizer quantum computation, non-Markovianity in multi-part quantum processes, nonlocality, and quantum correlations have all been studied as resource theories.  Even more foundational objects such as contextuality and Bell non-locality have been envisioned as resources within quantum information theory.

There are multiple benefits to framing a given quantum phenomenon in terms of a resource, and here we highlight four of them.  
\begin{enumerate}
\item Resource theories can be desirable from a practical perspective as they often restrict attention to quantum operations that reflect current experimental capabilities.  For example, one advantage of capturing entanglement as a resource under LOCC is that, relative to the challenges of faithfully transmitting quantum systems across large distances, classical communication is easy.  Practically speaking then, it is very reasonable to consider what information processing tasks can be performed when restricted to LOCC.  More generally, a resource theory can be associated to any experiment where the free operations are those that can be performed within the experimental degrees of freedom inherent to the particular setup (e.g., rotation of the inhomogeneous magnetic field in the Stern-Gerlach experiment).

\item Resource theories provide the foundation to rigorously compare the amount of resource held in different quantum states (or quantum channels).  Operationally speaking, one state possesses at least as much resource as another if it is possible to transform the former into the latter using the free operations of the resource theory.  This is simply because whatever tasks can be accomplished using the transformed state can also be accomplished using the original one.  By considering convertibility under the free operations, a preordering is established on the set of quantum states.  With a resource theory, measures can also be constructed so that it becomes possible to say ``how much'' resource is in a given state.  While the specific numerical value of these measures can have various operational meanings such as transformation probability or conversion rate, all meaningful measures of resource are monotonically decreasing under the free operations of the resource theory.  This captures the intuitive notion that a resource is something precious and its value cannot be freely increased.

\item Resource theories enable a fine-grained analysis of what fundamental processes and properties drive a certain phenomenon.  By placing restrictions on the allowed operations, one can pin-point precisely the essential physical requirements for performing some information-processing task.  Quantum teleportation provides a beautiful example of how, when restricting to LOCC, entanglement emerges as the essential ingredient for transmitting quantum information from one physical location to another.  By decomposing a given task in terms of free operations and resource consumption, one can further consider resource trade-offs.  For certain tasks, it may be advantageous to expand the set of free operations in order to reduce the overall consumption of resource.

\item By capturing a particular object of interest within a quantum resource theory, it becomes possible to identify structures and applications that are common to resource theories in general.  Problems that are challenging or not even recognized when approached internally take on a new light when approached externally, from the more general resource-theoretic perspective.  For example, elegant solutions to the notoriously difficult problem of entanglement reversibility emerge when drawing resource-theoretic connections to thermodynamics.  As another example, new areas of research in classical information theory open after recognizing that certain features of quantum entanglement can also be observed in the classical setting of private and public correlations.

\end{enumerate}

This article surveys the subject of quantum resource theories.  There are already a number of very nice reviews on individual resource theories: entanglement \cite{Plenio-2007a, Horodecki-2009a}, quantum reference frames and asymmetry \cite{Bartlett-2007a}, quantum thermodynamics \cite{Gour-2015a, Goold-2016}, coherence \cite{Streltsov-2016a}, nonlocality \cite{Brunner-2014a}, non-Gaussianity \cite{Weedbrook-2012a}, non-Markovianity \cite{Rivas-2014a}, and quantum correlations \cite{Modi-2012a, Adesso-2016a}.  The purpose of this article is to review the plethora of features that unite all these theories together under a common resource-theoretic framework, similar to the approach taken in \textcite{Horodecki-2013a}.  Early in its development, quantum entanglement was seen to possess many formal similarities to thermodynamics \cite{Popescu-1997a, Horodecki-1998b, Horodecki-2002c}, and connections between different resource theories have been investigated ever since \cite{Horodecki-2013a, Brandao-2015a, Zi-Wen-2017a, Gour-2016a, Anshu-2017a, Sparaciari-2018a}.  Many of the same mathematical tools and techniques can be applied across a wide variety of resource theories.  Examples include majorization theory, entropic quantities and their properties, results from convex analysis like the hyperplane separation theorem, and optimization techniques such as cone programming.  Furthermore, recently it was recognized that a resource theory can be formulated as a symmetric monoidal category~\cite{Coecke-2016,Fritz-2015}. This abstract formulation recognizes that the structure of resource theories goes far beyond quantum physics and has the potential to be useful in many other areas of science.  Here, however, we will focus only on resource theories admitting the structure of quantum mechanics.

As one of its primary goals, this article outlines the general framework for constructing a quantum resource theory and discusses the typical questions that emerge in its development.  After describing the different approaches to answering these questions, the article provides a comparative review of the more well-known resource theories.  A broad overview of tasks, measures, and analytic techniques is then conducted over the course of three sections.  The article closes with an overview of open problems and future research directions.

\section{Notation and Preliminaries}

\label{Sect:Preliminaries}

Here we introduce the notation that will be used throughout the paper, and we quickly review some of the basic concepts in quantum information theory that are relevant to quantum resource theories.  More detailed expositions of this introductory material can be found in \textcite{Nielsen-2000a,Wilde2013, Watrous-2018a}.  We will denote Hilbert spaces by $\mH^A$, $\mH^B$, etc., where the superscripts indicate the physical systems associated with these Hilbert spaces.  Composite systems will be denoted by $\mH^{AB}$, $\mH^{ABC}$, etc.  In some cases we will want to envision systems $A$, $B$, $C$, etc. as being held by generic agents Alice, Bob, Charlie, etc., but an association with personal agents is not necessary.  The set of bounded operators acting on a Hilbert space $\mH^A$ will be denoted by $\mB(\mH^{A})$, or simply $\mB(A)$.  Positive semi-definite operators in $\mB(A)$ will be typically denoted by lowercase Greek letters (e.g. $\rho,\sigma,\omega$), and we write $\rho\geq 0$ to indicate that all eigenvalues are non-negative. The set of quantum states (i.e., density matrices) in $\mB(A)$ consists of positive semi-definite matrices with trace one, and it will be denoted by $\mS(A)$.  Two common ways to quantify ``how close'' a given state $\rho$ is to another $\sigma$ is the trace distance, given by $D_{\tr}(\rho,\sigma)=\frac{1}{2}\Vert\rho-\sigma\Vert_{1}$ (where $\Vert X\Vert_1=\tr\sqrt{X^\dagger X})$, and the fidelity, given by $F(\rho,\sigma)=\Vert\sqrt{\rho}\sqrt{\sigma}\Vert_1$ \cite{Uhlmann-1976a}.  The two can be related using the inequalities \cite{Fuchs-1999a}
\begin{equation}
\label{Eq:trace-fidelity-relation}
1-F(\rho,\sigma)\leq D_{\tr}(\rho,\sigma)\leq \sqrt{1-F(\rho,\sigma)^2}.
\end{equation}  

Linear maps, or super-operators, that act on $\mB(A)$ will be denoted by capital Greek letters (e.g. $\Phi,\;\Lambda,\;\Gamma$), and the identity map will be denoted by $\id^A:\mB(A)\to\mB(A)$.  The set of all bounded linear maps from $\mB(A)$ to $\mB(B)$ will be denoted by $\mc{B}(\mc{H}^A\to\mc{H}^B)$, or simply $\mB(A\to B)$.
Linear maps that represent a physical evolution of a (possibly open) system must take density matrices to density matrices.  We say that a linear map $\Phi:\mB(A)\to\mB(B)$ is positive if $\Phi(\rho)\geq 0$ for all $0\leq \rho\in \mB(A)$, $k$-positive if $\Phi\otimes\id^C$ is positive  with $\dim(\mH^C)=k$, and completely positive (CP) if it is $k$-positive for all $k$. It is known that if a linear map is $k$-positive with $k\geq \dim(\mH^A)$ then it is completely positive. A physical evolution is therefore represented by a CP map. Moreover, since density matrices can only evolve to density matrices, a physical evolution must preserve the trace. Such completely positive trace-preserving (CPTP) maps are called \emph{quantum channels}. The set of all quantum channels in $\mB(A\to B)$ will be denoted by $\mQ(A\to B)$.

Quantum mechanics allows for stochastic processes and quantum measurements.  For measurements with a discrete set of outcomes, a quantum state $\rho$ is converted to another quantum state $\sigma_i$ with some probability $p_i$.  The average post-measurement description of both the classical measurement register and the quantum system can be given by a \emph{quantum-classical} (QC) state of the form:
\be
\sigma^{QX}=\sum_{i}p_i\sigma_i\otimes|i\lr i|^{X}.
\ee
Therefore, the entire measurement can be modeled by a deterministic process, $\Phi$, converting $\rho$ to $\sigma^{QX}$.  The map $\Phi$ in this case is a particular type of a quantum channel, often called a measurement map, that has the form $\Phi(\cdot)=\sum_i \Phi_i(\cdot)\otimes |i\lr i|^X$, with each $\Phi_i$ being CP and $\sum_i\Phi_i$ being trace-preserving.  By appending the classical ancillary system $X$ we can thereby consider \emph{trace-preserving} maps even when discussing stochastic processes.  Throughout this paper we will adopt the convention that $\mc{H}^X$ denotes a (classical) system whose states are always dephased in some \textit{a priori} fixed orthonormal basis $\{\ket{i}\}_i$.

Any quantum channel has three important representations that are frequently used in the field of quantum information science, and all three representations play a crucial role in quantum resource theories as well.  The most physically intuitive one is related to the Stinspring Dilation Theorem \cite{Stinespring-1955a}. In this representation, the evolution of an open quantum system $A$ is modeled by a unitary interaction $U^{AE}$ of the joint system A with the environment (represented by system $E$). When the environment is initially in some uncorrelated state $\op{0}{0}^E$, the reduced-state dynamics of system $A$ is described by the quantum channel
\be\label{unitary}
\rho^A\mapsto\Phi(\rho^A)=\tr_{E'}\left[U^{AE}\left(\rho^{A}\otimes|0\lr 0|^E\right)U^{\dag AE}\right],
\ee  
where $E'$ need not be the same system $E$.  It turns out that for every CPTP map $\Phi$, there exists such a unitary representation $U^{AE}$ in which Eq.~\eqref{unitary} holds for all $\rho^A$.  This can be interpreted as saying that every physical evolution is essentially a unitary evolution on the joint system and environment, and CPTP maps provide only an \emph{effective} description of the evolution due to the inaccessibility of the environment's degrees of freedom. 

The second representation of a quantum channel is known as the \emph{operator-sum representation}.  It states that the action of any quantum channel $\Phi$ can be written as $\Phi(\rho^A)=\sum_{j}K_j\rho^A K_{j}^{\dag}$, where $\{K_j\}_j$ is a set of complex matrices (known as Kraus operators) satisfying $\sum_jK_{j}^{\dag}K_j=\mbb{I}^{A}$, with $\mbb{I}^A$ being the identity in $\mB(A)$.  If we relax the trace-preserving condition on $\Phi$ to be just trace non-increasing, then the Kraus operators satisfy $\sum_jK_{j}^{\dag}K_j\leq\mbb{I}^{A}$.  Just as the unitary $U^{AE}$ in Eq.~\eqref{unitary} is not unique for each $\Phi$, the set of Kraus operators $\{K_j\}_j$ is not unique and is defined up to a unitary mixing~\cite{Nielsen-2000a}.  While the operator-sum representation at first seems very mathematical, translating physical constraints into Kraus operator constraints is often a very convenient way to characterize the allowed (i.e., free) operations of a QRT, such as in entanglement.

The last representation of a quantum channel that we consider involves an isomorphism between bipartite positive operators and CP maps.  At first glance, the mathematical structure of quantum channels appears to be more complex than that of density matrices.  However, the two objects are actually equivalent.  To establish this, we first let $\ket{\phi^+_{d_A}}=\sum_{j=1}^{d_A}|j\ra^{A}|j\ra^{A'}$ denote a canonical unnormalized maximally entangled vector acting on $\mc{H}^A\otimes\mc{H}^{A'}$, with $\mc{H}^A\approx\mc{H}^{A'}$.  When the dimension of the system is clear, we will omit the subscript $d_A$.  Then consider the action of the CP (but not necessarily trace-preserving) map $\Phi:\mB(A')\to\mB(B)$ when acting on one half of the maximally entangled vector $\ket{\phi^+}^{AA'}$.  This produces the bipartite operator 
\be\label{choi1}
J_\Phi^{AB}:= \id^{A}\otimes\Phi(\phi^+),
\ee 
where $\phi^+=\op{\phi^+}{\phi^+}$.  The operator $J_{\Phi}^{AB}$ is called the \textit{Choi matrix} of $\Phi$, and when the context is clear, we will omit the subscript $\Phi$ and simply write $J^{AB}$ or just $J$.  Any CP map $\Phi$ corresponds to such a bipartite positive semi-definite operator $J_\Phi^{AB}$ via Eq.~\eqref{choi1}, and conversely, any bipartite positive semi-definite operator $J^{AB}$ corresponds to a CP map $\Phi$ given by
\be\label{choi2}
\Phi_J(\rho)=\tr_{A}\left[J^{AB}\left(\rho^T\otimes I^B\right)\right],
\ee
where $\rho^T$ indicates the matrix transpose w.r.t. some fixed basis of $\mc{H}^A$.  If this basis is chosen to be the same as that used in the definition of $\ket{\phi^+}^{AA'}$, then it is easy to see that $\Phi_{J_\Phi}=\Phi$.  The relations in~\eqref{choi1} and~\eqref{choi2} therefore define an isomorphism between the set of CP maps in $\mc{B}(A\to B)$ and the set of bipartite positive semi-definite operators in $\mc{B}(AB)$, typically called the Choi-Jamio\l kowski Isomorphism.  If we further require $\Phi$ to be trace-preserving (i.e., a quantum channel), then the system $A$ reduced operator in $J_\Phi^{AB}$ of Eq.~\eqref{choi1} is the identity, i.e., $J_\Phi^A=\mbb{I}^A$.  Conversely, if the reduced operator of $J^{AB}$ is the identity in Eq.~\eqref{choi2}, then $\Phi_J$ will also be trace-preserving.  Thus, there is a bijection between the set of quantum channels and the set of positive semi-definite operators with the system-$A$ marginal being the identity.

Each of the three representations above plays an important role in QRTs. As we will see, different representations fit more naturally in the analysis of different QRTs.  Some, more physical QRTs, lend themselves best to a unitary representation, while others, more mathematical in nature, allow for easier analysis using the Choi representation.  

We finally draw attention to a special class of CP maps that act invariantly on the identity; i.e., $\Phi(\mbb{I})=\mbb{I}$.  These are called \textit{unital} maps, and they are closely related to the dual of a quantum channel.  Namely, for every CP map $\Phi\in\mc{B}(A\to B)$, its dual $\Phi^\dag\in\mc{B}(B\to A)$ is the adjoint map fixed by the Hilbert-Schmidt inner product; that is, it is the unique map $\Phi^\dag$ satisfying
\begin{equation}
\label{Eq:adjoint}
\tr\left(X\Phi^\dag(Y)\right)=\tr\left(\Phi(X)Y\right)
\end{equation}
for all hermitian $X\in\mc{B}(A)$ and hermitian $Y\in\mc{B}(B)$.  One can verify that a CP map $\Phi$ is trace-preserving if and only if its dual $\Phi^\dag$ is a unital CP map.  To see this, it is perhaps easiest to substitute the operator-sum representation $\Phi(A)=\sum_jK_j AK^\dagger_j$ directly into Eq.~\eqref{Eq:adjoint}, which reveals $\{K_j^\dagger\}$ to be the Kraus operators of $\Phi^\dag$.  In terms of its Choi matrix, $J_{\Phi^\dag}^{AB}$ for a channel $\Phi$ has the property that its system $B$ reduced state is the identity; i.e., $J_{\Phi^\dag}^B=\mbb{I}^B$.  Compare this to the condition $J_\Phi ^A=\mbb{I}^A$ for $J^{AB}_\Phi$ mentioned above.

\section{The General Structure of Quantum Resource Theories}

As discussed in the introduction, the structure of resource theories goes far beyond quantum physics.  For example, the set of all shapes that can be generated by a compass and a ruler could represent ``free states'' of a resource theory, with the action of the compass and ruler being the free operations.  Therefore, in this resource theory, all the shapes that cannot be generated by a compass and ruler are considered as resources\footnote{We credit Rob Spekkens for this simple example of a non-quantum resource theory.}.  However, the type of resource theories that we will consider here focus on quantum phenomena such as entanglement and coherence. Therefore, in what follows, QRTs will be defined with respect to a given Hilbert space so that the structure of quantum mechanics is prominent.

\subsection{Definition of a Quantum Resource Theory (QRT) and Tensor-Product Structures}

Already in the early stages of its development, it was clear that quantum information is a theory of interconversions among different resources~\cite{Bennett-2004a, Devetak-2008a}.  These resources are diversely classified as classical or quantum, noisy or noiseless, and static or dynamic.  However, the term ``resource theory" appeared much later. Originally, it was coined by Schumacher in 2003 (unpublished), and later in a 2008 paper on the resource theory of quantum reference frames~\cite{Gour-2008a}. The latter provided one of the first explicit constructions of a QRT that is different from entanglement theory, although the framework for a QRT of information had already been investigated in a series of earlier papers \cite{Oppenheim-2002a, Horodecki-2003a, Horodecki-2005a}.  Since then, many other resource theories have been developed, and a precise mathematical definition of a resource theory was given in~\cite{Coecke-2016, Fritz-2015} as a symmetric monoidal category. However, this definition involves terms from category theory and goes beyond the scope of this review. We therefore start with a mathematically less general definition of quantum resource theories, yet probably more accessible to a reader with a physics background.  
While in this paper we will consider mostly finite-dimensional Hilbert spaces over the complex field (in which case the Hilbert spaces will be isomorphic to $\mbb{C}^d$ for some integer $d$), the definition below can also be applied to infinite-dimensional Hilbert spaces.

\begin{definition}\label{maindef}  Let $\mc{O}$ be a mapping that assigns to any two input/output physical systems $A$ and $B$, with corresponding Hilbert spaces $\mH^A$ and $\mH^B$, a unique set of CPTP operations $\mO(A\to B)\equiv\mO(\mH^A\to\mH^B)\subset\mQ(A\to B)$.  Let $\mc{F}$ be the induced mapping $\mc{F} (\mc{H}):=\mc{O}(\mbb{C}\to\mc{H})$, where $\mc{H}$ is an arbitrary Hilbert space.  Then the tuple $\mR=(\mF,\mO)$ is called a \emph{quantum resource theory} (QRT) if the following two conditions hold: 
\begin{enumerate}
\item For any physical system $A$ the set $\mO(A):=\mc{O}(A\to A)$ contains the identity map $\id^A$. 
\item For any three physical systems $A$, $B$, and $C$, if $\Phi\in\mO(A\to B)$ and $\Lambda\in\mO(B\to C)$ then $\Lambda\circ\Phi\in \mO(A\to C)$.
\end{enumerate} 
In a QRT, the set $\mF(\mH)\subset\mS(\mH)$ defines the set of \emph{free states} acting on $\mc{H}$, and the elements belonging to $\mS(\mH)\setminus\mF(\mH)$ are called \textit{resource states} or \emph{static resources}.  Likewise the CPTP maps in $\mO(A\to B)$ are called \emph{free operations} and the CPTP maps that are not in $\mO(A\to B)$ are called \emph{dynamical resources}.
\end{definition}
\noindent As before, for Hilbert spaces $\mc{H}^A$, $\mc{H}^B$, etc., we will often denote sets of free states and free operations in terms of system labels, e.g., $\mc{F}(A):=\mc{F}(\mc{H}^A)$, $\mc{F}(AB):=\mc{F}(\mc{H}^A\otimes\mc{H}^B)$, $\mc{O}(A\to B):=\mc{O}(\mc{H}^A\to\mc{H}^B)$, $\mc{O}(A):=\mc{O}(\mc{H}^A\to\mc{H}^A)$, etc..  With a slight abuse of terminology, when the underlying Hilbert spaces are clear, we will often refer to $\mc{F}$ as the free states and $\mc{O}$ the free operations.  Also, here and throughout the paper we use the notation $\subset$ to indicate a generic set inclusion, which may or may not be strict.  

The physical interpretation of Definition \ref{maindef} is as follows.  Consider a quantum system held by one agent or distributed to a group of parties.  A QRT models what the parties can physically accomplish given some restrictions or constraints that result from technical or experimental limitations, the rules of some game, or simply the laws of physics.  What operations the agents can still perform given these restrictions is mathematically described by $\mO(A\to B)$, which is typically much smaller than the set of all quantum channels.  The first condition in Definition~\ref{maindef} simply says that the identity map (i.e., doing nothing) is free, an obvious requirement for any meaningful QRT.  Condition two says that $\Lambda\circ\Phi$ is free whenever $\Phi$ and $\Lambda$ are both free.  This ensures that the operations belonging to $\mO$ are indeed free in the sense that they can be performed freely any number of times and in any order.  A consequence of condition two is that the free operations cannot convert any state in $\mc{F}$ to one not belonging to $\mc{F}$.  More formally,
\begin{quote}
For any two physical systems $A$ and $B$, if $\Phi\in\mO(A\to B)$ and $\rho\in\mF(A)$, then $\Phi(\rho)\in\mF(B)$.
\end{quote}
This can be referred to as the \textit{golden rule of QRTs}, and it justifies the terminology of ``resource'' for states in $\mc{S}(\mc{H})\setminus\mc{F}(\mc{H})$.  

In the literature, the free states and free operations are typically presented on equal footing.  However here we identify the free operations as being more fundamental.  If free states are special objects that an experimenter can work with, then he/she must be able to prepare or obtain them.  Such an initial preparation is thus identified by an operation in $\mc{O}(\mbb{C}\to\mc{H})$.  Nevertheless, note that by keeping the set $\mc{O}(\mbb{C}\to\mc{H})$ fixed, different QRTs can be defined with the same set of free states.  We will therefore treat the free states as its own component of a QRT, even though they emerge from the definition of free operations.  Then unless otherwise stated, when we speak of ``free operations,'' we will always mean maps acting on input spaces having dimension at least two.

The golden rule of QRTs does not imply that resource states fail to play a functional role in the theory.  On the contrary, if the agents/parties do happen to have access to a resource state (perhaps prepared separately by a third party), it could possibly be used to circumvent (at least partially) the restrictions on the allowed operations.  That is, for some $\sigma\not\in\mc{F}(B)$, there may exist maps $\Phi\in\mc{O}(AB)$ and $\Lambda\not\in\mc{O}(A)$  such that $\Phi(\rho\otimes\sigma)=\Lambda(\rho)$ for all $\rho\in\mc{S}(A)$ (or perhaps just a subset of $\mc{S}(A)$).  In this case, the state $\sigma$ is literally functioning as a resource for the simulation of an otherwise restricted operation $\Lambda$ (see Sections \ref{Sect:processes} and \ref{Sect:Task-simulation}).  The most celebrated example of this is quantum teleportation in entanglement theory \cite{Bennett-1993a}.  


QRTs correspond to physical models.  As such, Hilbert spaces represent the state space of specific physical systems.  Therefore, a mathematical isomorphism between two Hilbert spaces, such as $\mbb{C}^2\otimes\mbb{C}^2\cong\mbb{C}^4$, does not necessarily translate into the same set of free states (or free operations) for the two spaces.  That is, $\mF(\mbb{C}^2\otimes\mbb{C}^2)$ can be very different from $\mF(\mbb{C}^4)$ since the two Hilbert spaces can represent different physical systems.  For instance, the space $\mbb{C}^2\otimes\mbb{C}^2$ might represent two spatially separated spin-$1/2$ particles, while $\mbb{C}^4$ may correspond to a \emph{single} particle with four spin or energy levels. Therefore, in the assignments of $\mF$ and $\mO$ for a given QRT, one must carefully consider what physical scenario the QRT is attempting to model.  On the other hand, a relabeling of tensor-product spaces does not change the free states and free operations of a QRT.  That is, if $\mc{H}^A$ is a Hilbert space for system $A$ and $\mc{H}^B$ is a Hilbert space for system $B$, then density matrices acting on $\mc{H}^A\otimes\mc{H}^B$ represent the same physical states as density matrices acting on $\mc{H}^B\otimes\mc{H}^A$.

While Definition \ref{maindef} stipulates the minimal mathematical requirements of a QRT, in practice there are other natural properties that one might desire in a QRT.  The most obvious of these can be collected together in what will be referred to as a tensor-product structure.
\begin{definition}\label{tensor}
A QRT $\mR=(\mF,\mO)$ is said to admit a\textit{ tensor-product structure} if the following three conditions hold:
\begin{enumerate}
\item The free operations are ``completely free'': For any three physical systems $A$, $B$, and $C$, if $\Phi\in\mO(A\to B)$ then $\id^C\otimes\Phi\in\mO(CA\to CB)$, where $\id^C$ is the identity map on $\mB(C)$.
\item Appending free states is a free operation: For any given free state $\sigma\in\mF(B)$, the CPTP map $\Phi_{\sigma}(\rho):=\rho\otimes\sigma$ is a free map, i.e., it belongs to $\mO(A\to AB)$.
\item Discarding a system is a free operation: For any Hilbert space $\mH$, the set $\mO(\mc{H}\to \mbb{R})$ is not empty.
\end{enumerate}
\end{definition} 

\begin{remark}
Note that $\mB(\mH\to \mbb{R})$ contains only one CPTP map which is given by the trace. Therefore, the statement that $\mO(\mH,\mbb{R})$ is not empty, is equivalent to the statement that the trace of a system is a free map.
\end{remark}

These conditions are highly intuitive and are to be expected in most physical models.  The first says that a free operation remains free when acting on just one part of any joint system.  Such maps are called ``completely free'' analogous to the notion of ``completely positive'' maps (see Section \ref{Sect:k-RNG}).  As a consequence of the first condition, if $\Phi\in \mO(A\to B)$ is free and $\Phi'\in\mO(A'\to B')$ is free, then $\Phi\otimes\Phi'$ must be free as well.  This follows from the fact that $\Phi\otimes\Phi'=\left(\id^B\otimes\Phi'\right)\circ\left(\Phi\otimes\id^{A'}\right)$ is a composition of two free operations.  The second and third conditions in Definition ~\ref{tensor} state that appending a free ancillary system and discarding a system are both free.  Both are very natural properties to suppose of a QRT.  In particular, the ability to append arbitrary free states to any system reflects the situation where free states are \emph{really} free to generate.

The defining conditions of a tensor-product structure are not completely independent, and they have several interesting consequences.  The first consequence is that the partial trace $\tr\otimes\id$ is a free operation, a fact that follows immediately from the first and third properties in Definition \ref{tensor}.  A second consequence is that every replacement channel $\Phi_\sigma\in\mQ(A\to B)$ of the form $\Phi_\sigma(X):=\tr[X]\sigma$, with a fixed $\sigma\in\mF(B)$, is free (i.e., belonging to $\mO(A\to B)$).  This can be seen by combining the partial trace with the second property in Definition \ref{tensor}.  In particular, if $\rho$ and $\sigma$ are two free states in a QRT with tensor-product structure, then $\rho$ can be converted to $\sigma$ by free operations, and vice versa.  Finally, a third consequence is that if both $\rho\in\mF(A)$ and $\sigma\in\mF(B)$ are free, then $\rho\otimes\sigma\in\mF(AB)$. This property follows from the previous consequence and the first condition of Definition ~\ref{tensor}.  The intuition behind this property is that if $\rho$ and $\sigma$ are free to prepare separately, then their joint state $\rho\otimes\sigma$ is also free to prepare.  This justifies the terminology of tensor-product structure since it implies that $\mF(A)\otimes\mF(B)\subset\mF(AB)$ for any two Hilbert spaces $\mH^A$ and $\mH^B$.  Note that a partial converse of this inclusion also holds in the sense that if $\rho^{AB}\in \mF(AB)$, then its marginals are also free; i.e., $\rho^A\in\mF(A)$ and $\rho^B\in\mF(B)$. This holds in light of the above observation that the partial trace is a free map.

Most of the physically motivated and previously studied QRTs admit a tensor-product structure, such as the QRTs of entanglement, coherence, asymmetry, and athermality.   However, there are less intuitive but still important QRTs that do not possess a tensor-product structure.  For example, certain models of Bell nonlocality do not admit such a structure and lead to examples of states $\rho$ and $\sigma$ that are free, even though $\rho\otimes\sigma$ is not \cite{Palazuelos-2012a}.  In such cases, the QRT is said to demonstrate a ``super-activation" of resource.
 
\subsection{Consistent QRTs for a given set of free operations}

\label{Sect:Consistent-given_free_operations}

When attempting to model some quantum phenomenon using a QRT, often physical constraints dictate the appropriate choice for either the free states or more generally the free operations.  For example, in many quantum information problems, multiple spatially separated parties share a composite quantum system and LOCC emerges as the natural choice of free operations.  On the other hand, in the resource theory of coherence, it is more natural to first turn to the free states and identify these as being the collection of density matrices that are diagonal in some fixed basis.  With either the free states or free operations given, the other must then be consistently specified so that the golden rule of QRTs is satisfied.  We now explore this specification in more detail.  As we will first see, when the free operations are given, the choice of free states is often unique.


For a given Hilbert space, suppose that the free operations of a physical system $A$ are fixed by the physical constraints, and let us consider the structure of QRTs that are consistent with these free operations $\mO(A)$.  First observe that the set $\mO(A)$ imposes a \emph{preorder} on the set of density matrices $\mS(A)$ (see Section \ref{Sect:Quasi-Order}).  That is, given two arbitrary states $\rho,\sigma\in\mS(A)$, we can write $\rho\toO\sigma$ if there exists a free operation $\Phi\in \mO(A)$ such that $\sigma=\Phi(\rho)$.  If both $\rho\toO\sigma$ and $\sigma\toO\rho$ we write $\rho\eqO\sigma$. Clearly, the relation $\toO$ is a preorder on $\mS(A)$ since for any $\rho,\sigma,\gamma\in\mS(A)$, if $\rho\toO\sigma$ and $\sigma\toO\gamma$ then also $\rho\toO\gamma$.  For any set of operations $\mO(A)$, we can then define the associated minimal set of free states 
$\mF_{\min}(A)$ as follows:
\begin{align}
\label{fmin}
\mF_{\min}(A)\equiv\big\{\rho\;:\;\forall \sigma\in &\mS(A)\; \; \exists \Phi\in\mc{O}(A) \notag\\
&\;\text{such that}\; \rho=\Phi(\sigma)\big\}\;.
\end{align}
In other words, $\rho\in\mF_{\min}(A)$ if it can be freely generated starting from \text{any} other state.  For a QRT with free operations $\mO(A)$ and any non-empty set of free states $\mF(A)$ consistent with $\mO(A)$, we must have that $\mF_{\min}(A)\subset\mF(A)$, a relationship expressing the sense in which $\mc{F}_{\min}(A)$ is a ``minimal'' set.  This is because if $\sigma\in\mF(A)$ and $\rho\in\mF_{\min}(A)$, then by definition $\sigma\toO\rho$, which means that $\rho$ can be obtained from a free state by a free operation; hence, $\rho\in\mF(A)$.

Suppose further that the QRT has the property that any two free states on the same space can be converted from one to the other using the free operations.  That is, $\rho,\sigma\in\mF(A)$ implies $\rho\eqO\sigma$.  When this condition holds, then we must have that $\mF_{\min}(A)=\mF(A)$. This remarkable observation follows easily from the fact that $\mF_{\min}(A)\subset\mF(A)$, which by assumption means that $\rho\toO\sigma$ for any $\sigma\in\mc{F}(A)$ and any $\rho\in\mF_{\min}(A)$.  But by the definition of $\mF_{\min}(A)$, it holds that $\omega\toO\rho$ for all $\omega\in\mc{S}(A)$.  Hence, we also have that $\sigma\in\mF_{\min}(A)$ since $\omega\toO\sigma$, which establishes the equality $\mF_{\min}(A)=\mF(A)$.  

In what type of QRTs can one free state always be converted to any other using the free operations?  Clearly the property holds for any QRT that admits a tensor-product structure.  But more generally, it suffices for the QRT to allow both discarding a system (i.e., ``trash'') and preparing any free state.  Combining these yields a replacement channel $\Phi_\sigma(X)=\tr[X]\sigma$ for some free state $\sigma$, and thus the set of free states must be $\mc{F}_{\min}(A)$ for these QRTs.

In conclusion, given a set of free operations $\mO(A)$, if one desires a QRT in which any two free states are freely interconvertible and discarding systems is allowed, then $\mF_{\min}(A)$ is the \textit{only} set of free states that is consistent with the set $\mO(A)$.  This demonstrates clearly how any natural physical constraint on the set of quantum processes leads to a unique QRT.

\subsection{Consistent QRTs for a given set of free states}

\label{Sect:Consistent_Operations}

As we discussed above, certain types of quantum phenomena can be identified directly on the level of states without involving constraints on quantum processes.  This is the case for coherence and some models of Bell nonlocality.  When characterizing such phenomena within a QRT framework, the task then becomes to identify sets of free operations that are consistent with the given set of free states.  Unlike the conclusion reached in the previous section - that fixing the free operations leads to a unique set of free states $\mc{F}_{\min}$ under reasonable assumptions - here there exists much greater freedom in choosing a consistent set of free operations for a fixed set of free states, even for QRTs admitting a tensor-product structure.

Often some physical consideration will motivate a certain choice of free operations. But even in this case, it is valuable to study different classes of free operations for the same set of free states.  This is because different classes may have an easier or more elegant mathematical structure than the physically-motivated class of operations.  This is the case, for example, in entanglement theory where LOCC is a notoriously difficult class of operations to characterize.  To avoid the technical difficulties that arise when using these operations, much work has been devoted to the study of entanglement theory under larger and more analytically-friendly sets of operations such as separable operations, non-entangling operations, and more \cite{Rains-1997a, Rains-1999b, Vedral-1998a, Pankowski-2013a, Chitambar-2018b}.  All of these resource theories have in common that the set of separable states is the set of free states.  Studying more powerful operations can lead to proving no-go results for the weaker yet more natural choice of free operations.  Indeed, any quantum information task that cannot be performed by the more powerful class cannot be performed by the weaker one.  In this subsection, we survey different consistent sets of free operations in general QRTs, highlighting their various physical motivations and properties.  

\subsubsection{RNG, $k$-RNG, and Completely RNG Operations}

\label{Sect:k-RNG}

Associated with any set of free states, there will always be a  maximal set of operations that is allowed by the definition of a QRT. This is the set of resource non-generating operations which is defined as follows:
\begin{definition}
Let $\mF$ be as in Definition~\ref{maindef}.  For any two physical systems $A$ and $B$, the set of \textit{resource non-generating} (RNG) operations $\mO_{\max}(A\to B)$ consists of \emph{all} quantum channels $\Phi\in\mQ(A\to B)$ having the property that $\Phi(\rho)\in\mF(B)$ whenever $\rho\in\mF(A)$. 
\end{definition}

Since the defining property of QRTs is that resource states cannot be generated from free states, it is obvious that if $\mO$ is any other assignment of free operations that is consistent with $\mF$, then it must be that $\mO\subset \mO_{\max}$ (meaning $\mO(A\to B)\subset\mO_{\max}(A\to B)$ for any input/output systems $A$ and $B$).  In this sense, $\mO_{\max}$ is justified in being called the \emph{maximal} assignment of operations for $\mF$.


One might wonder whether the maximal set of operations $\mc{O}_{\max}(A)$ is related to the minimal set of states $\mc{F}_{\min}(A)$ given by Eq. \eqref{fmin}.  In fact, there is a connection based on the discussion following Eq.~\eqref{fmin}.  From its definition, the RNG operations $\mc{O}_{\max}(A)$ for a given set of free states $\mc{F}(A)$ can transform any free state to any other.  Therefore, by the conclusion of Section \ref{Sect:Consistent-given_free_operations}, we must have $\mc{F}(A)=\mc{F}_{\min}(A)$, and it is the unique set of free states that is consistent with $\mc{O}_{\max}(A)$.

We can extend the notion of RNG maps to the setting where $\mc{H}^A$ and $\mc{H}^B$ represent subsystems of some large system.  By Definition~\ref{maindef} of a QRT, the identity map is considered free. Therefore, if $\Phi\in\mO(A\to B)$ is a free CPTP map, one would intuitively expect that $\id^C\otimes\Phi$ is also a free operation, where $\id^C$ is the identity map on $\mB(C)$.  Condition 1 of a tensor-product structure has identified any free map $\Phi$ having this property as being ``completely'' free.  Note that if $\Phi$ is not completely free, then there exists a bipartite free state $\sigma^{CA}\in\mF(CA)$ such that $\id^{C}\otimes\Phi(\sigma^{CA})$ is not free.  More generally, we can restrict the dimension of system $C$ and see whether $\id^C\otimes\Phi$ is still resource non-generating.  This leads to the following family of operational classes defined for a given set of free states.
\begin{definition}
Let $\mF$ be as in Definition~\ref{maindef}. A map $\Phi\in\mO_{\max}(\mc{H}^A\to \mc{H}^B)$ is \textit{$k$-resource non-generating} ($k$-RNG) if $\id_k\otimes\Phi\in\mO_{\max}(\mbb{C}^k\otimes\mH^A\to\mbb{C}^k\otimes\mH^B)$. Moreover, $\Phi$ is \emph{completely resource non-generating} (CRNG) if it is $k$-RNG for all $k$. We denote by $\mO_{k\max}(A\to B)$ the set of $k$-RNG operations, and by $\mO_{\text{c}\max}(A\to B)$ the set of CRNG operations.
\end{definition}

This definition generalizes the concepts of $k$-positivity and complete-positivity to QRTs. Particularly, if we take the free set $\mF(A)$ to be the set of \emph{all} density matrices acting on $\mH^A$, then maps that are $k$-RNG and completely-RNG are equivalent to maps that are $k$-positive and completely positive, respectively. Moreover, the set of $k$-RNG maps with $k=1$ is simply the set of RNG maps, and similar to $k$-positivity, it is known that if a CPTP map in $\mQ(A\to B)$ is $k$-RNG with $k\geq d_A:=\dim(\mc{H}^A)$ then it is completely RNG. We therefore have the following inclusion relations:
$$
\text{RNG}=1\text{-RNG}\supset2\text{-RNG}\supset\cdots\supset d_{A}\text{-RNG}=\text{CRNG}\;.
$$
In some QRTs the inclusions above are all equalities while in others they are all strict.  For example, if the free states $\mc{F}$ are separable states, then the set $\mc{O}_{\text{c}\max}$ is precisely the set of separable maps, while if the free states are those with a positive partial transpose (PPT), then $\mc{O}_{\text{c}\max}$ is the set of (completely) PPT-preserving maps \cite{Rains-1999a}.  In both these cases, the set $\mc{O}_{\text{c}\max}$ is strictly smaller than $\mc{O}_{\max}$.  In contrast, for the QRT of (speakable) coherence, the two sets of operations are equivalent.  Note that if a QRT $\mR=(\mF,\mO)$ admits a tensor-product structure, then any CPTP map in $\mO(A\to B)$ must be completely RNG.   


\subsubsection{Physically Implementable Operations}

\label{Sect:Physically_Implementable}

The use of CPTP maps and generalized measurements in quantum information science is so common that their physical implementations are often taken for granted.  As described in Section \ref{Sect:Preliminaries}, the Stinespring Dilation Theorem ensures that every CPTP map on system $A$ can be implemented by applying a unitary evolution on joint system $A+E$, where $E$ represents the environment that has inaccessible degrees of freedom and which is initially uncorrelated with $A$.  True as this may be, often the joint unitary identified in this theorem may not be physically implementable under the physical constraints of the QRT.  For example, in any QRT with locality constraints (such as entanglement), joint unitaries cannot be applied across the spatially separated subsystems.  This means that the only consistent unitary dilations are those that factor into independent dilations on each of the subsystems.  If no resource is drawn from the environment, the generated CPTP maps would then likewise factor into a product of independent maps.  However, if classical communication is allowed between the subsystems, then there will be free CPTP maps not having this form.  Does this mean that operations like LOCC cannot be implemented in a way that is consistent with the constraints of the QRT?  If quantum mechanics only permitted unitary evolution, then this would indeed be the case.  Yet standard quantum mechanics also allows for projective measurements as a distinct physical process, and this should be combined with the application of a joint unitary on system $A+E$ when considering physical implementations.

From a QRT perspective, it is natural to suppose that the free operations can be generated by a sequence of unitary evolutions (possibly on composite systems), projective measurements, and processing of the classical outcomes, where each element in the sequence is itself a free action. In particular, the classical processing would encompass classical communication between subsystems, if this were allowed in the QRT. Without such consistency, the QRT would identify certain maps as being free with no way to physically implement these processes using free operations \cite{Chitambar-2016c,Marvian-2016a}.  We formalize this idea in the following definition.
\begin{definition}
\label{Defn:Physically-consistent}
Let $\mR=(\mF,\mO)$ be a QRT in which appending free states and discarding subsystems are free operations.  A CPTP map $\Phi\in\mO(A\to B)$ is said to be \textit{physically implementable} if it can be decomposed into a composition of CPTP maps each acting on QC states according to
\begin{equation}
\label{Eq:physical-step}
\sum_ip_i\rho_i\otimes\op{i}{i}^X\mapsto \sum_{i,j,k}q_{k|i,j}p_i\Phi_{j|i}(\rho_i)\otimes\op{k}{k}^X,
\end{equation}
where $q_{k|i,j}$ is a family of conditional probability distributions, and each CP map $\Phi_{j|i}$ has the form
\begin{align}
\label{formpi}
\Phi_{j|i}&:\mB\left(\mc{H}^{A_i}\right)\to\mB\left(\mc{H}^{A_{j|i}}\right)\notag\\
\Phi_{j|i}&:\rho\mapsto\tr_{E_i'}\left[(\mbb{I}^{A_i'}\otimes P_{j|i})U_i\left(\rho\otimes\gamma_i\right)U_i^\dagger\right],
\end{align}
in which $\gamma_i^E\in\mc{F}(E_i)$ is a free state, $U_i\in\mO(A_iE_i\to A'_{i}E_i')$ is a free unitary, and $\{P_{j|i}\}_{j}$ constitutes a complete set of orthonormal projectors for a free projective measurement on system $E_i'$.  The resource theory $\mR$ is called \textit{physically implementable} if all the free CPTP maps are physically implementable.
\end{definition}
In this definition, it is assumed that appending free states and discarding subsystems are both free operations.  This is to ensure that the CPTP maps of Eq.~\eqref{Eq:physical-step} are indeed free. 

\begin{figure}[t]
\includegraphics[scale=.35]{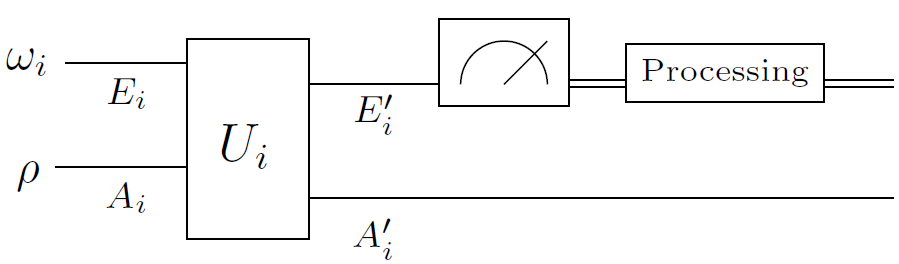}
\caption{\label{Fig:physical-implement}  A physically implementable CPTP map is one that can be realized by a sequence of channels, each having the form depicted in the figure.  The four steps - (i) appending ancilla state $\omega_i$, (ii) applying a unitary $U_i$, (iii) performing a projective measurement, and (iv) classically post-processing the measurement outcome - all must be free operations in the QRT.}
\end{figure}

In more detail, the transformation in Eq.~\eqref{Eq:physical-step} is generated as follows (see Fig. \ref{Fig:physical-implement}).  Conditioned on a classical input $i$, which might represent the outcome of a previous measurement, the experimenter introduces the free state $\omega_i$ on system $E_i$ and applies a joint unitary $U_i$ across $A_i$ and $E_i$.  A projective measurement $\{P_{j|i}\}_j$ is then performed on a subsystem $E_i'$, which may be larger or smaller than the original ancilla system $E_i$.  At this point in the process, the generated CPTP map has the form $\sum_{i,j}\Phi_{j|i}\otimes\op{j}{j}^X$.  The final step involves sending the classical register through the classical channel $\op{j}{j}\to\sum_k p_{k|i,j}\op{k}{k}$.  Doing so generates a CPTP map described by Eq.~\eqref{Eq:physical-step}. 

For a given designation of free states $\mc{F}$, it is possible to construct a unique physically implementable QRT that admits a tensor-product structure.  Simply define the free operations to be any composition of (i) appending arbitrary free states, (ii) discarding subsystems, (iii) CRNG unitaries and projective measurements, and (iv) all free classical post-processing maps.  
For the overall input/output spaces $\mH^{A}/\mH^{B}$, we denote this set of operations as $\mO_{\min}(A\to B)$.  By design, $(\mc{F},\mO_{\min})$ is physically implementable and has tensor-product structure.  Moreover, $\mO_{\min}$ is the minimal set of free operations that is consistent with $\mF$, when considering QRTs such that all the isometries in $\mO_{\max}$ are completely free.  The class $\mO_{\min}$ fits into the hierarchy of operations as follows:
$$
\mO_{\min}\subset \mO_{\text{c}\max}\subset\mO_{\max}.
$$

In a general resource theory, the set of CRNG unitaries can be strictly smaller than the set of RNG unitaries and strictly larger than the set of free unitaries.  
In several theories such as the QRT of athermality, the set of free operations is defined precisely as in Eq.~\eqref{formpi}.  One starts by identifying the set of free unitaries and then proceeds with the definition of free operations as in~\eqref{formpi} (in the case of thermodynamics, there would be no projective measurement).  However, in general, since the set $\mO_{\min}$ can be very small, this procedure often leads to a degenerate QRT where it is almost always impossible to convert one resource state to another using physically implementable operations.  This is the case in the QRT of (speakable) coherence (see Section \ref{Sect:Coherence}).

\subsubsection{Other classes of free operations} 

\label{Sect:Other_Classes_Operations}

Beyond those just discussed, there are many other types of operations that one can consider for a given set of free states.  Here we describe three more that have been explored in the literature.  The first is the class of \emph{dually resource non-generating operations}, and it has only been studied recently in QRTs.  This set of operations, denoted by $\mO_{\text{dual}}$, consists of all RNG operations for which their dual is also RNG. More precisely, $\Phi\in\mO_{\text{dual}}$ if $\Phi\in\mO_{\max}$ and for any free state $\rho$, the state $\Phi^{\dag}(\rho)/\tr[\Phi^{\dag}(\rho)]$ is also free.
Here we must normalize the state since $\Phi^\dag$ is not necessarily trace preserving.

By definition, dually RNG operations are a subset of RNG, and in certain QRTs this inclusion is strict, with dually RNG operations being a subset of even CRNG operations. 
For example, in the QRT of (speakable) coherence RNG=CRNG whereas the set of dually RNG operations coincides with the set of dephasing covariant incoherent operations~\cite{Chitambar-2016c,Marvian-2016a}, which is a strict subset of CRNG. In entanglement theory, on the other hand, all separable maps are CRNG, and it is simple to see that the dual of a separable map is also separable.  In fact, the set CRNG in this theory is precisely the set of separable maps, and it can be shown to be a strict subset of the set of dually non-entangling operations~\cite{Chitambar-2018b}.  In general, dually RNG operations provide a better approximation to the physically-motivated set of free operations than the full set of RNG operations.  Their usefulness arises in QRTs where the problem of state interconversion can be solved by semi-definite programming (SDP) for RNG operations but not the physically-motivated ones (this occurs, for example, in some models of coherence). In such QRTs, state interconversion under the dually RNG operations can also be solved with SDP, thereby providing a better approximation of what is feasible using the physically-motivated free operations.

The replacement channel $\Phi(X)=\tr[X]\sigma$ with $\sigma\in\mF(A)$ is RNG. However, it is not necessarily dually RNG. To see why, observe that its dual is given by $\Phi^\dag(Y)=\tr[Y\sigma]\mbb{I}$. Therefore, the replacement channel is dually-RNG if and only if the maximally mixed state is considered free in the resource theory. 

Another family of free operations is \emph{stochastically resource non-generating} operations, which we denote by $\mO_{\text{stochastic}}$.  This set of operations has been used heavily in the QRT of coherence \cite{Baumgratz-2014a}, and it is defined in terms of a Kraus operator decomposition.  A CPTP map $\Phi\in\mc{O}(A\to B)$ belongs to 
$\mO_{{\text{stochastic}}}$ if there exists an operator-sum representation $\Phi(\cdot)=\sum_jK_j(\cdot)K_{j}^{\dag}$ such that for any free state $\rho\in\mc{F}(A)$, it holds that
\begin{equation}
\label{Eq:stochasticRNG}
\frac{K_j\rho K_{j}^{\dag}}{\tr\left[K_j\rho K_{j}^{\dag}\right]}\in\mc{F}(B)\qquad\forall j.
\end{equation}
That is, any element in this particular operator-sum representation of $\Phi$ induces a resource non-generating transformation.  This is a much stronger requirement than $\Phi$ being RNG, and often this strengthening makes $\mO_{\text{stochastic}}$ simpler to work with.  Both LOCC and separable operations in entanglement theory are stochastically RNG.  
In terms of CPTP maps, stochastically RNG operations can be modeled by appending a classical register to each resource non-generating Kraus operator $K_j$ and summing over all outcomes, $\Phi(\cdot)=\sum_j K_j(\cdot)K_j^\dagger\otimes\op{j}{j}^X$.  This CPTP map is sometimes called a ``heralded'' or ``flagged'' measurement since the outcome $j$ can always be recovered after applying $\Phi$ by measuring the classical system $X$.  

A final class of operations for a given set of free states $\mc{F}$ actually violates the Golden Rule of QRTs.  Nevertheless, as discussed in Section \ref{Sect:Smooth_Entropies}, these operations have proven to be quite useful in the study of asymptotic resource convertibility.  For a fixed $\epsilon>0$, the class of \textit{$\epsilon$-resource generating} ($\epsilon$-RG) operations is the set of CPTP maps belonging to $\mQ(A\to B)$ such that 
\begin{equation}
\sup_{\rho\in\mc{F}(A)}\mc{R}_{\text{rob}}(\Phi(\rho))<\epsilon,
\end{equation}
where $\mc{R}_{\text{rob}}(\omega)=\min_{\sigma\in\mc{S}(\mc{H})}\{s\geq 0\;|\;\frac{\omega+s\sigma}{1+s}\in\mc{F}(\mc{H})\}$
is the \textit{resource robustness} for $\omega\in\mc{S}(\mc{H})$ \cite{Brandao-2015a} (see also Section \ref{Sect:Robustness}).

\subsection{Types of QRTs}

Definition~\ref{maindef} provides the general definition of a QRT, and it imposes very little structure on the theory.  We immediately introduced the tensor-product structure since it embodies a collection of highly natural properties that are possessed by almost all QRTs studied in literature.  In this section we review other types of mathematical structures that can arise in QRTs that are independent of the tensor-product structure.

\subsubsection{Convex resource theories}

\label{Sect:Convex}

Convexity is a very convenient mathematical property.  A QRT $\mc{R}=(\mc{F},\mc{O})$ is called \textit{convex} if $\mO(A\to B)$ is convex for any choices of Hilbert spaces (i.e., $p\Phi+(1-p)\Lambda\in\mO(A\to B)$ for any $\Phi,\Lambda\in\mO(A \to B)$, $p\in[0,1]$, and arbitrary $\mH^A$ and $\mH^B$).  In our formulation of QRTs, the free states $\mc{F}(\mc{H})=\mc{O}(\mbb{C}\to\mc{H})$ are defined as a special case of the free operations, and so convexity of the free operations implies that the set of free states $\mc{F}(\mc{H})$ is convex for every $\mc{H}$.  
The converse however is not true in general: a convex set of free states does not imply a convex set of free operations.  On the other hand, if the set of free states is convex, then any convex combination of free operations is a RNG map. In particular, the set of RNG maps is convex if and only if the set of free states is convex.

Many QRTs such as entanglement, coherence, asymmetry, and athermality, are all convex. The very rich mathematical tools from convex analysis can therefore be applied to these resource theories.  This is briefly discussed in Section \ref{Sect:Convex_analysis}.

There are subtleties in associating physical meaning to the convexity of a QRT.  It is often argued that convex QRTs do not allow for the generation of resource simply through the act of forgetting.  Specifically, suppose a free state $\rho_i$ is prepared with probability $p_i$, for $i=1,\cdots,n$.  Then, as the reasoning goes, forgetting which state was prepared leads to the mixture $\sum_{i}p_i\rho_i$, which should also be a free state.  However, as pointed out by Plenio and Virmani, this model for forgetting information is too simplistic \cite{Plenio-2007a, Plenio-2005a}.  First, the described transformation of an ensemble of states $\{\rho_i,p_i\}_{i}$ to the mixed density matrix $p\rho+(1-p)\sigma$ is not a CPTP map, and it therefore lies outside the resource-theoretic framework described here.  But more importantly, ``forgetting information'' involves discarding classical information, and this process needs to be considered.  A more precise model begins by describing a probabilistic state preparation through the introduction of some classical randomness $\sum_i p_i\op{i}{i}^X$, which is a density matrix diagonal in the computational basis $\{\ket{i}\}_i$.  State preparation maps are then performed, conditioned on the classical system $X$.  This leads to the QC state $\sum_i p_i \rho_i^A\otimes\op{i}{i}^X$.  The act of forgetting is then modeled by discarding the classical system $X$, which corresponds to the transformation
\begin{equation}
\label{Eq:Forget-info}
\sum_i p_i \rho_i^A\otimes\op{i}{i}^X\to\sum_i p_i \rho_i^A
\end{equation}
\cite{Plenio-2007a, Plenio-2005a}.  Thus, if the the classically ``flagged'' mixture $\sum_i p_i \rho_i^A\otimes\op{i}{i}^X$ is free, one naturally expects that the ``unflagged'' mixture $\sum_i p_i \rho_i^A$ is also free.  In this model, the intuition that ``forgetting should not generate quantum resource'' is made precise in the condition that ``discarding classical information does not generate a quantum resource.'' 

While every ensemble-to-mixture transformation 
\begin{equation}
\{\rho_i,p_i\}_{i}\to\sum_ip_i\rho_i
\end{equation}
is resource non-generating if and only if the QRT is convex, whether or not the transformation in Eq.~\eqref{Eq:Forget-info} is resource non-generating has nothing to do with the QRT being convex.  Rather, it depends entirely on whether discarding classical systems is a free operation in the theory.  Nevertheless, in most non-convex QRTs discarding classical systems can indeed be resource generating, which explains why non-convexity is often associated with the phenomenon of ``resource generation by forgetting,'' even though the precise model described here is usually overlooked. 


\subsubsection{Affine resource theories}

Many QRTs such as quantum thermodynamics, coherence, and asymmetry satisfy a stronger condition than convexity, which can be described as the \emph{affine} condition.  Recall that $\mF(A)$ is convex if any convex combination of free states $\sigma_j\in\mc{F}(A)$, i.e., $\sum_{j}t_j\sigma_j$, is itself free.  Here the numbers $t_j$ are all non-negative and sum to one.
In an affine combination of states, the non-negativity condition on the coefficients is relaxed while still requiring that $\sum_j t_j=1$.  As a result, affine combinations of states are not necessarily positive semi-definite, but their trace is still one.

A QRT $\mc{R}=(\mc{F},\mc{O})$ is called \textit{affine} if, for any two physical systems $A$ and $B$, any CPTP map that can be written as an affine combination of elements in $\mO(A\to B)$ is itself in $\mO(A\to B)$.  Like convexity, this definition implies that the set of free states is closed under affine combinations that result in a valid density matrix.  That is, if $\sigma=\sum_{j}t_j\sigma_j\in\mS(A)$ with $t_j\in\mbb{R}$ and $\sigma_j\in\mF(A)$, then $\sigma\in\mF(A)$.  This property alone does not ensure that the QRT is affine.  However, the set of free states is affine if and only if the set of RNG maps is affine, as well as the set of dually-RNG operations. Finally, we remark that the set of free operations is affine if and only if their Choi matrices form an affine set.


In entanglement theory, \textit{every} bipartite quantum state (even an entangled one) can be written as an affine combination of pure product states.  Therefore, in this sense, entanglement theory is maximally non-affine. On the other hand, the sets of free states in the QRTs of quantum thermodynamics, coherence, and asymmetry, are all affine.  In thermodynamics it is affine because there is only one free state, namely, the Gibbs state, while in the QRT of coherence it is affine because the free states are all the diagonal states and any linear combination of diagonal states is still diagonal.


As exemplified by entanglement theory, the affine condition is much stronger than convexity. Consequently, certain mathematical tools used to study affine QRTs cannot be applied to convex QRTs \cite{Gour-2016a}. If a QRT has an affine free set of states, but non-affine free operations, then one can replace the free operations with an affine one (e.g. the set of RNG or dually-RNG) to obtain a fully affine QRT.  This can often provide a good approximation to the original QRT.  

\subsubsection{QRTs with a resource-destroying map}

\label{Sect:RDM}

While the structure of a general affine QRT is simpler than a convex one, it can still be very rich and complex thereby making certain information-theoretic tasks difficult to analyze. In fact, many of the highly studied QRTs (such as quantum thermodynamics, coherence, asymmetry) have an additional structure that is not captured by the affine condition.  These theories all possess what is called a resource-destroying map~\cite{Zi-Wen-2017a}.

\begin{definition}
Given a QRT $\mR=(\mF,\mO)$ and Hilbert space $\mH$, a \emph{resource-destroying map} is a (not necessarily linear) map $\Delta:\mB(\mH)\to\mB(\mH)$ with the following two properties:
\begin{enumerate}
\item It maps any free state $\rho\in\mF(\mH)$ to itself; i.e., $\Delta(\rho)=\rho$.
\item It maps any (possibly not free) state $\rho\in\mS(\mH)$ to a free state; i.e., $\Delta(\rho)\in\mF(\mH)$.
\end{enumerate}
\end{definition}

From its definition, it is not clear that a resource-destroying map exists for a given QRT. However, for quantum correlations (specifically quantum discord), \textcite{Zi-Wen-2017a} describe a non-linear resource-destroying map and study some of its applications in the QRT of quantum discord.  They further show that non-linearity is a necessary condition of a resource-destroying map in any non-convex QRT, such as quantum discord.  In contrast, for the QRT of coherence, a linear resource-destroying map indeed exists and plays a central role in the theory.  Namely, it is the completely dephasing map which removes all the off-diagonal terms from the input density matrix (with respect to some fixed basis).  

One necessary condition for the existence of a (linear) CPTP resource-destroying map is that the set of free states $\mF(\mH)$ be affine~\cite{Gour-2016a}. This follows from the simple observation that if $\sigma=\sum_jt_j\sigma_j\geq 0$ is an affine combination of free states then $\Delta(\sigma)=\sum_jt_j\Delta(\sigma_j)=\sum_jt_j\sigma_j=\sigma$, which implies that $\sigma$ is free since the output of a resource-destroying map is always free. However, the affine condition alone is not sufficient to determine if there exists a CPTP resource-destroying map. Particularly, there exists affine QRTs that do not have a CPTP resource-destroying map.  The full necessary and sufficient conditions for the existence of a CPTP resource-destroying map were derived in~\cite{Gour-2016a}.

\subsubsection{Non-Convex resource theories}

While convexity is a mathematically convenient property to have, there are many QRTs that are not convex. For example, consider a bipartite quantum system consisting of two subsystems $A$ and $B$ held by Alice and Bob, respectively. If the parties cannot communicate (not even classically) due to some constraint, then their physical capabilities amount to applying local quantum channels $\Phi$ and $\Lambda$, resulting in an overall CPTP map of the form $\Phi\otimes\Lambda$. Therefore, in this scenario every allowed operation is a tensor product of two CPTP maps, and similarly every free state is a tensor-product state, $\rho^A\otimes\rho^B$. The sets of free states and free operations are not convex, which makes this QRT mathematically difficult to study. The resources in this model are bipartite states having either classical or quantum correlations. It is therefore a QRT of total correlations, distinguished from resource theories of \emph{quantum} correlations, such as discord.

\begin{figure}[t]
\includegraphics[scale=0.30]{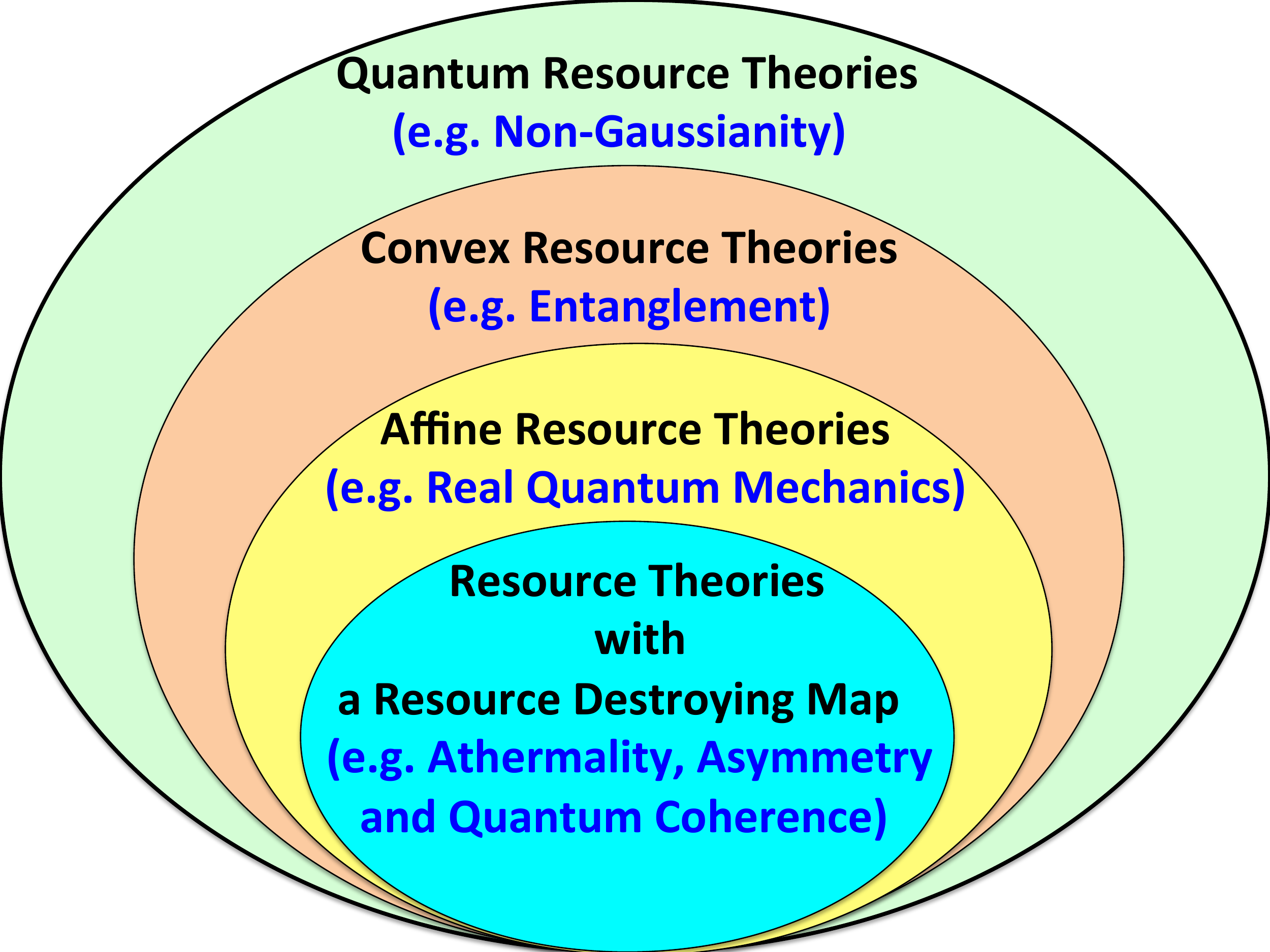}
\caption{An heuristic diagram of QRTs, classified according to the properties of their set of free states. Non-Gaussianity is an example of a QRT with non-convex set of free states. Entanglement theory is an example of a QRTs that is convex but not affine. Real (vs complex) quantum mechanics ~\cite{Hickey2018} is an example of an affine QRT that does not have a resource-destroying map, and athermality, asymmetry, and coherence, are examples of QRTs with a resource-destroying map.}
\label{fig3}
\end{figure}

There are other important non-convex QRTs. For example, in quantum optics, Gaussian operations are relatively easier to implement than non-Gaussian operations, and therefore one can construct a QRT in which Gaussian states and Gaussian operations are free. This QRT is neither convex nor finite-dimensional, making results relatively difficult to obtain (see Section~\ref{Sect:Non-Gaussianity} for recent progress).

One general strategy for studying a non-convex theory is to enlarge the set of free states and free operations by taking their convex hulls.  Then the standard techniques of convex analysis can be employed.  In some cases, like the QRT of total correlations, this relaxation is too strong, and all states become free.  On the other hand, this strategy leads to non-trivial results in the QRT of non-Gaussianity \cite{Lami2018, Zhuang2018}.

\subsubsection{Resource Theories of Quantum Processes}

\label{Sect:processes}

So far we have discussed QRTs involving interconversions among resource states. That is, the resources have been identified as all states not belonging to the set of free states $\mF(A)$.  However, quantum states are \textit{static} objects, and not all resources are static in nature.  To generalize the concept of resource, recall first that every resource state can be identified as a special type of CPTP map $\Psi:\mc{B}(A)\to\mc{B}(B)$, in which $\mc{B}(A)=\mbb{C}$.  More generally, one then constructs a QRT in which the resources are CPTP maps $\Psi:\mc{B}(A)\to\mc{B}(B)$ for different input spaces $\mc{B}(A)$.  We will say that $\Psi$ is a \textit{dynamical} resource if $d_A=\dim(\mH^A)>1$. 

Already in its early days, see e.g.,~\cite{Bennett-2004a,Dev2004,Devetak-2008a,Wilde2013}, it was recognized that the whole field of quantum information can be viewed as a theory of interconversions among different resources, where the resources can be classified as being static or dynamic, classical or quantum, noisy or noiseless. This diverse classification of resources leads to the very rich field of quantum Shannon theory.

From this more general perspective in which resources are not limited to quantum states, any QRT must also specify what type of transformations a given dynamical resource $\Psi$ can undergo.  To answer this question, we first need to understand what is the most general yet still physical transformation that a quantum channel can undergo. To answer this question, let $\mB(A\to B)$ be the space of all linear maps from $\mB(A)$ to 
$\mB(B)$. The space $\mB(A\to B)$ is itself a (finite-dimensional) Hilbert space with inner product given by 
$$
\la\Phi,\Psi\ra\eqdef\tr\left[(J_{\Phi})^{\dag}J_\Psi\right],
$$
for all linear maps $\Psi,\Phi:\mB(A)\to\mB(B)$ having Choi matrices $J_{\Phi}$ and $J_\Psi$ respectively. Similarly, let $\mB(C\to D)$ to be the Hilbert space of all linear maps from $\mB(C)$ to $\mB(D)$, and let $\Theta:\mB(A\to B)\to\mB(C\to D)$ be a linear map. Note that the spaces $\mB(A\to B)$ and $\mB(C\to D)$ contain the subsets of all CPTP maps. Therefore, if $\Theta$  represents a physical transformation, then we must have $\Theta[\Psi]\in\mQ(C\to D)$ if $\Psi\in\mQ(A\to B)$. In this case we say that $\Theta$ is positive. Moreover, $\Theta$ is said to be completely positive if $\1^{A'B'}\otimes\Theta:\mB(A'\to B')\otimes\mB(A\to B)\to\mB(A'\to B')\otimes\mB(C\to D)$ is positive for all systems $A'$ and $B'$.  Here we used the symbol $\1^{A'B'}$ to denote the identity map from $\mB(A'\to B')$ to itself.  Since transformations restricted to subsystems are physically possible, we conclude that any physical transformation $\Theta:\mB(A\to B)\to\mB(C\to D)$ must be completely positive, and we call these objects superchannels.

\begin{figure}[h]
\includegraphics[scale=0.40]{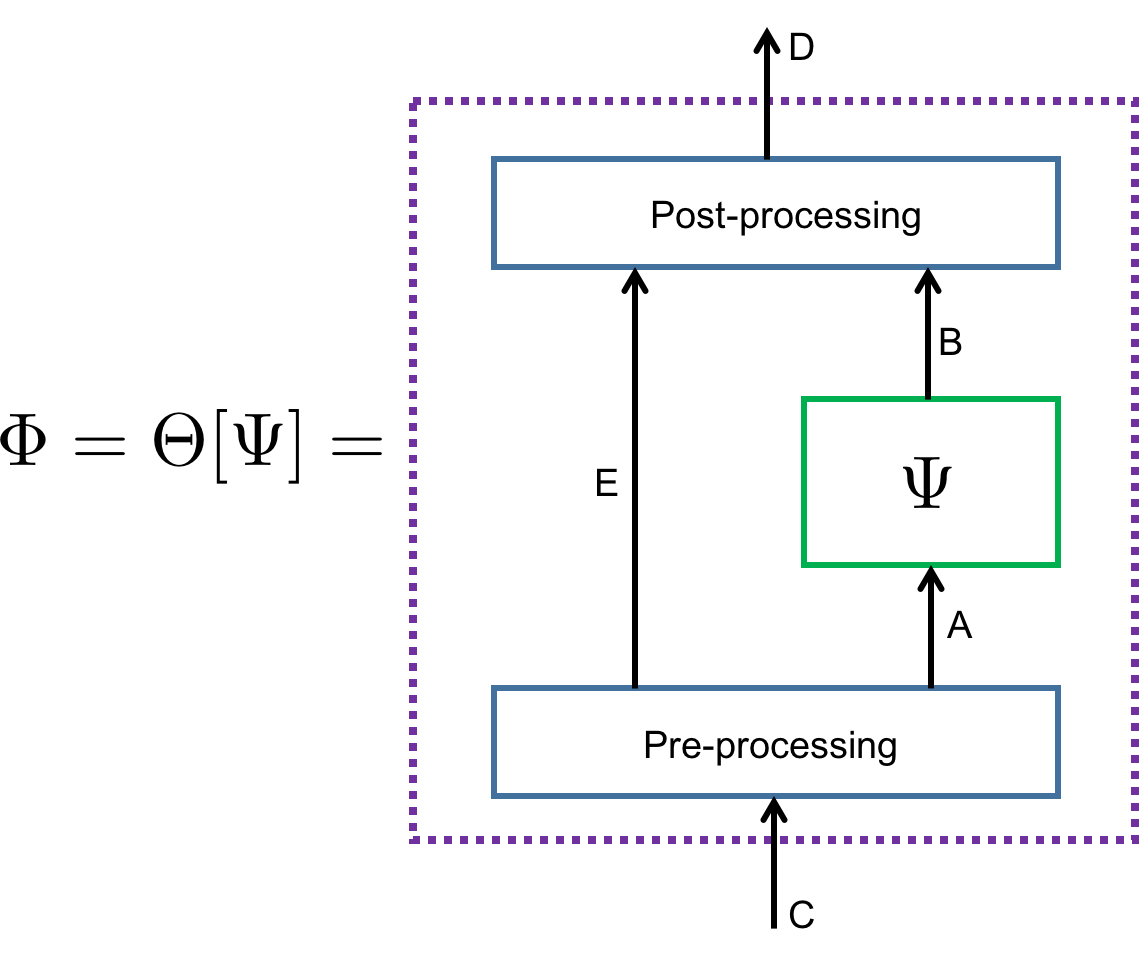}
\caption{Realization of a superchannel.}
\label{fig2}
\end{figure}

In~\cite{Pavia1} it was shown that any superchannel $\Theta:\mB(A\to B)\to\mB(C\to D)$ can be realized as follows.  Denoting $\Phi^{C\to D}\eqdef\Theta[\Psi^{A\to B}]$ for an arbitrary input $\Psi^{A\to B}$, the action of $\Theta$ decomposes as
\be\label{superchannel}
\Phi^{C\to D}=\Gamma_{\text{post}}^{BE\to D}\circ\left(\Psi^{A\to B}\otimes\id^E\right)\circ\Gamma_{\text{pre}}^{C\to AE}
\ee
where $\Gamma_{\text{pre}}^{C\to AE}\in\mQ(C\to AE)$ is a pre-processing CPTP map, $\Gamma_{\text{post}}^{BE\to D}\in\mQ(BE\to D)$ is the post-processing CPTP map, and system $E$ corresponds to a possible side channel (see Fig.~\eqref{fig2}) \footnote{In fact, $\dim(E)$ can be taken to be no greater than $\dim(A)\cdot\dim(C)$, and $\Gamma_{\text{pre}}$ can also be taken to be an isometry.}.

Fig.~\eqref{fig2} depicts the most general evolution that a quantum channel $\Psi$ can undergo. In general, the only restriction on the pre- and post-processing is that they are quantum channels (i.e., CPTP maps). However, in each QRT, the set of free channels consists of only a subset of all CPTP maps. Hence, a dynamical resource $\Psi:\mQ(A\to B)$ can be converted by free operations to another dynamical resource $\Phi: \mQ(C\to D)$ if and only if there exists an auxiliary system $E$, a \emph{free} pre-processing map
$\Gamma_{\text{pre}}^{C\to AE}\in\mO(C\to AE)$, and a \emph{free} post-processing map $\Gamma_{\text{post}}^{BE\to D}\in\mO(BE \to D)$ 
such that the relation~\eqref{superchannel} holds. Note that we can view this interconversion as a \emph{simulation} of the quantum channel $\Phi$ with the dynamical resource $\Psi$.

The free superchannels, acting on dynamical resources as described above, cannot generate a resource from a free channel. That is, from the definition of a QRT, if $\Psi\in\mO(A\to B)$ then the map $\Phi^{C\to D}$ as given in~\eqref{superchannel} with free pre- and post-processing is also a free channel (i.e., $\Phi\in\mO(C\to D)$). However, in analogy with RNG channels, the set of all RNG superchannels (i.e., superchannels that do not generate a dynamical resource from a free resource) can, in general, be larger than the set of free superchannels. One can therefore define different types of free superchannels in analogy with those defined in the previous sections.  To the authors' knowledge, little work has been done in this direction thus far, although see \textcite{Theurer-2018a} for some recent results in the resource theory of coherence.

\begin{figure}[h]
\centering
    \includegraphics[width=0.35\textwidth]{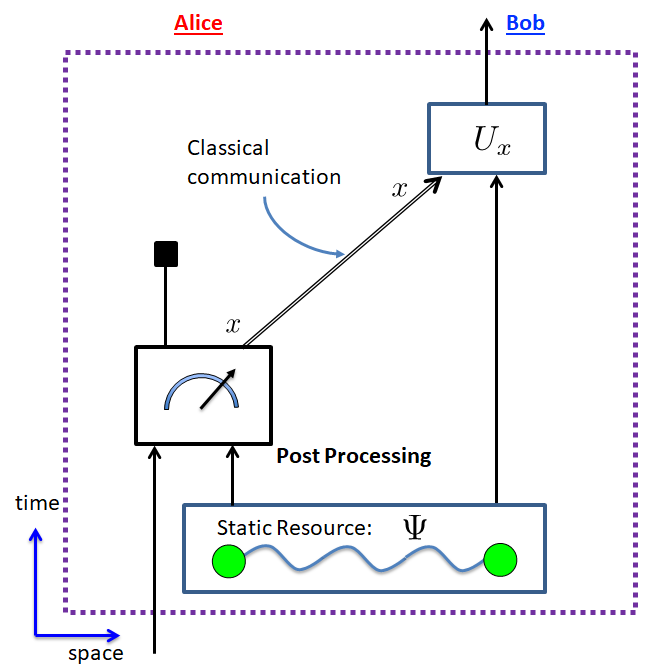}
  \caption{\linespread{1}\selectfont{\small Quantum teleportation \cite{Bennett-1993a}. Single (resp., double) line arrows corresponds to quantum (resp., classical) communication.
  The static resource, a maximally entangled state, is converted to a dynamical resource  via LOCC.}}
  \label{tele}
\end{figure}

The above formalism of superchannels enables one to consider the most general conversion of a resource channel $\Psi^{A\to B}$ into another channel $\Phi^{C\to D}$. In the special case that both $d_A=1$ and $d_C=1$, $\Psi^{A\to B}$ and $\Phi^{C\to D}$ can be viewed as density matrices, and in this case the free superchannel that achieves this transformation becomes equivalent to a free channel in $\mO(B\to D)$. Another interesting special case is that in which $d_A>1$ but $d_C=1$. That is, a dynamical resource (a quantum channel) is used to generate a static resource (a quantum state). Conversely, when
$d_A=1$ but $d_C>1$, a \emph{static} resource is used to generate a dynamical resource. Quantum teleportation is a perfect example of such an interconversion (see Fig.~\ref{tele}).

The resource $\Psi^{A\to B}$ can be classified into different types.
In~\textcite{Dev2004} a very useful notation was introduced to account for the different types of resources that are typically encountered in quantum information. 
We follow that convention here in letting the letter $c$ denote classical systems, and $q$ for quantum ones. In addition, square brackets represent noiseless resources while curly brackets indicate noisy ones.  Finally, an arrow $\to$ is used to distinguish between a dynamical resource and a static resource. With these rules, $[q\to q]$, for example, corresponds to an ideal qubit channel $\Psi^{A\to B}=\id^{A\to B}$ with $d_A=d_B=2$. The symbol
$[qq]$ stands for an ebit, i.e., a unit of a static noiseless resource consisting of a maximally entangled state of two qubits.  Similarly, $[c\to c]$ stands for a classical bit channel capable of transmitting perfectly one classical bit, and $[cc]$ corresponds to one bit of uniform shared randomness. An arbitrary noisy quantum channel $\Psi^{A\to B}$ is denoted by $\{q\to q\}$ and an arbitrary classical channel by $\{c\to c\}$.  In addition, a preparation of a quantum system is denoted by $\{c\to q\}$ and a quantum measurement by $\{q\to c\}$. 

With these symbols, all the protocols in quantum Shannon theory can be characterized as conversions between different resources~\cite{Wilde2013}. For example, the rate at which classical communication can be communicated reliably over a quantum channel is the supremum ratio $\frac{n}{m}$ such that
\be\label{qqcc}
m\{q\to q\}\geq_\epsilon n[c\to c]\qquad \forall \epsilon\in (0,1].
\ee
This inequality is to be understood as meaning that $m$ uses of the channel $\{q\to q\}$ are able to simulate $n$ uses of the channel $[c\to c]$ with $\epsilon$ error~\cite{Dev2004,Devetak-2008a}.  A comprehensive table of many different types of resource interconversions can be found in~\cite{Dev2004}.  From this perspective, the whole field of quantum Shannon theory can be viewed as a resource theory.  Continuing with the previous example, in the Holevo-Schumacher-Westmoreland (HSW) Theorem~\cite{HSW1,HSW2}, which identifies the optimal conversion in Eq. ~\eqref{qqcc}, one assumes that the side channel $E$ in Fig.~\ref{fig2} is not allowed since only local operations are free.  That is, in the setting of the HSW theorem, one can only apply coding (pre-processing) and decoding (post-processing) to many copies of a channel $\Psi^{A\to B}$, thereby converting it to many copies of a classical bit channel $\id^{X\to Y}$ (here $X$ and $Y$ stand for classical bit systems on Alice and Bob's sides, respectively).  

Another good example to illustrate the usefulness of this notation is quantum teleportation \cite{Bennett-1993a}. 
  In the process of teleportation, one ebit plus two uses of a classical one-bit channel simulate a qubit channel (Fig. \ref{tele}). In the resource calculus symbols, this can be characterized as the inequality
\be\label{r1}
[qq]+2[c\to c]\geq [q\to q]\;,
\ee
where here $\geq$ indicates zero-error simulation.  Note that this is an inequality, rather than equality, since a single use of a quantum channel cannot generate both a maximally entangled state \emph{and} two uses of a classical channel.

For superdense coding, an ebit plus one use of a quantum channel can be used to simulate two uses of a classical channel. This can be expressed as the resource inequality
\be\label{r2}
[qq]+[q\to q]\geq 2[c\to c]\;.
\ee
Note that if entanglement is not considered as a resource, that is, if the parties are supplied with unlimited singlet states, then we can remove the ebit cost $[qq]$ in~\eqref{r1} and~\eqref{r2} and get that 
$
2[c\to c]\geq [q\to q]
$
for teleportation and 
$
[q\to q]\geq 2[c\to c]
$ for superdense coding.  This makes teleportation and superdense coding dual protocols of each other, and in this case we can say that $[q\to q]= 2[c\to c]$.

In almost all practical scenarios, however, entanglement is an expensive resource that can be difficult to generate over long distances because of its high sensitivity to decoherence and noise. 
The question then whether the protocols of teleportation and superdense coding can be modified slightly to make them more symmetric, in the sense that the two resource inequalities of~\eqref{r1} and~\eqref{r2} become a single resource equality. This is indeed possible if we replace $2[c\to c]$ in the RHS of~\eqref{r2} with two uses of an isometry channel, 
denoted by $[q\to qq]$ and known as the coherent bit  (cobit) channel~\cite{Harrow,Wilde2013}.

The cobit channel is given in terms of an isometry $V^{A\to AB}$, defined with respect to some fixed basis $\{|x\ra^A\}_{x=0,1}$ as: 
$$
V|x\ra^A=|x\ra^A|x\ra^B\quad x=0,1
$$ 
Hence, the unit resource $[q\to qq]$ is highly nonlocal as it can be used to generate an ebit (i.e., $[qq]$) from the state $|+\ra^A\eqdef(|0\ra^A+|1\ra^A)/\sqrt{2}$. We therefore have
$$
[q\to qq]\geq [qq]\;.
$$
It is also straightforward to see that the cobit is more resourceful than one use of a classical bit channel; i.e.,
$$
[q\to qq]\geq [c\to c]\;.
$$
In~\cite{Harrow} it was shown that
$$
[qq]+[q\to q]=2[q\to qq]\;.
$$ 
That is, one ebit and one use of a qubit channel can be used to simulate two cobit channels (a process known as coherent superdense coding), and conversely, two cobit channels can be used to simulate a qubit channel along with one ebit (a process known as coherent teleportation). This result demonstrates that superdense coding is not the most efficient protocol since it simulates only two uses of a classical bit channel, whereas in coherent superdense coding, two cobit channels are simulated with the exact same resources.

The above discussion reflects the perspective that all of quantum (and classical) information theory can be viewed as a theory of interconversions among different types of resources.  Moreover, it reveals how a resource-theoretic framework can not only help determine if a protocol is optimal, but it can also help to motivate novel and innovative protocols. Most of the current literature has focused on QRT of states, and much less work has been conducted on resource theories of quantum processes. 
Nevertheless, among the QRTs of processes are the recent works on Markovian evolution~\cite{Waka,Plenio2012}, quantum memories~\cite{Rosset}, incompatibility of quantum measurements~\cite{GHS}, simulability of quantum measurements~\cite{Acin1,Acin2}, steering~\cite{Gallego-2015}, quantum coherence beyond states~\cite{Winter2017}, and the amortized resource of a channel in a general QRT \cite{Kaur-2018a}.


\section{Examples of Specific Resource Theories}

\label{Sect:Specific_QRTs}

\subsection{Convex Resource Theories}

\subsubsection{Entanglement}

The theory of quantum entanglement is an exemplar of a QRT.  
In this resource theory, the free operations capture the physical scenario where spatially separated parties freely exchange classical information, but all quantum information is processed locally through CPTP maps on the individual subsystems.  The global maps that can be implemented under this restriction constitute the class of LOCC, and this represents the free operations in the QRT of entanglement \cite{Bennett-1996b, Bennett-1996a, Vedral-1997a}.

The general structure of an LOCC map is quite complex.  Every LOCC operation is built from an interactive protocol in which each round of the protocol involves a local measurement by one of the parties followed by a global broadcast of the measurement outcome.  By concatenating the Kraus operators for each round, it is then not too difficult to see that every LOCC map $\Lambda$ will have the form 
\begin{equation}
\label{Eq:LOCC-form}
\Lambda(\cdot)=\sum_{k}\left(\otimes_{i=1}^N M^{A_i}_{k,i}\right)(\cdot)\left(\otimes_{i=1}^N M^{A_i}_{k,i}\right)^\dagger,
\end{equation}
where $M^{A_i}_{k,i}$ acts on the Hilbert space of party $A_i$. In other words, $\Lambda$ has a Kraus operator decomposition in which each Kraus operator is a tensor product \cite{Bennett-1999a, Donald-2002a}.  Structurally, LOCC is a physically implementable set of operations, and the dual of every LOCC map also has the form of Eq.~\eqref{Eq:LOCC-form}.  However, LOCC is not a closed set of operations in the sense that there exists a sequence of protocols, each increasing in round number, that converge to a map which cannot be implemented by either finite-round or unbounded-round LOCC \cite{Chitambar-2011a, Chitambar-2012a, Chitambar-2014a}.  
Hence the number of interactive communication rounds can also be seen as resource in the LOCC framework, and the general topic of LOCC round complexity is an active area of research \cite{Owari-2008a, Nathanson-2013a, Wakakuwa-2016a, Chitambar-2017a}.

Turning to the free states, one can easily identify $\mc{F}_{\min}(\mc{H})$, the minimal set of free states for LOCC on $N$-partite state space $\mc{H}=\bigotimes_{k=1}^N\mc{H}^{A_k}$.  From its definition in Eq.~\eqref{fmin}, this set consists of all density matrices on $\mc{H}$ that can be generated from any other state using LOCC.  Since LOCC involves local operations coordinated by global classical communication, the transformation $\sigma^{A_1\cdots A_N}\to\rho^{A_1\cdots A_N}$ is achievable by LOCC for any $\sigma^{A_1\cdots A_N}$ and any $\rho^{A_1\cdots A_N}$ of the form
\begin{equation}
\label{Eq:SepState}
\rho^{A_1\cdots A_n}=\sum_{k=1}^np_k\rho_{1,k}^{A_1}\otimes \rho_{2,k}^{A_2}\otimes\cdots\otimes\rho_{N,k}^{A_N},
\end{equation} 
where $\rho_{i,k}^{A_i}$ is an arbitrary state for party $A_i$.  Indeed, the parties can just discard the state $\sigma$ and then locally generate their respective state according to a globally shared probability distribution $p_k$ \cite{Werner}.  Any state having the form of Eq.~\eqref{Eq:SepState} is called \textit{separable}, and we denote the set of separable states by \text{SEP$(\mc{H})$}, where $\mc{H}$ has a fixed tensor-product structure.  Furthermore, from Eq.~\eqref{Eq:LOCC-form}, it can be directly seen that LOCC leaves \text{SEP$(\mc{H})$} invariant; hence $\mc{F}_{\min}(\mc{H})=\text{SEP$(\mc{H})$}$.  Any state not belonging to $\text{SEP$(\mc{H})$}$ is called \textit{entangled}.

The set $\text{SEP$(\mc{H})$}$ is closed, a fact that follows from the continuity of certain entanglement measures, such as the entanglement of formation \cite{Nielsen-2000b}.  Also, since every convex combination of separable states is again separable, entanglement theory is a convex QRT.  However, it is non-affine.  In fact, it is maximally non-affine in the sense that \textit{every} state can be expressed as an affine combination of free states.  This can be seen by taking an Hermitian tensor product basis of $\mc{B}(A_1A_2\cdots A_N)$ and then spectrally decomposing each of the Hermitian operators into a linear combination of eigenspace projectors.

Deciding membership of $\text{SEP$(\mc{H})$}$ is a notoriously difficult problem.  In fact, the problem is NP-Hard \cite{Gurvits2003, Gharibian-2010a}.  To cope with this difficulty, one strategy involves relaxing the separability constraint to encompass a more computationally manageable and experimentally verifiable set of states.  Traditionally this approach has consisted in identifying separability criteria, which are necessary (but not) sufficient conditions for a state to be separable \cite{Horodecki-2009a}.  The most famous separability criterion is positivity of the partial transpose (PPT) \cite{Peres-1996a}, which says that $\rho^{\Gamma_i}\geq 0$ for any separable state $\rho$, where $\Gamma_i$ indicates a partial transpose on system $A_i$.  Satisfying the PPT criterion is also sufficient for separability in $2\otimes 3$ and $3\otimes 2$ bipartite systems \cite{Horodecki-1996a}.  In higher dimensions, PPT entangled states are known to exist and are called ``bound'' entangled states since no pure-state entanglement can be asymptotically distilled from them (see Section \ref{Sect:Tasks-Asymptotic}) \cite{Horodecki-1998a}.  One of the most prominent open problems in entanglement theory is determining whether the converse is true; i.e., whether non-PPT (NPT) bound entangled states exist.

The complexity of LOCC presents formidable challenges for understanding its precise operational capabilities.  For example, it is currently unknown how to decide whether a given map $\Lambda$ belongs to LOCC.  While having the form of Eq.~\eqref{Eq:LOCC-form} is a necessary condition for LOCC maps, it is not sufficient.  Any map having tensor-product Kraus operators belongs to the class of \textit{separable} maps \cite{Vedral-1997a, Rains-1997a}, and LOCC represents a strict subset of the separable maps.  Separable maps that cannot be implemented by LOCC were originally described as demonstrating ``nonlocality without entanglement'' \cite{Bennett-1999a}.  This expression is perhaps best understood from a QRT perspective.  For the set of separable states $\text{SEP$(\mc{H})$}$, separable operations are precisely the class of completely RNG operations, $\mc{O}_{\text{c}\max}(\mc{H})$ \cite{Cirac-2001a}.  Therefore, ``nonlocality without entanglement'' in this context refers to non-LOCC maps lacking the ability to generate entanglement, even when acting on one part of some larger system.

One can move beyond separable maps to consider other classes of free operations that are consistent with the set of separable states.  Even though LOCC is the physically-motivated class of operations in the study of entanglement, its operational power is sharply limited when the number of parties exceeds two.  Specifically, for a generic state $\ket{\psi}^{A_1A_2\cdots A_N}$ with $N>3$, the set of states into which it can transform using LOCC constitutes a measure zero set in state space \cite{Sauerwein-2017a}.  Hence to obtain interesting QRTs, one needs to consider free operations more powerful than LOCC.  

Starting with the set $\mc{O}_{\max}$, the RNG or \textit{non-entangling} operations consist of all CPTP maps that map separable states to separable states.  Non-entangling operations were first proposed by \textcite{Harrow-2003a} in the study of quantum computation.  Later, Brand{\~{a}}o and Plenio invoked such maps and their asymptotic variant in the study of entanglement reversibility \cite{Brandao-2008a, Brandao-2010a}.  Non-asymptotic studies on the power of general non-entangling maps have been conducted by \textcite{Brandao-2011a} and \textcite{Chitambar-2018b}.  The simplest example of a bipartite non-entangling quantum operation that has no LOCC implementation is the SWAP operator $\mbb{F}$.  Since $\mbb{F}\ket{\alpha}^A\ket{\beta}^B=\ket{\beta}^A\ket{\alpha}^B$ for any two states $\ket{\alpha}$ and $\ket{\beta}$, the map is clearly non-entangling.  However, by considering $\mbb{F}$ on subsystems $A$ and $B$ in the unentangled state $(\ket{00}+\ket{11})^{A'A}\otimes(\ket{00}+\ket{11})^{B'B}/2$, we see that entanglement is generated across the bipartite cut $A'A:B'B$.  In fact, SWAP is not even a $2$-RNG operation.  

Other types of relaxations on LOCC can be obtained by considering the Choi matrix.  In the study of multipartite QRTs, it is helpful to define the Choi matrix of a map $\Lambda:\mc{B}(\mc{H}^{A_1}\otimes\cdots\mc{H}^{A_N})\to \mc{B}(\mc{H}^{A'_1}\otimes\cdots\mc{H}^{A_N'})$ in terms of the underlying tensor product structure:
\begin{equation}
J_\Lambda=\id^{A_1\cdots A_k}\otimes \Lambda[(\phi^+)^{A_1'A_1}\otimes\cdots\otimes(\phi^+)^{A_k'A_k}]
\end{equation}
Because separable operations are completely RNG, the Choi matrix of a CPTP map $\Lambda$ is separable, and the converse is also true \cite{Cirac-2001a}.  Consequently, any separability criterion on the set of density matrices can be applied to the Choi matrix $J_\Lambda$ as a separable criterion on the set of CPTP maps.  For example, a condition of $k$-symmetric extendibility has been integrated on the level of maps to study entanglement distillation and quantum communication \cite{Pankowski-2013a, Kaur-2018b}.  Similarily, the use of entanglement witnesses on the level of Choi matrices was recently investigated in \cite{Chitambar-2018b}.  

The operations most frequently employed beyond LOCC are the so-called PPT operations.  As originally defined by \textcite{Rains-1999b}, a CPTP map is called PPT (or completely PPT-preserving) if its Choi matrix is PPT for any party $A_i$. 
This is equivalent to the condition that $\Gamma_{i}\circ\Lambda\circ\Gamma_{i}$ is completely positive, where $\Gamma_i$ is the partial transpose map for party $A_i$.  A strictly weaker condition is that the map be PPT-preserving, i.e., $\Lambda(\rho)^{\Gamma_i}\geq 0$ whenever $\rho^{\Gamma_i}\geq 0$.  The distinction between these two operational classes can be elegantly characterized by considering a QRT of NPT entanglement.  Letting the free states be all density matrices with positive partial transpose, then the PPT-preserving maps correspond to the set of RNG operations, while Rains' original PPT maps correspond to the set of completely RNG operations \cite{Horodecki-2001a, Matthews-2008a}.  The QRT of NPT entanglement under completely PPT-preserving operations has been studied in the literature \cite{Audenaert-2003a, Ishizaka-2005a, Matthews-2008a, Wang-2017b}, primarily motivated by the insights it offers on the nature of entanglement and the limitations of LOCC.  

\subsubsection{Quantum Reference Frames and Asymmetry}

\label{Sect:Asymmetry}

In Shannon theory, information is modeled as having a \emph{fungible} nature.  Information can be encoded into any degree of freedom of any physical system, and the information content is independent of the choice of encoding.  For example, a simple yes/no message can be equivalently transmitted through a 5/0-volt potential difference across a circuit element, or through a heads/tails orientation of a coin.  Information of this sort is called \textit{speakable information}, and it is characterized by its ability to be communicated verbally or through a string of symbols.  

On the other hand, there also exists non-fungible types of information like a direction in space, the time of some event, or the relative phase between two quantum states in a superposition state.  Information of this sort is called \textit{unspeakable information} since it cannot be communicated verbally without first having a shared coordinate system, a synchronized clock, or a common phase reference \cite{Peres-2002a, Bartlett-2007a, Chiribella-2012a}.  For example, in the absence of, say, a common gravitational field or stellar background, directional information can be transmitted between two parties only through the exchange of some physical system whose state represents the direction itself, such as a classical gyroscope.  

Unspeakable information becomes speakable in the presence of a reference frame~\cite{Bartlett-2007a}, and this applies to both classical and quantum information.  Furthermore, even though speakable information is fungible, two or more parties must first establish how this information is to be encoded/decoded in some physical system, and this implicitly requires a common reference frame.  Thus, one always assumes a shared reference frame in the background of any quantum information processing task, and the absence of this greatly limits what can be accomplished.

A lack or degradation of a shared reference frame is therefore a natural constraint that often arises in the physics of multiple systems.  As such, it leads to a \textit{resource theory of reference frames}.  For simplicity, let us consider two parties (Alice and Bob) who do not share a reference frame, and we mathematically represent the information about the frame by an element $g$ of a compact group $G$.  For instance, $g\in G$ could correspond to a particular orientation in space, clock synchronization, phase information, etc.  Each element $g\in G$ is represented by a unitary matrix $U_g$ such that if $\rho\in\mS(\mH)$ is the density matrix assigned to some quantum system relative to Alice's reference frame, then 
\begin{equation}
\label{Eq:Unitary_conjugation}
\mU_g(\rho)\eqdef U_g\rho U_{g}^{\dag}
\end{equation}
is the state of the same physical system as described in Bob's frame.  On the other hand, since Alice lacks the information of $g$, her description of Bob's density matrix is obtained by averaging over all the possible values of $g$. Denoting by $dg$ the uniform Haar measure over the group G, this average can be expressed as:
$$
\mG(\rho)\eqdef\int dg\;\mU_g(\rho)\;.
$$ 
The averaging CPTP map $\mG$ is called the $G$-twirling map. If the group $G$ is finite then the integral is replaced with a discrete sum over the $|G|$ elements of the group; i.e., $\mG(\rho)=\frac{1}{|G|}\sum_{g=1}^{|G|}\mU_g(\rho)$. The lack of a shared reference frame hence imposes a restriction on what type of states Alice can prepare relative to Bob's reference frame.  Specifically, she can prepare states only of the form $\mG(\rho)$, and these constitute the free states in the QRT of reference frames,
\begin{equation}
\mF(\mH)\eqdef\left\{\mG(\rho)\;:\;\rho\in\mS(\mH)\right\}\;.
\end{equation}

The free states in this QRT have a very particular structure. First, note that $\rho\in\mF(\mH)$ if and only if it is G-invariant, meaning that $\mU_g(\rho)=\rho$ for all $g$.  In particular, $\mG(\rho)=\rho$ for all $\rho\in\mc{F}(\mc{H})$.  Combining this with the definition of $\mc{F}(\mc{H})$ implies that $G$-twirling is a resource-destroying map (see subsection~\ref{Sect:RDM}).  Additionally, one can characterize the above free states using 
techniques from representation theory.  This is accomplished by decomposing the underlying Hilbert space $\mH$ into its irreducible representations (irreps) of $\mc{G}$~\cite{Bartlett-2007a}:
$$
\mH=\sum_q\mH_q\equiv\sum_q\mathcal{M}_q\otimes\mathcal{N}_q
$$
where $q$ labels the irreps of $G$ (with the sum over $q$ being a direct sum), and $\mathcal{M}_q$ and $\mathcal{N}_q$ are the $q$-representation space and the $q$-multiplicity space, respectively.
Note that $\mathcal{M}_q$ and $\mathcal{N}_q$ are virtual (mathematical) subspaces which do not correspond to physical subsystems. With this notation, any free state has the form
\be\label{super1}
\mG(\rho)=\sum_{q}\Delta_{\mathcal{M}_q}\otimes\id_{\mathcal{N}_q}\left(\Pi_{\mH_q}\rho\Pi_{\mH_q}\right)\;,
\ee
where $\Delta_{\mathcal{M}_q}$ is the completely decohering map in the space $\mathcal{M}_q$, $\id_{\mathcal{N}_q}$ the identity map on the space $\mathcal{N}_q$, and $\Pi_{\mH_q}$ the projection onto $\mH_q$.

As an example, consider the group $G=U(1)$ that corresponds to an optical
phase reference. In this case, the unitary representation of $G$ is given by $U_{\theta}=e^{i\hat{N}\theta}$, where $\theta\in U(1)$ and $\hat{N}$ is the total number operator.
All the irreps of $U(1)$ are one-dimensional and can be labeled by the eigenvalues of $\hat{N}$ (i.e., nonnegative integers). In this case, $q= n\in\mbb{N}$, and the action of $G$-twirling on a pure state $|\psi\ra=\sum_n\sqrt{p_n}|n\ra$ therefore produces the state:
\begin{equation}\label{super2}
\mathcal{G}(|\psi\lr\psi|)=\sum_{n}p_n|n\lr n|\;.
\end{equation}

More generally, we can see from the forms of the free states in Eqs.~\eqref{super1}--\eqref{super2} that the lack of a shared reference frame imposes a \textit{superselection rule} on the type of states that Alice can prepare.
This superselection rule is manifested by the fact that coherent superposition between states in different irrep-subspaces $\mathcal{H}_q$ and $\mathcal{H}_{q'}$ are not possible. For example, with $U(1)$ states in a coherent superposition among eigenstates of the number operator are not free and cannot be prepared by Alice. 

The set of free operations in the QRT of reference frames can be defined in a way similar to the free states.  Consider an arbitrary density matrix $\sigma\in\mc{S}(\mc{H})$ of system $\mc{H}$ described relative to Bob's reference frame.  Suppose now that Alice performs a quantum operation on this system, with the operation being described by the CPTP map $\Phi:\mB(\mH)\to\mB(\mH)$ relative to her frame.  How would this operation be described relative to Bob's frame?  If he knows that their reference frames are related by an element $g\in G$, then $\mU_g^\dag(\sigma)$ is Alice's description of the initial state and $\Phi(\mU_g^\dag(\sigma))$ is her description of the final state.  Hence the final state relative to Bob's system is given by
$
\mU_g(\Phi(\mU_g^\dag(\sigma))),
$
and his description of Alice operation would be $\mc{U}_g\circ\Phi\circ\mc{U}_g^\dagger$.  However, if Bob does not know how their frames are related, he averages over $G$, and the resulting CPTP map has the form
\begin{equation}
 \int dg \;\mU_g\circ\Phi\circ\mU_{g}^{\dag}\;.
\end{equation}
Alice and Bob will have a similar description of the CPTP map Alice performs only if her operation has this form.  Quantum channels of this sort are called \textit{$G$-covariant}, and they constitute the free operations in the QRT of reference frames. Similar to $G$-invariant states, a quantum channel is $G$-covariant if and only if it commutes with $\mU_g$ for all $g\in G$.
Therefore, the set of free operations in the QRT of reference frames can be expressed as
\begin{equation}
\mO(\mH)=\left\{\Phi\in\text{CPTP}\;:\;[\Phi,\mU_g]=0\quad\forall\;g\in G\right\}.
\end{equation}
where $[\Phi,\mU_g]\eqdef\Phi\circ\mU_g-\mU_g\circ\Phi$ (see Fig. \ref{symmetry} below). Note also that $\Phi$ being $G$-covariant is equivalent to the condition that $
 \Phi=\int dg \;\mU_g\circ\Phi\circ\mU_{g}^{\dag}$.  In~\cite{Gour-2008a}, it was shown that $G$-covariant operations can be expressed in terms of Kraus operators, each being 
an irreducible tensor operator as defined in nuclear and atomic physics (see e.g.~\cite{Sakurai}).

To summarize, we found that in the QRT of reference frames, the set of free states is the set of \emph{symmetric} states (i.e., those states that commute with $U_g$ for all $g\in G$), and the set of free operations is the set of \emph{symmetric} operations (i.e., those operations that commute with $\mU_g$ for all $g\in G$).  Symmetric evolutions are very common in physics and can occur in different contexts, other than those arising from the lack of a shared reference frame. Therefore, the set of $G$-covariant operations defines a resource theory that has applications far beyond quantum reference frames.  It can therefore be described as a QRT of \emph{asymmetry}, since in any QRT for which $\mO$ defines a family of $G$-covariant operations, asymmetric states and asymmetric operations are the resources of the theory.

So far we only considered $G$-covariant channels with the same input and output dimensions. More generally, a quantum channel $\Phi:\mB(A)\to\mB(B)$ 
is $G$-covariant with respect to two (unitary) representations of $G$, $\{U_g^A\}_{g\in G}$ and $\{U_g^B\}_{g\in G}$, if
$$
\Phi\circ\mU_{g}^{A}=\mU_g^B\circ\Phi\quad\forall g\in G.
$$
See Fig.~\ref{symmetry} for a heuristic depiction of $G$-covariant operations. 

\begin{figure}[h]
\centering
    \includegraphics[width=0.3\textwidth]{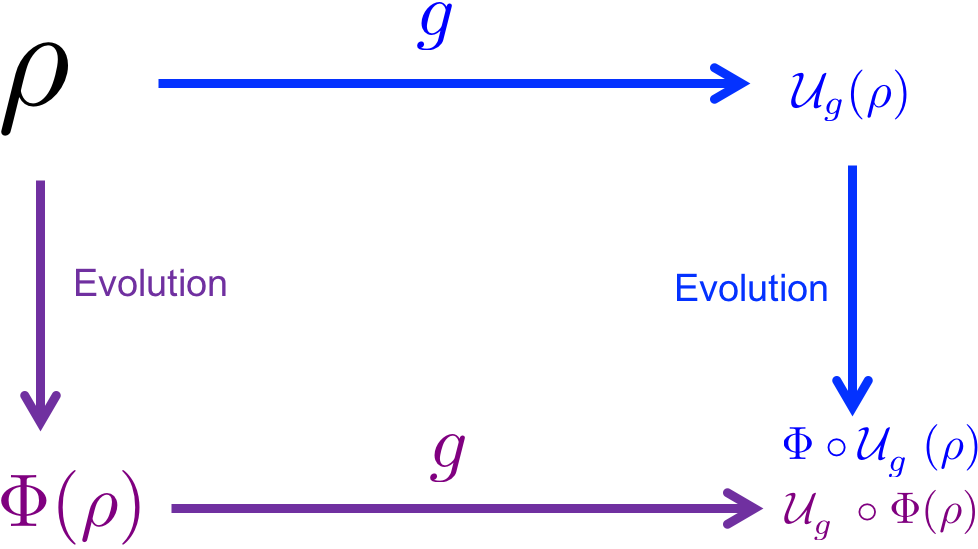}
  \caption{\linespread{1}\selectfont{\small Heuristic description of $G$-covariant operations. The channel $\Phi$ is $G$-covariant if the blue and purple pathways commute for all group elements.}}
  \label{symmetry}
\end{figure} 

The reformulation of symmetric dynamics in the context of a resource theory has profound implications. For example, quite often the dynamics of a system is so complicated that a complete characterization of its evolution becomes infeasible. Instead, by learning the symmetries of the Hamiltonian, one can gain partial information about its dynamics. Here, one can invoke Noether's Theorem, which states that a differentiable symmetry of the action of a physical system has a corresponding conservation law. However, Noether's theorem is not applicable to open systems, and it therefore does not capture all the consequences of symmetric evolution of mixed states. Recently, it was shown by \textcite{Marvian-2013, Marvian-2014} that the QRT of asymmetry provides a systematic way to capture all the consequences of symmetric evolution. The main idea is that the conserved quantities of closed systems can be replaced with resource monotones in open systems. These resource monotones quantify the amount of asymmetry in a quantum state, and they cannot increase under symmetric evolution (see Section \ref{Sect:Meaures-Monotones}).    


\subsubsection{Quantum Thermodynamics}

\label{Sect:Thermodnamics}

Thermodynamics and quantum mechanics represent two pillars of physics, and connections between them have been studied long before the advent of quantum information science.  However, recently quantum information theory has shed new light on some of the most fundamental questions in thermodynamics.  For example, by adopting a QRT perspective, the four Laws of Thermodynamics can be stated more precisely, and the relationships between them can be made more apparent \cite{Brandao-2015b, Masanes-2017a}.  This is not surprising since the ``golden rule'' of QRTs can be seen as expressing something akin to the Second Law of Thermodynamics, and similar connections between thermodynamics and the structure of general QRTs can also be formulated \cite{Sparaciari-2018a}.

Within quantum thermodynamics, several different QRTs have been studied, but almost all can be described using a similar framework.  The essential idea is to identify free operations as those that conserve certain extensive properties (such as energy, particle number, volume, etc.) when a given system interacts with a bath.  Here we describe the resource theory of \textit{athermality}, which involves just energy conservation, but its generalization to other conserved observables follows analogously \cite{Halpern-2016a, Halpern-2018a, Gour2018}, including non-commuting observables \cite{Horodecki-2016b, Guryanova-2016a, Lostaglio-2017a}.  One begins by characterizing a physical system not only by its underlying Hilbert space $\mc{H}$, but also by its Hamiltonian, since this corresponds to the property being conserved.  Hence a ``state'' in this QRT is represented by a pair $(\rho,H)$ consisting of a density matrix and a time-independent Hamiltonian.  For a heat bath held at some fixed inverse temperature $\beta=1/(k_BT)$, the free states consist of those that are in thermal equilibrium with the bath.  That is, the free states $\mc{F}(\mc{H})$ are given by $(\gamma_{H},H)$, where $\gamma_{H}=e^{-\beta H}/\tr[e^{-\beta H}]$ is the thermal equilibrium state (also called the Gibbs state) for the Hamiltonian $H$.  Below, a physical justification will be given for why these objects are considered free.    

As first introduced by \textcite{Janzig-2000a} and later extended in \textcite{Horodecki-2013b, Brandao-2013a}, the free operations in the QRT of athermality consist of all physical dynamics that conserve total energy as the system exchanges heat with the bath.  Any such process is called a \textit{thermal operation}, and it is constructed from three basic steps.  Let $S$ denote the primary system and $H_S$ its Hamiltonian.  First, the experimenter introduces an ancilla system $A$ in free state $(\gamma_{H_A},H_A)$.  Second, the primary and ancilla systems can interact via some unitary $U$ that commutes with the total Hamiltonian, i.e., $[U,H_{\text{tot}}]=0$ where $H_{\text{tot}}=H_S\otimes\mbb{I}+\mbb{I}\otimes H_A$.  
Third, a subsystem $B$ is discarded whose Hamiltonian enters the total Hamiltonian collectively; i.e., $H_{\text{tot}}=H_{SA\setminus B}\otimes\mbb{I}+ \mbb{I}\otimes H_{B}$, with $SA\setminus B$ denoting all subsystems other than $B$; note that $B$ may include part of the original system in addition to all or just part of the ancilla system $A$.  The composition of these three steps yields a CPTP map $\Phi\in\mc{B}(S\to SA\setminus B)$ having the form
\begin{equation}
\label{Eq:Thermo-Model}
\Phi(\rho)=\tr_{B}[U(\rho\otimes\gamma_{H_A}) U^\dagger].
\end{equation}

By construction, the QRT of athermality is a physically implementable resource theory.  In fact, having physically implementable free operations is essential to the overall objective of rigorously accounting for all dynamics in a system-bath exchange.  As a result, a thermodynamic transformation $(\rho, H)\to(\sigma,H')$ on a system becomes possible if and only if there exists a map having the form of Eq.~\eqref{Eq:Thermo-Model} such that $\Phi(\rho)=\sigma$ and $H'=H_{\text{tot}}-H_{B}$.  Whereas macroscopic state transformations via heat exchange are essentially governed by a decrease in free energy, in the quantum regime, more constraints dictate whether a given transformation is possible \cite{Horodecki-2013b, Brandao-2015b, Gour2018}. 

The free states in the QRT of athermality consist of Gibbs states $(\gamma_H,H)$, and there is strong operational justification for this.  First, the Gibbs state is the unique equilibrium state that a quantum system will evolve to under weak coupling with the bath \cite{Riera-2012a}.  Second, if, in the implementation of a thermal operation, one could freely introduce any other density operator $\sigma$ inequivalent to the Gibbs state of the ancilla system, then the QRT would become trivial.  More precisely, it would be possible to freely generate \textit{any} density matrix $\rho$ to arbitrary precision by consuming many copies of $\sigma$ \cite{Brandao-2015b, Halpern-2016a}.  The final, and perhaps most compelling reason for considering the Gibbs state to be free involves work extraction and the notion of passivity.  A thermodynamic state $(\rho, H)$ is called passive if $\tr[U\rho U^\dagger H]\geq \tr[\rho H]$ for all unitaries $U$.  Intuitively, if there exists some unitary for which this relation does not hold, then there exists a process in which energy can be drawn from the state to perform work.  For example, all states diagonal in the energy eigenbasis with eigenvalues decreasing with energy are passive.  A state $\rho$ is called completely passive if $\rho^{\otimes n}$ is passive for any $n$.  A classic result says that a state is completely passive if and only if it is the Gibbs state \cite{Pusz-1978a, Lenard-1978a, Jennings-2018a}.  Hence, the state $(\gamma_H,H)$ is the unique state from which work cannot be extracted, even after taking multiple copies.


Returning to Eq.~\eqref{Eq:Thermo-Model}, it is straightforward to verify that if the discarded system $B$ is the original ancilla system $A$, then every thermal operation acts invariantly on the Gibbs state.  That is,
\begin{equation}
\gamma_{H_S}=\Phi(\gamma_{H_S}),
\end{equation}
and all maps having this property are called \textit{Gibbs-preserving} with respect to the Hamiltonian $H$.  It is interesting to compare thermal operations with the more general class of Gibbs-preserving maps.  In the case that $\rho$ and $\sigma$ both commute with the Hamiltonian, if $\rho$ can be converted to $\sigma$ by some Gibbs-preserving map, then it can also be converted by thermal operations~\cite{Janzig-2000a, Horodecki-2013b, Korzekwa}.  However, for the convertibility between general states $\rho$ and $\sigma$ (i.e., those not commuting with the Hamiltonian), Gibbs-preserving operations are strictly more powerful \cite{Faist-2015a}.  The origin of this difference can be understood by invoking principles from the QRT of asymmetry.  With a time-independent Hamiltonian, states diagonal in the energy eigenbasis are symmetric under time evolution.  By introducing the one-parameter group of time translations $\{U_t:=e^{-iHt}:t\in\mbb{R}\}$,\footnote{Note this group is isomorphic to $U(1)$ provided the energy levels of $H$ are evenly spaced.} we see that $[\rho,H]=0$ if and only if $\mc{U}_t(\rho)=\rho$ for all $t$, where we have adopted the notation of Eq.~\eqref{Eq:Unitary_conjugation}.  States with full time-translation symmetry are often called \textit{quasiclassical} because, while having discrete eigenvalues, they lack coherence between the different energy eigenspaces.  In addition, from Eq.~\eqref{Eq:Thermo-Model} (with the system $A$ being discarded), it can be seen that every thermal operation is \textit{time-translation covariant}, i.e., $[\Phi,\mc{U}_t]=0$ for all $t$ \cite{Lostaglio-2015a}.  In contrast, a general Gibbs-preserving map need not satisfy this constraint.  Unlike thermal operations, Gibbs-preserving operations can break the time-translation symmetry of a state, and this is precisely what is demonstrated in the example of \textcite{Faist-2015a}.

In summary, every thermal operation with the same input/output system satisfies the two properties of being
\begin{enumerate}
\item[(i)] Gibbs-preserving: $\Phi(\gamma_H)=\gamma_H$;
\item[(ii)] Time-translation covariant; $[\Phi,\mc{U}_t]=0$ for all $t$.
\end{enumerate}
Maps satisfying these properties were identified as Gibbs-preserving covariant (GPC) maps in \textcite{Gour2018}, and they represent a strictly larger class than thermal operations.  It was shown that GPC maps are equivalent to the maps generated by so-called \textit{thermal processes}, which are more general physical processes than the ones leading to thermal operations in Eq.~\eqref{Eq:Thermo-Model} \textcite{Gour2018}.  Nevertheless, it remains an open problem whether there exists a state transformation $\rho\to\sigma$ that is possible by a GPC but not by a thermal map.  

A special type of QRT emerges when the Hamiltonians are required to  have a fully degenerate spectrum.  For instance, a paramagnetic system in the absence of an external magnetic field has complete degeneracy in its energy.  In this case, the Gibbs state of a $d$-dimensional system is the completely mixed state $\mbb{I}/d$, and \textit{all} unitaries commute with the total Hamiltonian.  Every free CPTP map then has the form
\begin{equation}
\Phi(\rho)=\tr_B[U(\rho\otimes\mbb{I}_d) U^\dagger],
\end{equation}
where $U$ is an arbitrary unitary and $B$ is an arbitrary subsystem.  Maps having this form are called \textit{noisy operations}, and they were originally proposed by \cite{Horodecki-2003a, Horodecki-2003b} as the free operations in the resource theory of \textit{purity}.  The latter is also called the QRT of \textit{nonuniformity} to better reflect the dimensional dependence in the resourcefulness of pure states \cite{Gour-2015a}.  The set of Gibbs-preserving maps here is precisely the set of unital maps, and all operations are trivially time-translation covariant.  \textcite{Haagerup2011} have shown that noisy operations form a strict subset of unital channels.  Yet, the two operational classes have the same conversion power since one density matrix can be converted to another by noisy operations if and only if the conversion can be achieved by a unital channel \cite{Gour-2015a}.

\subsubsection{Quantum Coherence}

\label{Sect:Coherence}

For a given quantum system, consider some quantum observable $T$ with eigenvectors $\{\ket{\lambda_n}\}_n$ and eigenvales $\{\lambda_n\}_n$.  Quantum mechanics allows the system to be prepared in a coherent superposition of eigenstates $\ket{\psi}=\cos\theta\ket{\lambda_i}+\sin\theta\ket{\lambda_j}$, and there are two ways in which such states can be viewed as a resource.  The first identifies a specific task in which $\ket{\psi}$ can assist in accomplishing the task, and the degree of its resourcefulness depends on the particular $\ket{\lambda_i}$ and $\ket{\lambda_j}$ forming the superposition.  For example, when $\lambda_n=n\in\mbb{N}$, the state $\cos\theta\ket{0}+\sin\theta\ket{1}$ can detect a phase shift of $\phi=\pi$ induced by the unitary $e^{-i\phi T}$ while the state $\cos\theta\ket{0}+\sin\theta\ket{2}$ cannot \cite{Marvian-2014a, Marvian-2016a}.  The second interpretation deems $\ket{\psi}$ as a resource simply because it is a coherent superposition in the eigenbasis $\{\ket{\lambda_n}\}_n$.  From this perspective, the eigenvectors $\ket{\lambda_i}$ and $\ket{\lambda_j}$ of $T$ appearing in the superposition are irrelevant, and the full nature of this resource is captured just in the wave components $\cos\theta$ and $\sin\theta$.

These two ways of characterizing $\ket{\psi}$ as a resource reflects the broader distinction between speakable and unspeakable information described in Section \ref{Sect:Asymmetry}.  The QRT of \textit{unspeakable coherence} is essentially a resource theory of asymmetry for translations generated by some quantum observable \cite{Marvian-2014a, Marvian-2016b, Marvian-2016a}.  Similar to the group of time translations discussed in the previous section, for an observable $T$, one defines the group of $s$-translations $\{U_s:=e^{-iT s}:s\in \mbb{R}\}$.  A map is called a \textit{translationally-covariant incoherent operation} (TIO) if it is covariant under the action of this group (i.e., $[\Phi,\mc{U}_s]=0$ for all $s$), and likewise a quantum state is identified as free in this QRT if it is invariant under the action of the group (i.e., $[\rho, \mc{U}_s]=0$ for all $s$).  Coherence here refers to nonzero off-diagonal elements of a density matrix when it is expressed in an eigenbasis of $T$.  This coherence is said to be unspeakable since it is needed when establishing a phase reference for $s$-translations, information which is unspeakable.

The second school of thought regards coherence as a resource in the context of processing speakable information.  These are QRTs of \textit{speakable coherence}, and most attention in the literature has focused on coherence theories of this sort. \textcite{Aberg-2006a} first proposed a resource theory of speakable coherence as an operational framework for quantifying superposition in general mixed states.  For a given system, one begins by fixing a direct sum decomposition of state space: $\mc{H}=\bigoplus_{i}\mc{H}_i$.  The set of free states $\mc{F}(\mc{H})$ then consists of all those for which $\rho=\sum_iP_i\rho P_i$, where $P_i$ is the projector onto $\mc{H}_i$.  The free operations, as originally considered by \"{A}berg, consists of the maximal set $\mc{O}_{\max}(\mc{H})$, i.e., the collection of all resource non-generating operations.  This is sometimes called the set of \textit{maximal incoherent operations} (MIO).

Note, the subspaces $\mc{H}_i$ in the direct sum decomposition of $\mc{H}$ could correspond to the eigenspaces of some observable $T$.  Then this QRT would resemble the theory of unspeakable coherence described above.  However, the key difference is that in \"{A}berg's theory, the index $i$ on the eigenspace $\mc{H}_i$ functions exclusively as a label, having no connection to the resource content of the state.  In contrast, the index is important in the QRT of unspeakable coherence since coherence across different eigenspaces of $H$ carry different physical meaning when the encoded quantum information is unspeakable.  Hence, transformations of the form $\frac{1}{\sqrt{2}}(\ket{0}+\ket{1})\to\frac{1}{\sqrt{2}}(\ket{0}+\ket{2})$ are forbidden using TIO but not by MIO.

A different QRT of speakable coherence was put forth by \textcite{Baumgratz-2014a}, which has now become the most frequently studied resource theory of coherence.  In the approach of Baumgratz \textit{et al.}, a complete orthonormal basis $\{\ket{i}\}_i$ is fixed for every Hilbert space.  This is called the incoherent basis, and a state is free (or ``incoherent'') if it is diagonal in this basis, a condition that can be expressed as \begin{equation}
\label{Eq:Incoherent_states}
\rho=\Delta(\rho),
\end{equation} where $\Delta(\cdot)=\sum_i\op{i}{i}(\cdot)\op{i}{i}$ is the completely dephasing map in the incoherent basis.  The free states in this QRT are thus defined in the same way as \"{A}berg's theory, except that the direct sum decomposition of $\mc{H}$ consists of one-dimensional subspaces.  For the free operations, Baumgratz \textit{et al.} define the class of \textit{incoherent operations} (IO), which consists of all stochastically resource non-generating operations.  In other words, $\Phi$ is free if it can be represented by Kraus operators $\{K_j\}_j$ such that for all $\rho$
\begin{equation}
\label{Eq:IO-Kraus}
K_j\Delta(\rho) K_j^\dagger=\Delta(K_j\rho K_j^\dagger)\qquad\forall j.
\end{equation}

It is not difficult to show that every Kraus operator satisfying Eq.~\eqref{Eq:IO-Kraus} has the form $K_j=\sum_{k}c_{j,k}\op{f(k)}{k}$, where $f$ is a function with domain and range being the labels of the incoherent basis vectors.  In particular, every coherence non-generating unitary belongs to IO since it has the form $U=\sum_{k}e^{i\phi_k}\op{\pi(k)}{k}$.  \textcite{Yadin-2016a} used this latter observation to consider what CPTP maps arise when such unitaries are used in a Stinespring dilation, i.e.,
\begin{align}
\label{Eq:SIO}
\Phi(\rho)&=\sum_j\tr_E[(\mbb{I}\otimes P_j)U(\rho^{S}\otimes\gamma^E)U^\dagger]\otimes\op{j}{j}^X\notag\\
&=\sum_j K_j\rho K_j^\dagger\otimes\op{j}{j}^X,
\end{align}
where $\gamma^E$ is an arbitrary incoherent state, $U$ is coherence non-generating on systems $SE$, and the $\{P_j\}_j$ form an arbitrary rank-one projective measurement on system $E$.  The Kraus operators $K_j$ can be shown to have the form $K_j=\sum_{k}c_{j,k}\op{\pi(k)}{k}$, where $\pi$ is a permutation on the labels of the incoherent basis vectors.  These operations have been referred to as \textit{strictly incoherent operations} (SIO) in \textcite{Yadin-2016a}, and they also received prior consideration in \cite{Winter-2016b} as a special subclass of IO.

Notice in the description of an SIO operation, the projectors $\{P_j\}_j$ on the ancilla system might be coherence-generating.  Consequently, the resource theory of coherence under SIO is not a physically implementable theory, as defined in Section \ref{Sect:Physically_Implementable}.  Starting from the set of incoherent states, which consists of those satisfying Eq.~\eqref{Eq:Incoherent_states}, the most general set of physical implementable operations can be constructed.  This is the set $\mc{O}_{\min}$, and in \cite{Chitambar-2016c}, it was shown that every $\Phi\in\mc{O}_{\min}$ has a Kraus operator decomposition of the form $K_j=U_jP_j=\sum_ie^{i\phi_i}\op{\pi_j(i)}{i}P_j$, where the $P_j$ are coherence non-generating projectors and the $\pi_j$ are permutations.  These \textit{physically implementable incoherent operations} (PIO) represent a highly restricted operational class, and in terms of state convertibility, almost any two pure states will not be interconvertible using PIO.

If one discards the classical system $X$ in Eq.~\eqref{Eq:SIO}, the resulting map has the property that
\begin{equation}
\Phi(\Delta(\rho))=\Delta(\Phi(\rho)).
\end{equation}
Any CPTP satisfying this equality is called a \textit{dephasing covariant incoherent operation} (DIO).  These operations were introduced independently in \cite{Chitambar-2016c} and \cite{Marvian-2016a}.  While not being physically implementable, DIO does represent a dually RNG set of operations, unlike IO.  A number of results have been established comparing the convertibility of states using different classes of incoherent operations \cite{Regula-2017a, Chitambar-2016b, Fang-2018a, Zhao-2018a, Zhao-2018b}.

Note that the set of incoherent states forms an affine set of density matrices, and therefore the QRT is affine.  Indeed this already follows from the fact that the completely dephasing map $\Delta(\cdot)$ is resource-destroying (see Section \ref{Sect:RDM}).  All operational classes discussed above can interconvert any two incoherent states, and they therefore identify $\mc{F}_{\min}(\mc{H})$ as the set of incoherent states acting on $\mc{H}$.  Other families of incoherent operations have also been studied in the literature \cite{deVicente-2017a}, and connections with stabilizer operations have been considered \cite{Mukhopadhyay-2018a}.

\subsubsection{Stabilizer Computation and ``Magic" States}

One of the most important questions in quantum information theory is to what extent quantum computers offer advantages over their classical counterpart.  Shor's algorithm provides one of the most celebrated examples of how, in principle, a quantum computer can perform a task exponentially faster than the best known classical algorithms \cite{Shor-1997a}.  This has motivated researchers to try and identify the certain features of quantum mechanics that appear to enable its superior computational capabilities.  Such features become a resource for the purposes of quantum computation.

To make this more precise, a resource-theoretic formalism can be adopted in which the free operations are quantum-computational processes that can be efficiently simulated using a classical computer.  Interestingly, the known results of such a resource theory depend on whether the underlying dimension of the system is even or odd.  Nevertheless, the basic framework is the same in both cases.  For prime dimension $d$, one first introduces the Heisenberg-Weyl operators
\begin{align}
T_{(a_1,a_2)}&=\begin{cases} 
iZ^{a_1}X^{a_2}\quad \text{for $d=2$}\\
e^{\frac{-\pi i a_1a_2}{d}}Z^{a_1}X^{a_2}\quad \text{for odd $d$}
\end{cases}
\label{Eq:Heisenberg-Weyl}
\end{align}
for $(a_1,a_2)\in\mbb{Z}_d\times\mbb{Z}_d$, where all arithmetic is done modulo $d$, and
\begin{align}
X\ket{j}&=\ket{j+1}&Z\ket{j}&=e^{\frac{2\pi i j}{d}}\ket{j}.
\end{align}
The $T_{(a_1,a_2)}$ generalize the Pauli operators in $d=2$ up to a modification in the overall phase, and extending to non-prime dimensions can be accomplished by taking tensor products of the $T_{(a_1,a_2)}$.  
The Clifford group $\mc{C}_d$ consists of operators that, up to an overall phase, transform the $T_{\mbf{u}}$ among themselves by conjugation.  In other words, $U$ is a Clifford operator on the $d$-dimensional space if $UT_{\mbf{b}}U^\dagger =e^{i\phi}T_{\mbf{b}'}$ for $\mbf{b},\mbf{b}'\in\mbb{Z}_d\times\mbb{Z}_d$ and arbitrary $\phi$.  The family of $d$-dimensional \textit{stabilizer states} are defined as
\begin{equation}
\mc{F}(\mc{H}_d)=\text{conv}\{U\op{0}{0}U^\dagger\;:\; U\in\mc{C}_d\},
\end{equation}
where $\text{conv}\{\cdot\}$ indicates the convex hull of the set.  These are the free states in the theory.  The free operations are called \textit{stabilizer operations}, and they consist of (i) preparing an ancilla in a stabilizer state, (ii) applying a Clifford unitary, (iii) measuring in the computational basis, and (iv) discarding subsystems \cite{Veitch-2014a}.  The QRT of stabilizer quantum computation for odd dimensions has been developed in \textcite{Veitch-2012a, Veitch-2014a}, while a multi-qubit treatment has been conducted in \textcite{Howard-2017a}.  The maximal QRT consistent with stabilizer states has also recently been studied by \textcite{Ahmadi-2017a}.

Stabilizer operations provide a fault-tolerant scheme for quantum computation \cite{Gottesman-1998b}, thereby making them an attractive candidate for implementing scalable quantum computation.  Unfortunately, the Gottesman-Knill Theorem stipulates that any stabilzier operation acting on pure stabilizer states can be efficiently simulated using classical computers \cite{Gottesman-1998a}.  Thus, to attain a computational speed-up using quantum computers, some additional ingredient beyond stabilizer states and operations is needed.  Perhaps the cleanest approach involves simulating non-stabilizer operations through stabilizer operations and the consumption of non-stabilizer states.  In fact, this technique is universal in that \textit{any} non-stabilizer operation can be implemented in such a way, provided the non-stabilizer states consumed belong to the class of so-called \textit{magic states} \cite{Gottesman-1999a, Bravyi-2005a, Knill-2005a}.  For example, the non-stabilizer qubit state $\cos(\pi/8)\ket{0}+\sin(\pi/8)\ket{1}$ can be used to realize the $\pi/8$ phase gate using stabilizer operations, which can then be subsequently used to perform any single qubit gate with arbitrary precision \cite{Nielsen-2000a}.

In realistic implementations, an experimenter is faced with noisy and non-ideal ancilla states.  One of the most important questions in this resource theory is whether a general mixed state can be freely converted into a magic state.  In many-copy form, this is a problem of resource transformation $\rho^{\otimes n}\toO_\epsilon \op{\phi}{\phi}$, where $\ket{\phi}$ is a magic state.  This task  has been coined \textit{magic state distillation} \cite{Bravyi-2005a, Reichardt-2005a, Campbell-2009a, Anwar-2012a}, and it is an analog to the task of ebit distillation in the QRT of entanglement.  

A natural question is whether all non-stabilizer states can be distilled into magic states.  For qubit states, \textcite{Campbell-2010a} have shown that undistillable resource states indeed exist if a finite limit is placed on the number of copies consumed in the distillation protocol.  For odd dimensions, \textcite{Mari-2012a, Veitch-2012a} demonstrated a stronger form of ``bound'' resource, in the sense that no magic state can be distilled from certain non-stabilizer states even in the asymptotic limit.  The proof of this result involves connecting the task of magic state distillation to the discrete Wigner function of a quantum state \cite{Wootters-1987a, Gross-2006a}.  It was shown by \textcite{Veitch-2012a} that the action of a stabilizer operation on any state with a non-negative discrete Wigner function can be efficiently simulated by a classical computer.  Consequently, magic states cannot be distilled from a state if its discrete Wigner function is non-negative.  Hence, \textit{positivity of the discrete Wigner function is analogous to the PPT distillability criterion in entanglement}.  The existence of bound resource states in the QRT of magic states then follows from the existence of non-stabilizer states with a non-negative discrete Wigner function \cite{Gross-2006a, Veitch-2012a}.

It remains an open problem whether, conversely, negativity of the discrete Wigner function is a sufficient condition for magic state distillability (analogous to the question of NPT bound entanglement).  However, a deep connection has been drawn between the QRT of magic states and the QRT of contextuality (see Section \ref{Sect:contextuality} for a description of the latter).  Any state having a negative discrete Wigner function can be used, in principle, to demonstrate contextuality in some family of stabilizer measurements \cite{Howard-2014a}.  In other words, a quantum state must have the capacity to reveal contextual effects using the free operations if it can generate universal quantum computation through magic state distillation.  This suggests that contextuality may be a key resource that empowers quantum computing.  However, this cannot be the full story in qubits, at least, since undistillable multiqubit states can nevertheless demonstrate contextuality \cite{Mermin-1990a, Howard-2013a}.

\subsection{Resource Theories in Quantum Foundations}

QRTs can transform the way we think about previously well-studied properties of physical systems, even those touching the foundations of quantum mechanics.  This includes Bell non-locality, contextuality, incompatibility of quantum measurements, steering, non-projectiveness, and more. Here we provide a very succinct description of some of these properties, focusing on how the set of free operations is defined in each of the theories.

\subsubsection{Bell Nonlocality}

Perhaps one of the most profound discoveries in the history of science is the incompatibility between quantum mechanics and the intuitive notion of locality, as first demonstrated by Bell in 1964 \cite{Bell-1964a}.  Since Bell's discovery, there have been numerous studies on the subject, and quantum nonlocality has emerged as a genuine resource in quantum communication \cite{Buhrman-2010a} and cryptography \cite{Ekert-1991a} tasks.  A comprehensive review on its state of the art can be found in a recent review by~\cite{Brunner-2014a}. 

\begin{figure}[t]
    \includegraphics[width=0.45\textwidth]{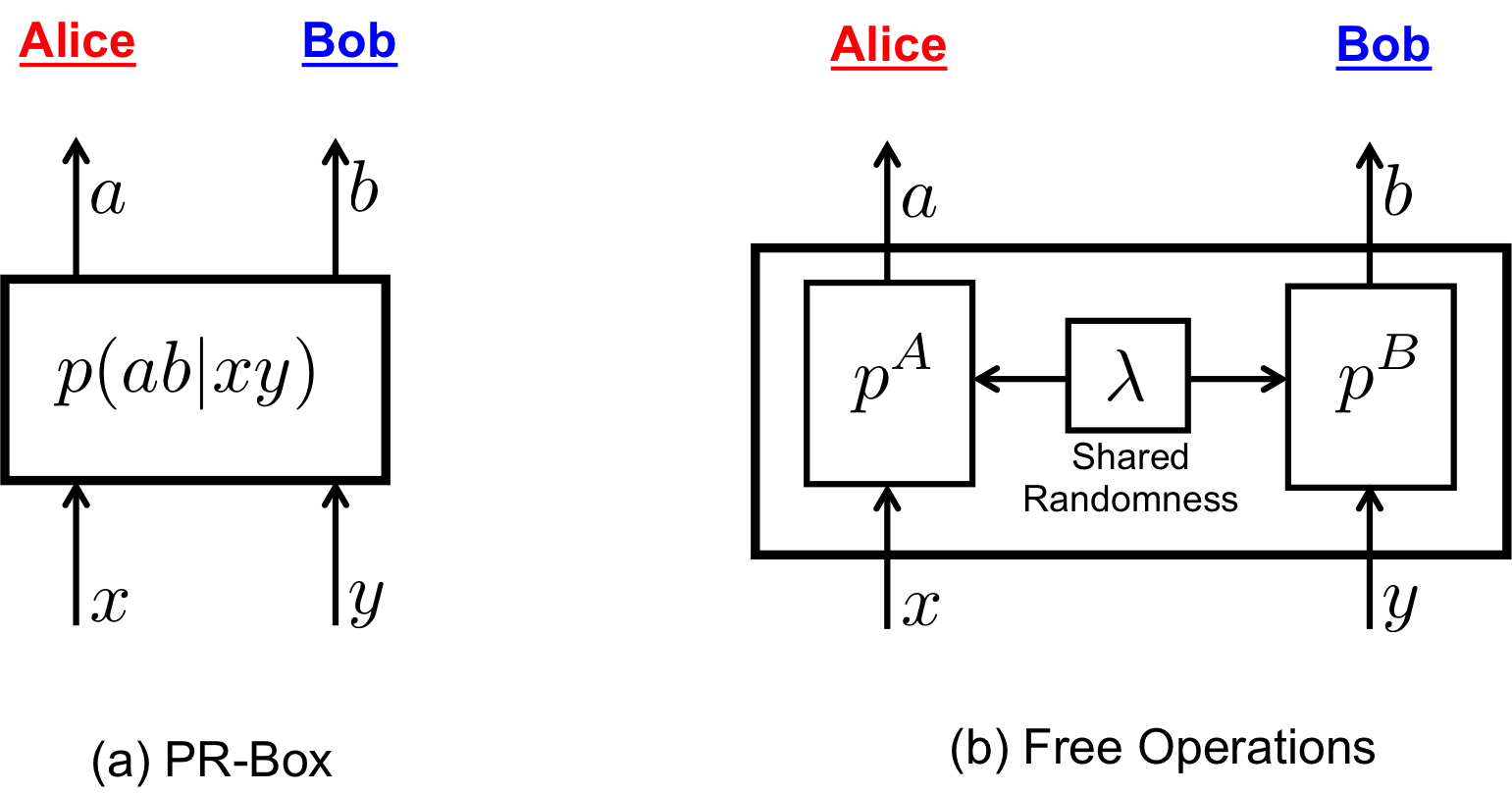}
  \caption{\linespread{1}\selectfont{\small Resource theory of Bell nonlocality.}}
  \label{box}
\end{figure}

\begin{figure}[h]
    \includegraphics[width=0.45\textwidth]{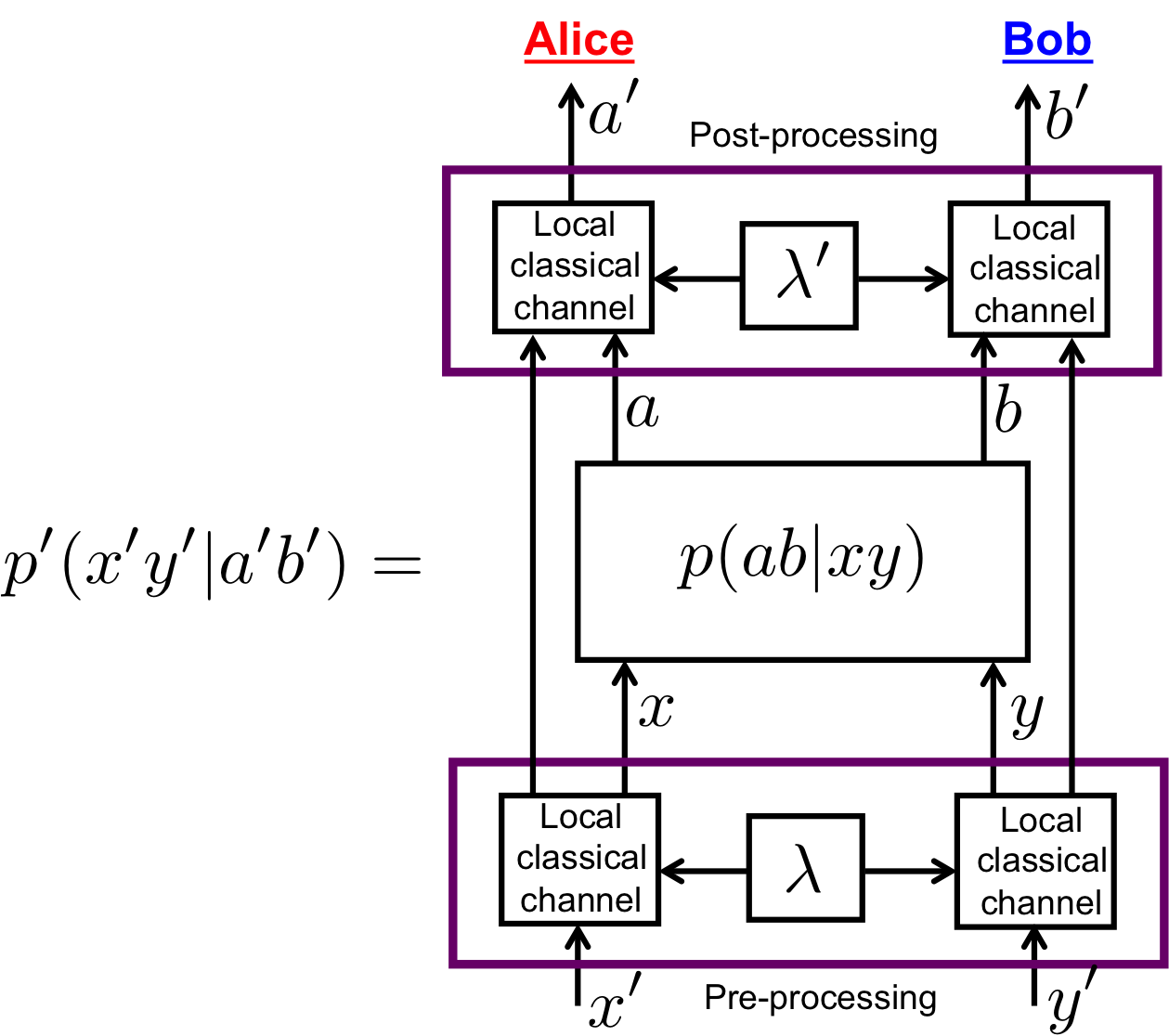}
  \caption{\linespread{1}\selectfont{\small Free superchannels.}}
  \label{losr}
\end{figure}

A resource theory of nonlocality can be formulated on different levels, and we begin by describing the most abstract, which is characterized entirely in terms of classical channels.  We focus just on bipartite systems, but its generalization to multipartite systems is straightforward.  Consider all possible classical channels $p(a,b|x,y)$ from input set $\mc{X}\times\mc{Y}$ to output set $\mc{A}\times\mc{B}$.  One can then construct a static resource theory in which states are bipartite probability distributions $p(x,y)$ and the allowed operations belong to some restricted set of channels.  A particular class of channels are those that are generated by local channels $p^A(a|x,\lambda)$ and $p^B(b|y,\lambda)$ for Alice and Bob, which are conditioned on some shared variable $\lambda$ (see Fig. \ref{box} (b)).  When averaging over $\lambda$ with some probability density function $q_\lambda$, the resulting channel is given by
\begin{equation}
\label{Eq:freebox}
p(a,b|x,y)=\int d\lambda  p^A(a|x,\lambda)p^B(b|y,\lambda)q_\lambda.
\end{equation} 
Channels of this form constitute the class of classical local operations and shared randomness (LOSR).  However, these operators are too powerful to define a static resource theory: any probability distribution $p(x,y)$ can be converted into any other $p(a,b)$ by a suitable LOSR channel.  To obtain a nontrivial theory, one must move a level higher and consider the dynamical resource theory of non-LOSR processes.  Applying the general discussion of Section \ref{Sect:processes}, the induced resource theory here is a resource theory of nonlocality, and it involves the conversion of one bipartite channel $p(a,b|x,y)$ (or ``box'') into another $p(a',b'|x',y')$ by LOSR superchannels \cite{Jones-2005a, Barrett-2007a, Gallego-2012a, deVicente-2014a}.   The free boxes in this theory are those satisfying Eq.~\eqref{Eq:freebox}, and the allowed conversions are built using three ingredients: pre-LOSR, post-LOSR, and local side channels (see Fig.~\ref{losr} and compare with the general construction of a free superchannel in Fig.~\ref{fig2} with free pre- and post- processing). 

The abstract resource theory of nonlocal boxes can be connected to a fully quantum resource theory of nonlocality by studying the boxes generated through local quantum measurements.  Consider a bipartite quantum state $\rho^{AB}\in\mS(AB)$, and suppose Alice and Bob perform one out of several possible positive-operator valued measures (POVMs) on their respective subsystems.  Letting $\Theta^A_x=\{M_{a|x}\}_a$ denote the POVM elements for Alice's $x$-POVM and $\Theta^B_y=\{N_{b|y}\}_b$ the POVM elements for Bob's $y$-POVM, the probability of outcome $(a,b)$ given measurement choice $(x,y)$ is computed using Born's rule:
\be\label{nonlocal}
p(a,b|x,y)=\tr\left[\rho^{AB}\left(M_{a|x}\otimes N_{b|y}\right)\right]\;.
\ee 
It is immediately straightforward to verify that separable states, i.e., states of the form $\rho^{AB}=\sum_{\lambda}q_\lambda\sigma_\lambda^A\otimes\omega_\lambda^B$, can only generate boxes having the structure of Eq.~\eqref{Eq:freebox}, although determining the minimum amount of shared randomness needed to simulate the quantum statistics can still be non-trivial \cite{Jebaratnam-2017a}.  Any state whose local measurements have outcomes that can always be described in the form of Eq. \eqref{Eq:freebox} are called \textit{Bell local} and said to admit a local hidden-variable model \cite{Augusiak-2014a}.

One very natural class of free operations in a QRT of Bell nonlocality consists of quantum LOSR maps.  These are bipartite CPTP maps of the form $\int_\lambda d\lambda q_\lambda \Phi_\lambda^A(\cdot)\otimes \Psi^B_\lambda(\cdot)$.  It is easy to see that such operations act invariantly on the set of Bell-local density matrices.  Thus, a QRT of nonlocality can be defined in which the free states are all Bell local.  In such a theory, a state $\rho$ is a quantum static resource if and only if it can be converted into a dynamical classical resource via Eq.~\eqref{nonlocal}.  Other approaches can be taken, however, such as the semi-quantum theory of \cite{Buscemi-2012a}, which identifies a state as being a static nonlocal resource if it can be converted into a dynamic quantum-classical resource.  Even more generally, one could consider the dynamic nonlocal resource of quantum channels themselves, which is manipulated under LOSR superchannels.  

The relationship between entanglement and Bell nonlocality is subtle. In~\cite{Werner}, it was shown that there exist bipartite entangled states that cannot generate a nonlocal box by local measurements.  Historically this gave the first indication that entanglement and nonlocality may be inequivalent resources, and the precise relationship between the two is still an area of active research \cite{Methot-2007a, Lipinska-2018a}.  Another topic of ongoing investigation involves the activation of nonlocality.  It was shown by \textcite{Palazuelos-2012a} that there exist bipartite quantum states $\rho$ and $\sigma$ such that each one of them cannot be converted into a nonlocal box, but when combined together, the joint state $\rho\otimes\sigma$ can generate a nonlocal box. This phenomenon occurs in some other QRTs as well, and it is known as resource \emph{activation} since the nonlocality of $\rho$ is activated by the state $\sigma$ and vice versa. In some cases, it is even possible to activate the nonlocality of $\rho$ with the same state $\rho$ or several copies of it, in which case it is called super-activation. 

\subsubsection{Contextuality}
\label{Sect:contextuality}

Bell nonlocality can be seen as a specific manifestation of contextuality in quantum mechanics.  In general, contextuality refers to the certain way in which a state is prepared, a transformation arises, or a measurement is performed \cite{Spekkens-2005a}.  Here we focus just on measurement contextuality.  

We begin by reviewing the notion of measurement incompatibility in quantum mechanics, and this discussion will be relevant to the QRTs described in the next section as well.  A family of POVMs is called compatible if its elements can be generated from a single ``mother'' POVM \cite{Acin1}.  More precisely, a collection of POVMs $\{\Theta_{x}\}_{x=1}^n$ with $\Theta_x=\{E_{a_x|x}\}_{a_x}$ is compatible if there exists a single POVM $\{F_\lambda\}_\lambda$ such that
\begin{equation}
\label{Eq:POVM-simulation}
E_{a_x|x}=\sum_{\lambda}p(a_x|x,\lambda)F_\lambda\qquad\forall a_x,x,
\end{equation}
where $p(a_x|\lambda,x)$ is some classical conditional probability distribution.  For an arbitrary state $\rho$, Eq.~\eqref{Eq:POVM-simulation} immediately yields a joint distribution for the measurement outcomes $\mbf{a}=(a_1,a_2,\cdots,a_n)$ of the POVMs $(\Theta_{1},\Theta_{2},\cdots,\Theta_{n})$ given by
\begin{align}
\label{Eq:POVM-simulation-joint}
p(\mbf{a}|\rho)&:=\sum_\lambda\prod_{x=1}^{n}p(a_x|x,\lambda)\tr[F_\lambda\rho]=\tr[G_{\mbf{a}}\rho],
\end{align}
where $G_{\mbf{a}}:=\sum_\lambda\prod_{x=1}^{n}p(a_x|x,\lambda)F_\lambda$ defines the elements of a multi-valued POVM $\overline{\Theta}=\{G_{\mbf{a}}\}_{\mbf{a}}$.  Note that $E_{a_x|x}=\sum_{y\not=x}\sum_{a_y}G_{\mbf{a}}$.  Thus, the single POVM $\overline{\Theta}$ is able to simultaneously measure all of the $\Theta_x$, and the $\Theta_x$ are therefore described as being ``jointly measurable'' \cite{Kraus-1983a, Lahti-2003a, Heinosaari-2008a}.  We use the terms compatible and jointly measurable interchangeably.  

Returning to contextuality, suppose that $\Theta$ is a POVM belonging to two different families of jointly measurable POVMs, $\mc{M}_1$ and $\mc{M}_2$.  We say that each of these families constitutes a \textit{context} for measuring $\Theta$.  This can be understood from Eq.~\eqref{Eq:POVM-simulation}, which shows $\mc{M}_1$ and $\mc{M}_2$ arising from two different ``mother'' POVMs, each one generating a different way, or context, for measuring $\Theta$.

This notion of contextuality can be used to distinguish and rule out certain hidden-variable theories of quantum mechanics.  Let $\mc{M}=(\Theta_1,\Theta_2,\cdots)$ be an arbitrary family of POVMs for some quantum system.  A noncontextual classical model for $\mc{M}$ on state $\rho$ is a probability distribution $f_\rho(\mbf{a})$ over all sequences $\mbf{a}=(a_1,a_2,\cdots,)$ of outcomes for the POVMs in $\mc{M}$.  To be consistent with the predictions of quantum mechanics, the model must have the correct marginals for any subset of jointly measurable POVMs.  That is, if $\mc{M}'\subset\mc{M}$ is one family of jointly measurable POVMs, i.e., one context, then 
\begin{equation}
\tr[G_{\mbf{a}'}\rho]=\sum_{a_i\not\in\mbf{a}'}f_\rho(a_1,a_2,\cdots),
\end{equation}
where $\mbf{a}'$ is any sequence of outcomes in $\mc{M}'$, the operator $G_{\mbf{a}'}$ is a POVM element for outcomes $\mbf{a}'$ as defined above, and the sum is over all outcomes for POVMs not in $\mc{M}'$.  The model is called \textit{noncontextual} because for every POVM in $\mc{M}$, the state $\rho$ is assigned an outcome distribution with no regard to contexts.  Hence in such a model, the measurement outcomes for any POVM only depend on the state $\rho$ and not on the way the measurement is carried out.

The famous Bell-Kochen-Specker theorem shows that a full description of quantum mechanics cannot be attained by a noncontextual hidden-variable theory \cite{Bell-1966a, Kochen-1967a}.  In other words, there exist families of POVMs $\mc{M}$ that require context-dependent hidden-variable models to accurately match the predictions of quantum mechanics.  Since context-dependent effects are purely a quantum phenomenon, it is possible to consider quantifying the amount of noncontextuality in a quantum measurement scenario, as a numerical signature of nonclassicality \cite{Svozil-2012a, Kleinmann-2011a, Fagundes-2017a, Grudka-2014a}.  

Additionally, steps have been taken to develop a resource theory of contextuality \cite{Horodecki-2015a, Abramsky-2017a, Amaral-2018a}.  In this resource theory, one first fixes a collection of POVMs $\mc{M}$.  A state is then a family of probability distributions $\{p(\mbf{a}'|\mc{M}')\}_{\mc{M}'}$ called a ``box,'' with a distribution defined for every jointly measurable subset $\mc{M}'\subset\mc{M}$.  A consistency condition requires that if two contexts $\mc{M}_1'$ and $\mc{M}_2'$ share a common POVM, then $p(\mbf{a}_1'|\mc{M}_1')$ and $p(\mbf{a}_2'|\mc{M}_2')$ must have the same reduced distributions for this POVM  \cite{Horodecki-2015a, Amaral-2018a}.  Free states, called noncontextual boxes, are boxes in which $p(\mbf{a}'|\mc{M}')$ is a marginal distribution of a single distribution $p(\mbf{a}|\mc{M})$ for every jointly measurable subset $\mc{M}'\subset\mc{M}$.  Regarding free operations, \textcite{Horodecki-2015a} have proposed all consistency-preserving transformations of boxes that act invariantly on the set of noncontextual boxes, i.e., the set $\mc{O}_{\max}$.  A more operational approach has been taken by \textcite{Amaral-2018a}, which involves modifying the formalism slightly to allow for composing, or ``wiring,'' of boxes.  Then the free operations are described analogously to Fig. \ref{losr} in the resource theory of nonlocality.  Namely, free operations are built by composing pre- and post- noncontextual boxes with a classical side channel extending from the latter to the former.

It is interesting to observe that within this resource-theoretic framework, Bell nonlocality can be characterized as a special case of contextuality \cite{Horodecki-2015a}.  Consider again the local POVMs $\Theta^A_x$ and $\Theta^B_y$ leading to Eq.~\eqref{Eq:freebox} for a given state $\rho$.  By the locality constraint, $\mc{M}_{x,y}:=(\Theta_x^A, \Theta^B_y)$ is a jointly measurable pair for every $(x,y)$.  It can then be seen that $\{p(a,b|\mc{M}_{x,y})\}_{x,y}$ forms a noncontextual box if and only if $p(a,b|x,y)$ is an LOSR channel (i.e., satisfying Eq.~\eqref{nonlocal}) \cite{Fine-1982a, Abramsky-2017a}.  Consequently, a state $\rho$ is Bell local if and only if it generates noncontextual boxes under any family of local POVMs.

\subsubsection{Incompatibility, Steering, and Projective Simulability}

\label{Sect:Steering}

\begin{figure}[h]
    \includegraphics[width=0.45\textwidth]{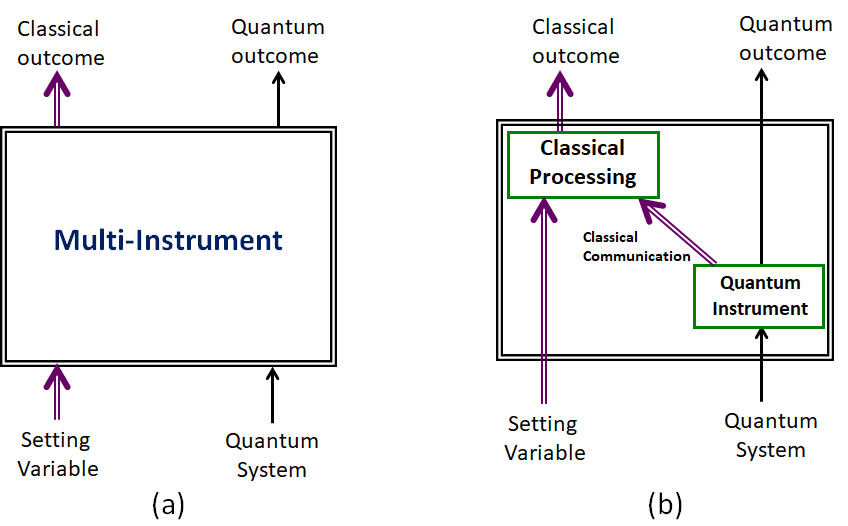}
  \caption{\linespread{1}\selectfont{\small (a) Multi-instrument: Quantum input/output in purple (double-line)  and classical input/output in black (single line) (b) Free (compatible) multi-instrument}}
  \label{incomp}
\end{figure}

Measurement incompatibility, as described in detail above, is a property of quantum mechanics that has been intensely studied since the early days of the subject.  This highly nonclassical feature can be characterized in terms of a quantum resource theory.  The main objects in this QRT are \emph{sets} of quantum measurements.  Formally, these can be characterized in terms of a special type of quantum channel, called a multimeter~\cite{Pusey15,GHS}, which has one classical input (the setting variable) that determines which measurement to perform, one quantum input upon which the measurement is performed, and one classical output corresponding to the measurement outcome. The notion of multimeter can be further generalized to include a quantum output of the measurements. In this case, the device is called a multi-instrument (see Fig.~\ref{incomp}a). Note that if we trace out the quantum output, then a multi-instrument reduces to a multi-meter, and if we remove the quantum input, then the device reduces to a multi-source.

In Fig.~\ref{incomp}b a free (i.e., compatible) multi-instrument is depicted, which is a single quantum instrument that simulates several instruments. Note that a compatible multi-instrument has the property that the quantum output is independent of the setting variable.  This is a property belonging to a more general class of multi-instruments called \textit{semicausal}~\cite{Pre2001,Wer2001,Pia2005}, and these can always be realized by replacing the classical communication and classical processing in Fig.~\ref{incomp}b with quantum communication and quantum processing.

With the above definition of free operations, the QRT of incompatibility of quantum instruments is well-defined. Since this is a resource theory of quantum processes, one resource (i.e., incompatible multi-instrument) can simulate another by a free superchannel as depicted in Fig.~\ref{fig2}, where the pre- and post- processing are compatible multi-instruments (see more details in~\cite{GHS}).

The QRT of compatibility also captures the notion of \emph{steering} as a special case, and it can therefore be used to define the QRT of steering~\cite{Gallego-2015}.
Steering is a process by which a bipartite quantum state $\rho^{AB}$ is used to remotely prepare an ensemble of quantum states in system $B$, by performing local measurements on system $A$ \cite{Wiseman-2007a, Jones-2007a, Cavalcanti-2017a}.  The objects in this resource theory are called ``assemblages,'' which are equivalent to multi-sources in the terminology used here (i.e., devices with classical input and both classical and quantum outputs).  That is, an assemblage has the form $\{p(a|x),\sigma_{a|x}\}_{a\in\mc{A},x\in\mc{X}}$ with $\{p(a|x),\sigma_{a|x}\}_{a\in\mc{A}}$ being an ensemble of quantum states for every $x\in\mc{X}$.  An assemblage is called \textit{unsteerable} if it admits a local hidden-state model:
\begin{equation}
p(a|x)\sigma_{a|x}=\sum_\lambda p(a|x,\lambda)\rho_{\lambda}q_\lambda \quad \forall a,x,
\end{equation}
and these are the free objects in the QRT of steering.  From Fig.~\ref{incomp}b it can be seen that unsteerable assemblages are precisely compatible multi-sources.  Moreover, the forward classical communication in semicausal multi-sources corresponds to the allowed one-way communication from Bob to Alice in the steering scenario.  Hence, the QRT of steering is equivalent to the QRT of incompatible (semicausal) multi-sources. 





One can think of other types of resources that are associated with quantum measurements. One such example is the degree in which a general quantum measurement or POVM differs from a projective von-Neumann measurement.  Since generalized quantum measurements and POVMs provide only an effective description of the measurement process, it is natural to ask how difficult it is to physically implement them, as in Section \ref{Sect:Physically_Implementable}.  Any implementation will involve projective measurements acting on a larger Hilbert space (a joint system + ancillary space). As joint projective measurements can be more challenging to realize, it is natural to consider a QRT in which such measurements are forbidden; this gives rise to a resource theory of joint measureability~\cite{Acin1,Acin2}.
In this model, the free operations are projective measurements assisted with classical processing and mixing. Simulability of one POVM (or generalized measurement) from another can be obtained as in Fig.~\ref{fig2} with the pre- and post-processing being the free operations.

\subsection{Non-Convex Resource Theories}

\subsubsection{Non-Gaussianity}

\label{Sect:Non-Gaussianity}

The QRT of non-Gaussianity, like entanglement, is another example of a resource theory that arises from natural constraints on the set of free operations. Gaussian states and Gaussian operations (including Gaussian measurements) are relatively easy to realize in experiments using lasers, phase-sensitive and phase insensitive optical amplifiers, and spontaneous parametric down conversion~\cite{Braunstein2005}. Consequently, Gaussian quantum information has been developed~\cite{Weedbrook-2012a, Serafini-2017a}, demonstrating that many quantum information processing tasks, such as QKD, can be implemented with only Gaussian states and Gaussian operations.  Analogous to entanglement-breaking channels, the structure of nonclassicality-breaking Gaussian channels has also been investigated \cite{Solomon-2013a, Sabapathy-2015a}.  Despite the success of Gaussian quantum information, it was also realized that many other important tasks such as entanglement distillation, quantum error correction, optimal cloning and more (see e.g.~\cite{Lami2018,Ryuji2018,Zhuang2018} and references therein) require non-Gaussian resources to be implemented. All this provides a strong motivation to develop a resource theory in which the free operations and free states are given in terms of Gaussian states and Gaussian operations.  The QRT of non-Gaussianity is different in two aspects from the QRTs we discussed above. First, it deals with continuous variable systems which are described by infinite-dimensional Hilbert spaces, and second, the sets of Gaussian states and Gaussian operations are not convex. 
 To overcome the challenge of non-convexity, one typically enlarges the set of free states and free operations to be the convex hull of Gaussian states and Gaussian operations~\cite{Lami2018,Ryuji2018,Zhuang2018, Albarelli-2018a}.

Some techniques from QRTs have been adopted to quantify non-Gaussianity, such as measures based on the relative entropy~\cite{Genoni2008,Genoni2010,Marian2013}, and very recently more systematic methods has been explored~\cite{Lami2018,Ryuji2018,Zhuang2018, Albarelli-2018a}.  For example, in~\cite{Albarelli-2018a} a state-based approach has been developed, where both quantum non-Gaussianity and the Wigner negativity have been identified as resources, depending on whether one chooses the set of free states to be the convex hull of Gaussian states or the set of states with positive Wigner function. Unlike entanglement theory, it was shown in~\cite{Albarelli-2018a} that there is no maximal resource state in this QRT. Non-Gaussian states have also been shown to be resource states for universal quantum computation~\cite{Ryuji2018}, and in~\cite{Zhuang2018} the entanglement-assisted non-Gaussianity generating power have been defined and proved to be a monotone under Gaussian operations.

In~\cite{Lami2018}, a broad resource-theoretic framework has been developed that encompasses Gaussian quantum information. 
Interestingly, it was shown that in all these models there are fundamental constraints on state manipulations, leading to a remarkable conclusion that no Gaussian quantum resource
can be distilled with free Gaussian operations.  Despite all of the very recent activities on the QRT of non-Gaussianity, there is much more room for development, and the theory is undoubtedly still in its infancy.

\subsubsection{Non-Markovianity}

\label{Sect:Markovianity}

Two markedly different resource theories of quantum non-Markovianity have been proposed in the literature.  We first recall the notion of Markovianity in the classical setting.  Three random variables $XYZ$ with joint distribution $p_{XYZ}$ form a (short) Markov chain, denoted by $X-Y-Z$, if 
\begin{equation}
\label{Eq:Markov-classical-static}
I(X:Z|Y)=0,
\end{equation}
where $I(X:Z|Y)=I(X:YZ)-I(X:Y)$.  Equivalently, the conditional distributions $p_{Z|X=x}(z):=\frac{p_{ZX}(z,x)}{p_X(x)}$ satisfy
\begin{equation}
\label{Eq:Markov-classical-dynamic}
p_{Z|X=x}(z)=\sum_yp_{Z|Y=y}(z)p_{Y|X=x}(y).
\end{equation}
Eq.~\eqref{Eq:Markov-classical-static} can be interpreted as a static condition on the variables $XYZ$, while Eq.~\eqref{Eq:Markov-classical-dynamic} can be interpreted as a dynamic condition on the induced transition matrices $p_{Z|Y}$ and $p_{Y|X}$.   These two sides of the same classical coin lead to two different generalizations of quantum Markovianity.

The first approach follows Eq.~\eqref{Eq:Markov-classical-dynamic} and involves classifying a quantum dynamical process as being either Markovian or non-Markovian.  In general, a time-parametrized evolution for a quantum system is represented by a family of trace-preserving dynamical maps $\{\Phi_{(t_2,t_1)}:\tau\geq t_2\geq t_1\geq 0\}$, with each map characterizing how the system transforms over time interval $[t_1,t_2]$.  Note that this encompasses the notion of ``process'' discussed in Section \ref{Sect:processes}, with the latter capturing a particular ``snapshot'' in some family of dynamical CP maps \cite{Wolf-2008a}.  An evolution is called Markovian if its dynamical maps are CP and satisfy the composition law $\Phi_{t_3,t_1}=\Phi_{t_3,t_2}\circ\Phi_{t_2,t_1}$ \cite{Rivas-2010a, Rivas-2014a}.  Markovian processes are non-convex in the sense that the CPTP map $\lambda\Phi_{t_2,t_1}+(1-\lambda)\Phi_{t_2,t_1}'$ need not arise within some Markovian evolution even if $\Phi_{t_2,t_1}$ and $\Phi_{t_2,t_1}'$ do.  Nevertheless, in the resource-theoretic spirit, several (non-convex) measures have been constructed to quantify the degree in which a particular quantum evolution is non-Markovian \cite{Wolf-2008a, Breuer-2009a, Chruscinski-2011a, Hall-2014a, Rivas-2014a, Bhattacharya-2018a}.  

The second type of resource theory follows Eq.~\eqref{Eq:Markov-classical-static} and characterizes quantum non-Markovianity in terms of tripartite static resources, i.e., tripartite quantum states $\rho^{ABE}$.   A quantum Markov state refers to any state $\rho^{ABE}$ whose conditional quantum mutual information vanishes.  That is, $I(A:B|E)_\rho=0$, where
\begin{align}
I(A:B|E)_\rho&:=S(AE)_\rho+S(BE)_\rho-S(ABE)_\rho-S(E)_\rho\notag
\end{align}
with $S(X)_\rho$ denoting the von Neumann entropy of systems $X$ in state $\rho$.  Non-Markov states have been shown to provide a resource for the tasks of quantum state redistribution \cite{Devetak-2008b}, secure communication by a one-time conditional pad \cite{Sharma-2017a}, and state deconstruction \cite{Berta-2018a}.  The conditional quantum mutual information of a state is also closely related to how well it allows for reconstruction from its bipartite reduced states \cite{Fawzi-2015a, Brandao-2015c}.  Note that $I(A:B|E)_\rho=0$ when the strong subadditivity bound is tight \cite{Lieb-1973a}, and the structure of such states has been determined by \textcite{Hayden-2004a}.  A QRT has recently been proposed by \textcite{Wakakuwa-2017a} in which Markov states are free and the free operations consist of LOCC between Alice and Bob, reversible quantum operations by Eve, and quantum communication from Alice and Bob to Eve.  Any combination of such actions leaves the set of Markov states invariant.  Moreover, these are natural operations to consider in cryptographic settings, such as the one-time pad, where Eve is an unwanted eavesdropper.  In fact, a full resource theory of secrecy involving two (or more) honest parties and one adversary can be constructed along these lines \cite{Horodecki-2005c}.  Classically, such a resource theory studies the processing of tripartite distributions $p_{XYZ}$ under local (classical) operations and public communication (LOPC) \cite{Collins-2002a, Christandl-2007a}.  A number of remarkable results have been obtained revealing analogous structures between quantum entanglement under LOCC and classical secrecy under LOPC \cite{Collins-2002a, Gisin-2002b, Renner-2003a, Oppenheim-2008a, Chitambar-2014d, Chitambar-2017a}.

\subsubsection{Quantum Correlations}

Traditionally, the term ``quantum correlations'' has been used in reference to the quantum entanglement in a multipartite quantum state.  However, as the subject of quantum information theory matured, quantum correlations became recognized as an arguably broader concept than just entanglement.  There are a variety of ways to quantify quantum correlations \cite{Modi-2012a, Horodecki-2013a, Adesso-2016a}, including those that measure the correlated dynamics in the evolution of a multi-part quantum system \cite{Rivas-2015a, Postler-2018a}.  Here we just focus on bipartite quantum discord \cite{Ollivier-2001a} and its associated resource theory.  As originally defined as \textcite{Ollivier-2001a} (see also \cite{Zurek-2000a}), the quantum discord from Bob to Alice is defined by
\begin{equation}
\label{Eq:Discord}
J(A|B)_\rho=I(A:B)_\rho-\max_{\{P_i^B\}_i}I(A:X)_{\rho'},
\end{equation}
where $I(A:B)_\rho$ is the quantum mutual information in $\rho$, the maximization is taken over all projective measurements on Bob's side, and $\rho'=\sum_i\tr_B(\mbb{I}^A\otimes P_i^B\rho)\otimes\op{i}{i}^B$ is a QC state.  From Eq.~\eqref{Eq:Discord}, the discord $J(A|B)$ can be interpreted as the correlations that remain when the classical correlations in $\rho$ are subtracted from its total correlations \cite{Henderson-2001a}.  An overview of different operational interpretations of discord can be found in \textcite{Modi-2012a, Streltsov-2015a, Berta-2018a}.

In a QRT of discord, the free states are characterized by the condition $J(A|B)_\rho=0$ and are said to be classically correlated.  It is not difficult to show that $\rho$ has vanishing discord if and only if it is a QC state $\rho=\sum_{i}p_i\rho_i^A\otimes\op{e_i}{e_i}^B$, where $\{\ket{e_i}\}_i$ is \textit{any} orthonormal basis for Bob's system \cite{Ollivier-2001a, Hayashi-2006a, Datta-2008b}.  An alternative characterization can be given by considering isometries of the form $U^{B\to BC}:\ket{e_i}^B\to\ket{e_i}^B\ket{e_i}^C$ applied to a given state $\rho^{AB}$.  It can be shown \cite{Datta-2017a} that $\rho^{AB}$ has vanishing discord if and only if there exists such an isometry for which $\hat{\rho}^{ABC}=(\mbb{I}\otimes U)\rho^{AB}(\mbb{I}\otimes U)^\dagger$ is a Markov state conditioned on system $B$.  In other words, we must have $I(A:C|B)_{\hat{\rho}}=0$ (see Section \ref{Sect:Markovianity}).  As for the free operations, it is obvious that any local CP map on Alice's side is resource non-generating.  On the other hand, Bob's actions must be restricted, and \textcite{Hu-2012a} have shown that a CPTP map on Bob's side is resource non-generating if and only if it is commutativity-preserving; i.e., $[\Phi(\eta),\Phi(\zeta)]=0$ whenever $[\eta,\zeta]=0$.  Other classes of free operations for discord have also recently been considered in \textcite{Zi-Wen-2017a}.

\section{Resource-theoretic Tasks}

In all QRTs, the same basic information-theoretic tasks can be studied, and often techniques used to solve a problem in one QRT can be applied to solve the analogous problem in another.  Here we review the most well-studied tasks in QRTs and highlight their general features.  

\subsection{Single-shot convertibility}

\label{Sect:Single-shot}

The most basic problem studied in any QRT is the conversion of one resource state to another using the free operations of the theory.  For any $\rho\in\mc{S}(A)$ and $\sigma\in\mc{S}(B)$, the question is whether there exists a CPTP map $\Phi\in\mc{O}(A\to B)$ such that $\Phi(\rho)=\sigma$.  If such a map exists, then we will write
\begin{equation}
\label{Eq:Single-shot-exact}
\rho\toO\sigma.
\end{equation}
Unfortunately, the task of exact state transformation is usually too strict.  That is, in most interesting QRTs, it will generally not be possible to perfectly transform one given state to another using the free operations, or vice-versa.  Furthermore, from an experimental perspective, exact transformations are artificial since any physical implementation will deviate from the theoretical ideal.

These considerations have motivated several variations to the problem of exact resource transformation.  The first involves relaxing the condition that the transformation $\rho\to\sigma$ be achieved deterministically.  Instead, one only seeks a ``flagged'' CPTP map $\Phi(\cdot)=\sum_j\Phi_j(\cdot)\otimes\op{j}{j}^X$ that is free and such that $\Phi_j(\rho)/p_j=\sigma$ for some $j$ with $p_j=\tr[\Phi_j(\rho)]\not=0$.  Starting with $\rho$, one then freely obtains $\sigma$ with probability $p_j$ by performing $\Phi$ and then measuring classical system $X$.  If such a map exists, $\sigma$ is said to be obtained from $\rho$ by a \textit{stochastic} or \textit{probabilistic} transformation, and we will denote this relationship by
\begin{equation}
\label{Eq:Single-shot-stochastic}
\rho\toSO\sigma.
\end{equation}
Note the only requirement in Eq.~\eqref{Eq:Single-shot-stochastic} is that $\rho$ be freely transformed to $\sigma$ with some nonzero probability, regardless of how small this may happen to be.  A more general yet typically more difficult question is to compute the greatest probability of transforming one state to another using the free operations.  That is, for a given $\sigma$ one can consider the problem of determining the value 
\begin{equation}
\label{Eq:Single-shot-stochastic-optimal}
P^{(\max)}_\rho(\sigma):=\sup_{\Phi_0}\left\{\tr[\Phi_0(\rho)] \:\bigg|\; \frac{\Phi_0(\rho)}{\tr[\Phi_0(\rho)]}=\sigma\right\},
\end{equation}
where the supremum is taken over all CP maps $\Phi_0$ such that $\Phi(\cdot)=\Phi_0(\cdot)\otimes\op{0}{0}^X+\Phi_1(\cdot)\otimes\op{1}{1}^X$ is a free CPTP map for some CP map $\Phi_1$.  Interest in stochastic convertibility first arose in the study of entanglement distillation where it was observed that every bipartite weakly entangled state can be transformed into a maximally entangled state with nonzero probability, a process sometimes called the \textit{Procrustean method} \cite{Bennett-1996b}.  But the idea of stochastic transformations applies to all QRTs in which classically ``flagged'' CPTP maps are free.

A second variation involves relaxing perfect fidelity in the target state.  One allows for an $\epsilon$-ball or an ``$\epsilon$-smoothing'' around $\sigma$ and deems the transformation a success if $\rho$ is transformed to any state within that ball.  More precisely, one first defines the set of density matrices $B_\epsilon(\sigma)=\{\sigma':F(\sigma,\sigma')\geq 1-\epsilon\}$ and then writes
\begin{equation}
\label{Eq:Single-shot-epsilon}
\rho\toO_\epsilon\sigma
\end{equation}
if $\rho\toO\sigma'$ for some $\sigma'\in B_\epsilon(\sigma)$.  
Transformations like Eq.~\eqref{Eq:Single-shot-epsilon} are the primary focus in \textit{one-shot} information theories, and they are operationally linked to the ``smooth'' entropic quantities (see Section \ref{Sect:Smooth_Entropies}).  
The single-shot transformations of Eqns. \eqref{Eq:Single-shot-exact}--\eqref{Eq:Single-shot-epsilon} are ostensibly different from the \textit{many-copy} or asymptotic transformations that are traditionally considered in information theory, and which we discuss next.


\subsection{Asymptotic Convertibility}

\label{Sect:Tasks-Asymptotic}

Even after relaxing the demand of exact transformation to allow for stochastic or approximate outcomes, most pairs of states in a typical QRT will still not be interconvertible.  Another approach is to abandon the one-shot scenario and consider transforming multiple copies of the same state.  The object of interest now becomes the optimal \textit{rate} of input to output states that is achievable using the free operations.  More precisely, a rate $R$ is said to be achievable in transforming $\rho$ to $\sigma$ if for every $R'<R$ and every $\epsilon\in(0,1]$, there exists an integer $n$ sufficiently large so that 
\begin{equation}
\rho^{\otimes n}\toO_\epsilon\sigma^{\otimes \lfloor nR'\rfloor}.
\end{equation}
The optimal rate is then denoted as $R(\rho\to\sigma)=\sup\{R\;|\;\text{such that $R$ is achievable under $\mc{O}$}\}$.

Computing $R(\rho\to\sigma)$ for arbitrary states is usually a formidable task.  Nevertheless, there are some general properties of $R(\rho\to\sigma)$ that can be observed.  Trivially if the QRT allows for the preparation of free states, then $R(\rho\to\sigma)=+\infty$ for any free state $\sigma$.  Likewise, one would expect that $R(\rho\to\sigma)=0$ for any free state $\rho$ and any resource state $\sigma$.  However, some care is needed because a nonzero asymptotic rate in this case would not automatically violate the Golden Rule of QRTs.  On the other hand, if (for any dimension) the set of free states is closed and discarding subsystems is RNG, then it is easy to show that indeed $R(\rho\to\sigma)=0$ for $\rho\in\mc{F}$ and $\sigma\not\in\mc{F}$.

A fundamental problem in any QRT is to compare the two directions of asymptotic convertibility for a given pair of states.  How does $R(\rho\to\sigma)$ compare to $R(\sigma\to\rho)$?  States $\rho$ and $\sigma$ are said to be (weakly) \textit{reversible} in a QRT if $R(\rho\to\sigma)R(\sigma\to\rho)=1$.  Roughly speaking, reversibility in this sense means that the transformation cycle $\rho\to\sigma\to\rho$ returns one copy of $\rho$ for each starting copy of $\rho$, in the asymptotic limit.  For the definition of optimal convertibility rate, it follows that $R(\rho\to\omega)\geq R(\rho\to\sigma)R(\sigma\to\omega)$ for any three states \cite{Horodecki-2003c}.  Consequently, if $\rho$ and $\sigma$ are reversible, and $\sigma$ and $\omega$ are also reversible, then we must have that $\rho$ and $\omega$ are reversible.  

Reversibility thus establishes an equivalence class on the set of all density matrices such that two states belong to the same family if and only if they are asymptotic reversible under the free operations of the QRT.  A canonical representative $\tau_0$ can be identified for each equivalence class, and to compute $R(\rho\to\sigma)$ for any two states in the class, it suffices to determine $R(\rho\to \tau_0)$ for all $\rho$ in the class.  For example, in bipartite entanglement theory, the maximally entangled state $\ket{\Phi^+}=\sqrt{1/2}(\ket{00}+\ket{11})$ is the natural choice for a class representative, and all pure states belong to the same reversibility class as $\ket{\Phi^+}$ \cite{Bennett-1996b}.  In this case, the quantity $R(\rho\to\op{\Phi^+}{\Phi^+})$ is called the \textit{distillable entanglement} of $\rho$ \cite{Bennett-1996a}, while $R(\op{\Phi^+}{\Phi^+}\to\rho)$ is called the \textit{entanglement cost} of $\rho$ \cite{Hayden-2001a}.  

In most QRTs, \textcite{Brandao-2015a} have shown that \textit{all} resource states become asymptotically reversible using operations that are asymptotically resource non-generating for the set of free states, and the optimal rate of convertibility is determined by $\mc{R}^\infty_{\text{rel}}$, the regularized relative entropy of resource (see Section \ref{Sect:Smooth_Entropies}).  Using strictly weaker classes of operations, reversibility between any two states was also shown to hold in the QRTs of purity \cite{Horodecki-2003b} and coherence \cite{Chitambar-2018a}.

However, in most QRTs, different reversibility classes will exist and $R(\rho\to\sigma)R(\sigma\to\rho)<1$ for states belonging to different classes.  In this case, the states $\rho$ and $\sigma$ demonstrate resource \textit{irreversibility} and the cycle $\rho\to\sigma\to\rho$ incurs nonzero loss per initial copy of $\rho$.  The strongest manifestation of irreversibility arises when $R(\rho\to\sigma)>0$ but $R(\sigma\to\rho)=0$ for some resource state $\sigma$.  This has been observed most prominently in the QRT of entanglement, a phenomenon known as \textit{bound entanglement} \cite{Horodecki-1998a}.  A bound entangled state $\rho$ is characterized by the conditions that $R(\op{\Phi^+}{\Phi^+}\to \rho)\in (0,+\infty)$ and $R(\rho\to\op{\Phi^+}{\Phi^+})=0$.  An analog to bound entanglement has also been discovered in the resource theories of thermodynamics \cite{Lostaglio-2015a}, ``magic state'' quantum computation \cite{Veitch-2012a}, speakable \cite{deVicente-2017a, Zhao-2018b} and unspeakable \cite{Marvian-2018a} coherence, and even multipartite secret key \cite{Acin-2004a}.

There is a stronger form of asymptotic reversibility that is not defined in terms of transformation rates.  Again, consider the transformation cycle $\rho\to\sigma\to\rho$, but instead of just demanding that one copy of $\rho$ returns for every initial copy, one demands that the output of the cycle have arbitrarily close fidelity to the input.  While second-order losses can be ignored when just focusing on rates, they become largely important when considering transformation fidelities.  A precise definition for this stronger form of resource reversibility has been proposed by \textcite{Kumagai-2013a} for the specific case of entanglement.  Under the stronger definition of reversibility, non-maximally entangled pure states can no longer be reversibly transformed into $\ket{\Phi^+}$.

\subsection{Catalytic Convertibility} 

\label{Sect:Catalyst}

In a chemical process, a reaction catalyst is a substance whose presence makes the process possible while remaining unaltered at the end of the process.  This effect can be incorporated within the framework of QRTs.  A state $\omega$ is called a \textit{resource catalyst} for the transformation of $\rho$ to $\sigma$ if 
\begin{equation}
\label{Eq:Single-shot-catalysis}
\rho\not\toO\sigma\qquad\text{but}\qquad \rho\otimes\omega\toO\sigma\otimes\omega.
\end{equation}
If $\rho$ can be transformed into $\sigma$ using some catalysis state, then we will write $\rho\tocO\sigma$.  

The fact that catalysts exist in QRTs is not obvious.  However, it was discovered by Jonathan and Plenio that entanglement transformations allow for catalysts \cite{Jonathan-1999b}, and since then nearly all QRTs have been found to demonstrate catalytic phenomena.  Analyzing catalyst-assisted transformations has been particularly illuminating in the QRT of quantum thermodynamics where a family of free energies have been shown to characterize convertibility from one state to another, thereby generalizing the second law of thermodynamics \cite{Brandao-2015b}.

Like the single-shot problem of Eq.~\eqref{Eq:Single-shot-exact}, variations to the question of exact catalytic convertibility can also be considered.  Stochastic catalytic transformations refer to pairs of states $\rho$ and $\sigma$ such that $\rho\not\toSO\sigma$ but $\rho\otimes\omega\toSO\sigma\otimes\omega$ for some $\omega$ \cite{Chen-2010a}.  A more general question is whether catalysts can increase the optimal success probability of a stochastic transformation.  In other words, does there exist an $\omega$ such that
\begin{equation}
P^{(\max)}_{\rho\otimes\omega}(\sigma\otimes\omega)>P^{(\max)}_{\rho}(\sigma),
\end{equation}
where $P^{(\max)}$ is defined in Eq.~\eqref{Eq:Single-shot-stochastic-optimal} \cite{Feng-2005a}?  

It is also possible to consider $\epsilon$ approximations in the catalytic transformation, i.e., $\rho\otimes\omega\toO_\epsilon\sigma\otimes\omega$.  Here, one can further stipulate that the output be an exact tensor product state with $\sigma$ obtained on the primary system and all the error in the final state occuring on the catalysis system.  Error in this case represents non-cyclic behavior for the catalysis, and often a more operationally-based measure of this acyclicity is better to use than just the fidelity between initial and final state \cite{Brandao-2015b}.  If one allows for $\epsilon$ error to occur across both the primary and the catalysis systems, then the problem often becomes trivial.  This is because of a powerful phenomenon known as \textit{resource embezzlement}, which refers to a family of states $\{\omega_n\}_n$ such that for any $\sigma\in\mc{S}(\mc{H})$ and any $\epsilon>0$, 
\begin{equation}
\label{Eq:Embezzlement}
\omega_n\toO_\epsilon\sigma\otimes\omega_n
\end{equation}
for all $n$ sufficiently large.  Entanglement embezzlement was first discovered by \cite{vanDam-2003a}, and it has found important applications in, for example, proving the Quantum Reverse Shannon Theorem \cite{Bennett-2014a, Berta-2011a}.  Outside of entanglement, resource embezzlement or variations to this idea have also been demonstrated in the QRTs of thermodynamics \cite{Brandao-2015b, Gour-2015a} and coherence \cite{Aberg-2014a}.

\subsection{Convertibility preordering}

\label{Sect:Quasi-Order}

The convertibility tasks described in the past three sections establish various preorderings of state space.  Recall that for a general set $S$, a preorder is a binary relation $\prec$ satisfying the properties of (i) reflexivity: $a\prec a$ for all $a\in S$, and (ii) transitivity: $a\prec b$ and $b\prec c$ implies $a\prec c$ for all $a,b,c\in S$.  Clearly the relations $\toO$, $\toSO$, and $\tocO$ form preorders.  Approximate convertibility fails to be transitive in general, but if $\rho\toO_\epsilon\sigma$, and $\sigma\toO_\epsilon\omega$, then $\rho\toO_{2\epsilon}\omega$.  Optimal rates of asymptotic convertibility also induce a preorder on state space, the details of such can be found in \textcite{Bennett-1999c}.



One advantage of studying preorders in a QRT is that it allows for a comparison of resources without first having to specify any sort of resource measure.  If $\rho\to\sigma$ under any of the aforementioned orders, then $\rho$ has no less resource than $\sigma$ in a truly operational sense.  However, comparing resources in this manner is limited since typically state convertibility will not form a total order on the set of density matrices, whether the convertibility is considered to be exact, asymptotic, or even with small $\epsilon$ error.  

There are two extremes that can arise in a convertibility preorder for a given QRT.  The first occurs when most if not all pairs of states fail to be ordered, i.e., one cannot be transformed into any other.  This is the case in the QRT of multipartite entanglement under LOCC \cite{Sauerwein-2017a}.  The other extreme is when any state can be converted into any other.  In this case the free operations are so powerful that all preorders collapse, and all states are essentially equivalent.  Both extremes are uninteresting to study from a QRT perspective.  An analytically rich QRT is one where the preorders are not trivial and resource hierarchies can be established for a wide range of physically relevant states.

\subsection{Simulation of non-free operations}

\label{Sect:Task-simulation}

One of the most important tasks in any QRT is overcoming the operational limitations intrinsic to the definition of the resource theory.  Given that the QRT only allows certain physical operations, how can the experimenter transcend this restriction and perform essentially non-free operations?   This question is especially important because often the operational constraints in a QRT reflect practical challenges that would not arise in ideal experimental setups.  In general, it is possible for non-free operations to be simulated by free operations at the cost of consuming a resource state.  More precisely, we say $\Phi_{\text{Resource}}\not\in\mc{O}(A\to B)$ is simulated by $\Phi_{\text{Free}}\in\mc{O}(AC\to B)$ and $\nu\not\in\mc{F}(C)$ if
\begin{equation}
\label{Eq:Simulate_Operations}
\Phi_{\text{Resource}}(\rho)=\Phi_{\text{Free}}(\rho\otimes\nu)
\end{equation}
for all $\rho\in\mS(A)$.  Such channels have been studied for LOCC under the name of $\sigma$-stretchable channels \cite{Pirandola-2017a}, and later for general resource theories under the name of $\nu$-freely-simulable channels \cite{Kaur-2018a}.  This equation can be seen simply as the conversion of a static resource $\nu$ into the dynamic resource $\Phi_{\text{Resource}}$ (see Section \ref{Sect:processes}).

For a given QRT, a natural question is whether all CPTP maps $\Phi\in\mc{O}(A\to B)$ can be simulated in such a manner.  If not, what is the largest set of channels that can be simulated?  Furthermore, for those channels that can be simulated, what is the minimal amount of resource needed?  Note that the task of channel simulation generalizes the task of catalytic convertibility.  

Quantum teleportation describes one of the most important examples of such a simulation, where general bipartite CPTP maps can be simulated using shared entanglement \cite{Bennett-1993a, Bennett-1996a, Horodecki-1999b}.  A similar result holds for coherence \cite{Chitambar-2016a}.  As another example, the underlying motivation for the QRT of magic states lies in the fact that magic states generate universal computation when consumed by stabilizer operations \cite{Gottesman-1999a, Bravyi-2005a, Knill-2005a}.  Finally, in the QRT of asymmetry, the state $\omega$ in Eq.~\eqref{Eq:Simulate_Operations} could represent a shared reference frame, or more generally one that breaks the underlying symmetry, thereby allowing for the simulation of asymmetric transformations \cite{Bartlett-2007a}.

\subsection{Erasing Resources}

In Section \ref{Sect:RDM} resource-destroying maps were discussed, and the existence of such maps was shown to be intimately linked to the overall structure of the particular QRT.  Although not all QRTs possess resource-destroying maps, the general task of erasing resource can be studied in any QRT.  By understanding the operational requirements to erase the resource in a given state, a fundamental connection is drawn between dynamic physical processes and the static resource held in the state.  This idea has its origins in Landauer's principle, which identifies the amount of information stored in a computer system as being proportional to the amount of work needed to reset (i.e., erase) the system into some fixed initial configuration \cite{Landauer-1961a}.

Landauer's principle can be extended to the erasing of resource in a general QRT having minimal structure.  Suppose maps of the form
\begin{equation}
\label{Eq:map_twirl}
\rho\mapsto \sum_{i=1}^Np_{i} U_i\rho U_i^\dagger\otimes \op{i}{i}^X
\end{equation}
are allowed, where the $U_i$ are free unitaries.  If discarding classical information is also permitted (as in Section \ref{Sect:Convex}), then Eq.~\eqref{Eq:map_twirl} could be continued to obtain the CPTP map $\Phi(\cdot)=\sum_{i=1}^Np_iU_i(\cdot)U_i^\dagger$.  Even though this process is deemed free by the QRT, Landauer's principle still associates a thermodynamical cost with the step of discarding classical information.  In the many-copy limit, the rate of physical work required to erase the classical memory is given by the entropy of the distribution $p_i$.  For a given resource $\rho$, one can minimize the entropy over all such channels such that $\sum_{i=1}^Np_i U_i(\rho)U_i^\dagger$ is a free state.  This then captures the asymptotic minimal work-cost to erase the resource in $\rho$ via such a protocol.  The minimum number of random unitaries needed to erase the resource in a given state (i.e., $N$ in the above sum) also provides a more conservative and often asymptotically tight measure of erasure cost \cite{Groisman-2005a}.  

One can also consider the task of erasing resources using catalysis.  Building on the work of \textcite{Majenz-2017a} involving catalytic decoupling, \textcite{Anshu-2017a} have recently shown that in convex QRTs admitting a tensor product structure, $\mc{R}^\infty_{\text{rel}}$ quantifies the optimal asymptotic rate of erasing resource using free unitaries and catalysis, where $\mc{R}^\infty_{\text{rel}}$ is the regularized relative entropy of resource (see Section \ref{Entropic measures}).  Roughly speaking this means that for any $R>\mc{R}^\infty_{\text{rel}}$, there exists free unitaries $U_i$ and a probability distribution $p_i$ such that
\begin{equation}
\Phi(\rho^{\otimes n}\otimes \omega)=\sum_{i=1}^{2^{nR}} p_i U_i\left(\rho^{\otimes n}\otimes \omega\right)U_i^\dagger\approx \sigma\otimes \omega,
\end{equation}
where $\sigma$ is a free state.  Specific cases of this task have been investigated in the QRTs of entanglement \cite{Groisman-2005a}, coherence \cite{Singh-2015a}, non-Markovianity \cite{Wakakuwa-2017a}, asymmetry \cite{Wakakuwa-2017b}, and more generally in scenarios involving state deconstruction \cite{Berta-2018a, Majenz-2017a}.  One-shot variations of the problem have also been proposed and extensively studied in \cite{Majenz-2017a, Berta-2017a, Anshu-2017a}.

\section{Quantifying Resources}

One of the most useful aspects of a QRT is that it generates precise and operationally meaningful ways to quantify a given physical resource.  Dedicated studies on the broad characterization and computation of resource measures in a general QRT have been conducted \cite{Zi-Wen-2017a, Bromley-2018a, Regula-2018a}.  Here we review a variety of resource measures that can be introduced in any QRT.  We first begin with an axiomatic approach which involves identifying some necessary and desirable properties that any resource measure should satisfy.  After that, we review different families of specific resource measures that can be used in the study of general QRTs.

\subsection{An Axiomatic Approach}

\label{Sect:Measures}

In its definition, a QRT is defined for any Hilbert space $\mc{H}$.  Therefore, a true resource measure should be able to quantify the resource of a density operator acting on any space; that is, we should consider non-negative functions of the form $f:\cup_\mc{H}\mc{S}(\mc{H})\to\mbb{R}_{\geq 0}$.  However, in practice one may be satisfied with restricting the domain of a measure and focusing on just a single input space $\mc{H}$.  The additional structure required of $f$ to be a resource measure can be cast in axiomatic form.  Below we state five axioms for a resource measure.  Lest this approach be too restrictive, we distinguish the first two axioms (vanishing for free states and monotonicity) as being essential, while the others (convexity, subadditivity and subextensivity, and asymptotic continuity) as being convenient and non-essential.

\subsubsection{Vanishing for Free States}

The first and most obvious axiom of a resource measure is that for a given system $A$
\begin{equation}
\label{Eq:Reource-axiom1}
\rho\in\mc{F}(A)\quad\Rightarrow\quad f(\rho)=0.
\end{equation}
This condition makes the statement ``no resource'' quantitatively precise.  Intuitively it may be tempting to require that the converse of Eq.~\eqref{Eq:Reource-axiom1} also holds.  This property is called \textit{faithfulness}, and a general function $f$ is called resource faithful if $f(\rho)=0$ implies that $\rho$ is free.  However, it may be that for a given task, certain resource states provide no operational advantage over free states.  Such states should then be assigned zero resource by any measure that quantifies the utility of a state for performing the given task.  For example, the distillable entanglement is an important measure of entanglement that vanishes for all bound entangled states.  Thus, while faithfulness is intuitively appealing, it is not required for a resource measure.

\subsubsection{Monotonicity}

\label{Sect:Meaures-Monotones}

 A more fundamental property of any resource measure is that its value cannot be increased using free operations.  This is called monotonicity, and it can be seen as encompassing the Golden Rule of QRTs.  A non-negative function $f:\cup_{\mc{H}}\mc{S}(\mc{H})\to\mbb{R}_{\geq 0}$ is called a \textit{resource monotone} if, for any $\Phi\in\mc{O}(A\to B)$ and $\rho\in\mc{S}(A)$, it holds that
\begin{equation}
\label{Eq:monotone1}
f(\rho)\geq f(\Phi(\rho)).
\end{equation}
For QRTs in which any two free states are interconvertible, such as those admitting a tensor product structure, monotonicity immediately implies that $f(\rho)=f(\sigma)$ whenever $\rho$ and $\sigma$ are free.  Thus, the axiom of vanishing for free states can always be satisfied by shifting the function $f$ so that $f(\rho)=f(\sigma)=0$.  

Quantum measurements represented in the form of QC maps $\Phi(\cdot)=\sum_i\Phi_i(\cdot)\otimes\op{i}{i}^X$ are not permitted in every resource theory, quantum thermodynamics being one such example.  However, in QRTs like entanglement and magic states, measurements are physically allowed, and they represent an important component of the theory.  We are thus typically interested in the behavior of a resource monotone when evaluated on QC states.  We say that such a function $f$ is convex linear on QC states if
\begin{equation}
f\left(\sum_ip_i\sigma_i\otimes \op{i}{i}\right)= \sum_i p_if(\sigma_i\otimes\op{i}{i})
\end{equation}
for every QC state $\sigma^{QX}=\sum_ip_i\sigma_i\otimes \op{i}{i}$.  For example, the von Neumann entropy and all Schatten $p$-norms have this property.  If $f$ is convex linear on QC states, then monotonicity obviously implies   
\begin{equation}
\label{Eq:monotone4}
f(\rho)\geq \sum_ip_i f\left(\sigma_i\otimes \op{i}{i}\right).
\end{equation}
Equation \eqref{Eq:monotone4} says that the function is non-increasing on average under any ``flagged-outcome'' quantum measurement.
In many QRTs appending or discarding classical flags is a free operation; i.e., $\rho\leftrightarrow\rho\otimes\op{i}{i}^X$ is allowed for any orthonormal set of vectors $\{\ket{i}\}_i$.  For such QRTs, all monotones must satisfy $f(\rho_i)=f(\rho_i\otimes\op{i}{i})$.  Then from Eq.~\eqref{Eq:monotone4} it follows that
\begin{equation}
\label{Eq:monotone2}
f(\rho)\geq\sum_i p_i f(\sigma_i),
\end{equation}
where $\rho\mapsto\sum_i \Phi_i(\rho)\otimes\op{i}{i}$ is any free QC measurement map, $\sigma_i=\Phi_i(\rho)/p_i$, and $p_i=\tr[\Phi_i(\rho)]$.  This property is sometimes referred to as \textit{strong monotonicity} \cite{Vidal-2000a}.
An intuitive justification for requiring strong monotonicity is to prevent $f$ from increasing on average when the experimenter can post-select or ``flag'' the multiple outcomes of a quantum measurement.  However, Eq.~\eqref{Eq:monotone2} does not precisely reflect this justification since the full description of a post-measurement quantum system with measurement outcome $i$ is the QC state $\sigma_i\otimes \op{i}{i}^X$.  When including the measurement outcome, the statement of $f$ being non-increasing on average is Eq.~\eqref{Eq:monotone4}, which always holds when $f$ satisfies Eq.~\eqref{Eq:monotone1} and demonstrates convex linearity on QC states.


Every nonnegative monotone satisfying Eq.~\eqref{Eq:monotone2} can be used to derive an upper bound on the stochastic convertibility of transforming one state $\rho$ into another $\sigma$ using the free operations.  For any stochastic transformation on $\rho$ generating outcomes $\rho_x$, Eq.~\eqref{Eq:monotone2} immediately implies that $\frac{f(\rho)}{f(\rho_x)}\geq p_x$.  Hence one obtains
\begin{equation}
\label{Eq:Monotone-prob-bound}
P_\rho^{\max}(\sigma)\leq \frac{f(\rho)}{f(\sigma)},
\end{equation}
with $P_\rho^{\max}(\sigma)$ defined in Eq.~\eqref{Eq:Single-shot-stochastic-optimal}.  In fact, $P_\rho^{\max}(\sigma)$ is itself a resource monotone \cite{Vidal-2000a}.

\subsubsection{Convexity}
Convexity of a resource measure says that 
\begin{equation}
f(\sum_i p_i \rho_i)\leq \sum_i p_i f(\rho_i)
\end{equation}
for any collection of density matrices $\rho_i$ and associated probability distribution $p_i$.  This is a very desirable property to have from a mathematical perspective when it comes to computing the value of some function for a given state \cite{Girard-2014a, Regula-2018a}.  A physical interpretation often associated with convex measures is that mixing states never increases the amount of resource.  However, since mixing in this sense describes a process of discarding information, care is needed when relating convexity to the physical process of mixing states (see Section \ref{Sect:Convex}).

\subsubsection{Subadditivity}

A function $f:\cup_{\mc{H}}\mc{S}(\mc{H})\to\mbb{R}$ is called subadditive if 
\begin{equation}
\label{Eq:subadd}
f(\rho\otimes\sigma)\leq f(\rho)+f(\sigma)
\end{equation}
for all $\rho,\sigma$.  While subadditvity is a natural property to suppose of a resource measure, Eq.~\eqref{Eq:subadd} will not hold for all measures in a general QRT.  In particular, for any QRT admitting superactivion, such as nonlocality \cite{Palazuelos-2012a} and quantum channel capacities \cite{Smith-2008a, Cubitt-2011a}, all faithful resource measures will not be subadditive.

The function $f$ is called additive when equality holds in Eq.~\eqref{Eq:subadd} for all states.  This is a strong property that most resource measures will not possess.  However, a procedure known as regularization allows for the general construction of functions that are additive on multiple copies of the same state.  For a function $f:\cup_{\mc{H}}\mc{S}(\mc{H})\to\mbb{R}$, its regularized version is defined by
\begin{equation}
f^\infty(\rho)=\lim_{n\to\infty}\frac{1}{n}f(\rho^{\otimes n}),
\end{equation}
provided the limit exists.  One sufficient condition for the existence of this limit is a weaker form of subadditivity given by $f(\rho^{\otimes(m+n)})\leq f(\rho^{\otimes m})+f(\rho^{\otimes n})$ for all $\rho$ and $m,n$ \cite{Donald-2002a}.  By definition, $f^\infty(\rho^{\otimes n})=nf^\infty(\rho)$, and furthermore, if $f$ is a resource monotone then so will be $f^\infty$.

\subsubsection{Asymptotic Continuity}

Continuity of measure is a reasonable property to expect for any resource measure having physical meaning.  If one state can be obtained from another through subtle perturbation, then one would naturally anticipate their resource content to be very similar.  Of course, the range of a resource measure should grow as the system dimension increases, and thus relative to the dimension, two states can have similar resource content while their absolute difference in resource measure is proportional to the dimension.  Asymptotic continuity is a notion of continuity that considers convergence relative to the dimension.  More precisely, a function $f$ is said to be asymptotically continuous if 
\begin{equation}
|f(\rho)-f(\sigma)|\leq K\epsilon\log[\dim(\mc{H})]+c(\epsilon)
\end{equation}
for all states $\rho$ and $\sigma$ having support on $\mc{H}$, where $K$ is some constant, $\epsilon=\frac{1}{2}\Vert\rho-\sigma\Vert_1$, and $c(\epsilon)$ is any function converging to zero as $\epsilon\to 0$ \cite{Synak-Radtke-2006a}.  For example, the von Neumann entropy $S$ is an asymptotic continuous function, as revealed by the Fannes-Audenaert Inequality
\[|S(\rho)-S(\sigma)|\leq \epsilon \log[\dim(\mc{H})-1]+h(\epsilon),\]
where $h(x)=-x\log x-(1-x)\log(1-x)$ \cite{Fannes-1973a, Audenaert-2007a}.  Asymptotic continuity plays a crucial role in the analysis of asymptotic state convertibility.  As we discuss in Section \ref{Sect:Smooth_Entropies}, the regularized version of all asymptotically continuous measures coincides on states in the same reversibility class.  

The regularization of an asymptotically continuous monotone can be used to bound the rate of any asymptotic transformation.  Suppose that $f$ is an asymptotically continuous monotone with regularization $f^\infty$, and consider the asymptotic convertibility of $\rho$ into $\sigma$.  If $R$ is an achievable rate, then for any $\delta>0$, there exists some $n$ such that $\rho^{\otimes n}\toO \sigma_n$ with $F(\sigma_n,\sigma^{\otimes\lfloor nR\rfloor})>1-\delta$.  In terms of the trace distance, this means $\frac{1}{2}\Vert\sigma_n-\sigma^{\otimes\lfloor nR\rfloor}\Vert<\sqrt{2\delta}$.  Then
\begin{align}
f(\rho^{\otimes n})&\geq f(\sigma_n)\notag\\
&\geq f(\sigma^{\otimes \lfloor nR\rfloor})-K'\sqrt{\delta} nR\log d-c(\delta),
\end{align} 
where the first inequality follows from monotonicity and the second from asymptotic continuity, with $d$ being the dimension of $\text{supp}(\sigma)$.  To obtain the regularizations, we divide both sides by $n$ and take the limit.  Noting that $\lim_{n\to\infty}\frac{1}{n}f(\sigma^{\otimes\lfloor nR\rfloor})\geq \lim_{\lfloor nR\rfloor\to\infty}\frac{R}{\lfloor nR\rfloor}f(\sigma^{\otimes\lfloor nR\rfloor})$, we have $f^\infty(\rho)\geq Rf^\infty(\sigma)-O(\sqrt{\delta})$.  Since this holds for all $\delta>0$ and any achievable rate $R$, the optimal asymptotic rate of conversion is bounded as \cite{Horodecki-2002c}
\begin{equation}
R(\rho\to\sigma)\leq\frac{f^\infty(\rho)}{f^\infty(\sigma)}.
\end{equation}
This can be seen as the asymptotic version of Eq.~\eqref{Eq:Monotone-prob-bound}.

\subsection{General Distance-Based Constructions} 

\label{Sect:Measures_General}

We now describe a general recipe for constructing measures in a general QRT.  The idea is to quantify the amount of resource in a quantum state by ``how far'' it is from the set of free states.  There are many well-defined measures that satisfy the mathematical requirements of distance between two positive operators in Hilbert space.  However from a physical perspective, the property of monotonicity offers a more useful foundation for quantifying distance than standard metric space approaches.   A function $d:\mc{S}(\mc{H}\otimes\mc{H})\to\mbb{R}_{\geq 0}$ is said to be \textit{contractive} under CTPP maps $\Phi$ if $d(\rho,\sigma)\geq d(\Phi(\rho),\Phi(\sigma))$ for arbitrary $\rho$ and $\sigma$.   If $d$ is such a function, then one can define for any QRT the resource measure
\begin{equation}
\mc{R}_d(\rho)=\inf_{\sigma\in\mc{F}(\mc{H})}d(\rho,\sigma).
\end{equation}
This is easily shown to be a resource monotone by the following argument. Let $\rho\in\mS(A)$ be a density matrix, and $\Phi\in\mO(A\to B)$ a free operation. Then,
\begin{align}
\mc{R}_d(\Phi(\rho))&=\inf_{\tau\in\mc{F}(B)}d(\Phi(\rho),\tau)\notag\\
&\leq \inf_{\sigma\in\mc{F}(A)}d(\Phi(\rho),\Phi(\sigma))\notag\\
&\leq \inf_{\sigma\in\mc{F}(A)}d(\rho,\sigma)=\mc{R}_d(\rho)\;.
\end{align} 
Under this construction strong monotonicity is not guaranteed to be satisfied, but for the measures considered below this will indeed be the case.  In general, $\mc{R}_d$ is subadditive, and it will be a convex measure for all convex QRTs.  We describe in the next few sections specific resource measures that are constructed using the approach described here.

\subsection{Entropic measures}\label{Entropic measures}

The starting point for most entropic measures is some quantum generalization of the relative R\'{e}nyi entropies from classical information theory \cite{Renyi-1961a}.  Two well-studied generalizations are the \textit{(quantum) relative R\'{e}nyi entropies} \cite{Petz-1986a}, defined by 
\begin{align}
D_\alpha(\rho\Vert\sigma)=\begin{cases}+\infty\quad\text{if $\alpha\not\in(0,1)\wedge\supp(\rho)\not\subset\supp(\sigma)$}\\\frac{1}{\alpha-1}\log(\tr\rho^\alpha\sigma^{1-\alpha})\quad\text{otherwise}
\end{cases}
\end{align}
for all $\alpha\in[0,+\infty)\setminus\{1\}$, and the \textit{quantum R\'{e}nyi divergences} \cite{Muller-Lennert-2013a} or \textit{sandwiched R\'{e}nyi entropies} \cite{Wilde-2014a}, defined by
\begin{align}
\label{Eq:Sandwiched_Renyi}
\wt{D}_\alpha(\rho\Vert\sigma)=\begin{cases}+\infty\quad\text{if $\alpha\not\in(0,1)\wedge\supp(\rho)\not\subset\supp(\sigma)$}\\
\frac{1}{\alpha-1}\log(\tr[(\sigma^{\frac{1-\alpha}{2\alpha}}\rho\sigma^{\frac{1-\alpha}{2\alpha}})^\alpha])\quad\text{otherwise}
\end{cases}
\end{align}
for all $\alpha\in(0,+\infty)\setminus\{1\}$.  Both families of entropies have found applications in quantum information theory, particularly in quantum hypothesis testing \cite{Mosonyi-2011a, Mosonyi-2015a, Hayashi-2016a}, and various strong converse proofs \cite{Konig-2005a, Wilde-2014a, Cooney-2016a, Leditzky-2016a}.   Of special interest are certain limiting cases of $D_\alpha(\rho\Vert\sigma)$ and $\wt{D}_\alpha(\rho\Vert\sigma)$.  For states $\rho,\sigma$ with $\supp(\rho)\subset\supp(\sigma)$, we have
\begin{align}
&D_0(\rho\Vert\sigma)= -\log\tr(\Pi_\rho\sigma)\\
\lim_{\alpha\to 1}&D_\alpha(\rho\Vert\sigma)\to S(\rho\Vert\sigma):=-\tr[\rho(\log\sigma-\log\rho)]
\end{align}
where  $\Pi_\rho$ is the projector onto $\supp(\rho)$ and $S(\rho\Vert\sigma)$ is the quantum relative entropy \cite{Vedral-2002a}.  
Similarly, we have
\begin{align}
\lim_{\alpha\to 1}&\wt{D}_\alpha(\rho\Vert\sigma)\to S(\rho\Vert\sigma)\\
\lim_{\alpha\to \infty}&\wt{D}_\alpha(\rho\Vert\sigma)\to D_{\max}(\rho\Vert\sigma):=\inf\{\lambda \;|\;\rho\leq 2^\lambda\sigma\},
\end{align}
where $D_{\max}(\rho\Vert\sigma)$ is called the \textit{max-relative entropy} \cite{Datta-2009a}.  Henceforth we will use the definitions $D_1(\rho\Vert\sigma)=\wt{D}_1(\rho\Vert\sigma):=S(\rho\Vert\sigma)$ and $\wt{D}_\infty(\rho\Vert\sigma):=D_{\max}(\rho\Vert\sigma)$.

The relative R\'{e}nyi entropy serves as building block for a number of other useful entropic quantities.  For example, if $\rho$ is a state whose support is contained in a $d$-dimensional subspace $\mc{H}$, then the R\'{e}nyi entropy of the eigenvalues of $\rho$ can be obtained by taking $\sigma=\mbb{I}_{\mc{H}}$:
\begin{equation}
S_\alpha(\rho):=-D_\alpha(\rho\Vert\mbb{I}_{\mc{H}})=\frac{-1}{\alpha-1}\log(\tr\rho^
\alpha).
\end{equation}
As described in Section \ref{Sect:Majorization-Entanglement}, the quantities $S_\alpha(\rho)$ determine catalytic convertibility of states in many resource theories.  For special choices of $\alpha$ we have
\begin{align}
\begin{cases}
H_{\min}(\rho)=-D_{\max}(\rho\Vert\mbb{I}_{\mc{H}})=-\log\Vert\rho\Vert_\infty,\\
S(\rho)=-D_1(\rho\Vert\mbb{I}_{\mc{H}})+\log d=-\tr[\rho\log(\rho)],\\
H_{\max}(\rho)=-D_{0}(\rho\Vert\mbb{I}_{\mc{H}})=\log \text{rank}[\rho]
\end{cases}
\end{align}
where $\Vert\rho\Vert_\infty$ is the largest eigenvalue of $\rho$.  


For operational purposes, one crucial property of the R\'{e}nyi relative entropies is that they are contractive under CPTP maps for certain ranges of $\alpha$, i.e., $D_\alpha(\rho\Vert\sigma)\geq D_\alpha(\Phi(\rho)\Vert\Phi(\sigma))$ for any CPTP map $\Phi$ and all states $\rho$ and $\sigma$.  This is also sometimes referred to as a data-processing inequality.  For the range $\alpha\in[0,2]$, the quantum relative R\'{e}nyi entropy is monotonic under CPTP maps \cite{Petz-1986a}, similarly the quantum R\'{e}nyi divergence behaves monotonically for $\alpha\in[1/2,\infty]$ \cite{Frank-2013a, Beigi-2013a, Wilde-2018a}.  It also holds that $D_\alpha(\rho\Vert\sigma)=0$ if and only if $\rho=\sigma$, and likewise for the equality $\wt{D}_\alpha(\rho\Vert\sigma)=0$.

Given these properties, one then obtains a whole family of resource measures for any QRT $(\mc{F},\mc{O})$.  These are functions $\mc{R}_\alpha$ and $\wt{\mc{R}}_\alpha$ whose values on any $\rho\in\mc{S}(\mc{H})$ are given by
\begin{align}
\mc{R}_\alpha(\rho):&=\inf_{\sigma\in\mc{F}(\mc{H})}D_\alpha(\rho\Vert\sigma)\qquad\text{for $\alpha\in[0,2]$}, \\
\label{Eq:Sandwiched_Renyi_Resource}
\wt{\mc{R}}_\alpha(\rho):&=\inf_{\sigma\in\mc{F}(\mc{H})}\wt{D}_\alpha(\rho\Vert\sigma)\qquad\text{for $\alpha\in[1/2,\infty]$}.
\end{align}
In particular, we define
\begin{align}
\mc{R}_{\max}(\rho):&=\wt{\mc{R}}_\infty(\rho)=\inf_{\sigma\in\mc{F}(\mc{H})}\!\{\lambda \;|\;\rho\leq 2^\lambda\sigma\}.\label{Eq:R-max}
\end{align}
Note that whenever the set of free states $\mc{F}(\mc{H})$ is closed, each infimum is attained by some free state $\sigma$.  These are true resource measures since $\mc{R}_\alpha(\rho)=0$ and $\wt{\mc{R}}_\alpha(\rho)=0$ if and only if $\rho\in\mc{F}(\mc{H})$.  Monotonicity under any free CPTP map $\Phi$ follows from the data-processing inequality and the discussion of Section \ref{Sect:Measures_General}.  Furthermore, $\mc{R}_\alpha$ and $\wt{\mc{R}}_\alpha$ demonstrate strong monotonicity for $\alpha\in[1,2]$.  To see this, consider any free transformation $\rho\to\Phi(\rho)=\sum_ip_i\rho_i\otimes\op{i}{i}$ and suppose for simplicity that $\mc{F}(\mc{H})$ is closed.  Then there exists some $\sum_i q_i\sigma_i\otimes\op{i}{i}\in\mc{F}(\mc{H}\otimes\mc{H}^X)$ attaining the minimum in the definition of $R_\alpha(\Phi(\rho))$.  Monotonicity or $R_\alpha$ then implies
\begin{align}
\mc{R}_\alpha(\rho) &\geq D_\alpha\left(\sum_ip_i\rho_i\otimes\op{i}{i}\bigg|\bigg|\sum_i q_i\sigma_i\otimes\op{i}{i}\right)\notag\\
&=\frac{1}{\alpha-1}\log\left(\sum_ip_i^\alpha q_i^{1-\alpha}\tr[\rho_i^\alpha \sigma_i^{1-\alpha}]\right)\notag\\
&\geq\frac{1}{\alpha-1}\sum_ip_i\log\left((q_i/p_i)^{1-\alpha}\tr[\rho_i^\alpha \sigma_i^{1-\alpha}]\right)\notag\\
&\geq \sum_i p_i\mc{R}_\alpha(\rho_i),
\end{align}
where the concavity of logarithm function has been used for $\alpha>1$, and the last inequality follows from the fact that $-\sum_i p_i\log(q_i/p_i)$ is nonnegative as it is the relative entropy of the distributions $(p_i)_i$ and $(q_i)_i$.  An analogous argument holds for $\wt{R}_\alpha$.  It is also easy to see that these measures are subadditive and subextensive in general, and they are convex for all convex QRTs.

For example, in the QRT of thermodynamics the Gibbs state $\gamma_H$ is the unique free state for a given Hamiltonian and bath temperature.  The quantities $\mc{R}_\alpha(\rho)=D_\alpha(\rho\Vert\gamma_H)$ and $\mc{R}_\alpha(\rho)=\wt{D}_\alpha(\rho\Vert\gamma_H)$ are then resource measures, and their monotonicity represents an $\alpha$-family of second laws of thermodynamics \cite{Brandao-2015b}.

The most important entropic measure emerges when $\alpha=1$, and it is called the \textit{relative entropy of resource}, denoted as
\begin{equation}
\mc{R}_{\text{rel}}(\rho)=\inf_{\sigma\in\mc{F}(\mc{H})}S(\rho\Vert\sigma)
\end{equation}
for $\rho\in\mc{S}(\mc{H})$.  It can be shown that $\mc{R}_{\text{rel}}$ is asymptotically continuous in any convex QRT for which the maximally mixed state is free \cite{Donald-1999a, Synak-Radtke-2006a} (see also \cite{Winter-2016a}).  As mentioned above, the regularization of $\mc{R}_{\text{rel}}$ is the key quantifier in most asymptotic resource tasks such as the asymptotic convertibility and erasing resources (see also \ref{Sect:Smooth_Entropies}).

One can also consider a relative entropy-type measure by switching the roles of $\rho$ and $\sigma$ \cite{Vedral-1998a, Eisert-2003a}.  That is, the function 
\begin{equation}
\mc{R}'_{\text{rel}}(\rho)=\inf_{\sigma\in\mc{F}(\mc{H})}S(\sigma\Vert\rho)
\end{equation}
provides a resource measure that vanishes on free states and behaves monotonically under free operations.  In addition, $\mc{R}'_{\text{rel}}$ has the relatively rare feature of being an additive function in QRTs admitting a tensor product structure.  This follows from the equality
\begin{align}
S(\sigma^{AB}\Vert\rho^A\otimes\rho^B)&=-\tr[\sigma^{AB}\log(\rho_1\otimes\rho_2)]-S(\sigma^{AB})\notag\\
&=S(\sigma^A\Vert\rho^A)+S(\sigma^B\Vert\rho^B)+I(A:B)_{\sigma}.\notag
\end{align}
This is minimized by taking $\sigma^{AB}=\sigma^A\otimes\sigma^B$.  In QRTs with a tensor product structure, $\sigma^A\otimes\sigma^B$ is free whenever $\sigma^{AB}$ is free.  Hence, additivity holds: $\mc{R}_{\text{rel}}'(\rho^A\otimes \rho^B)=\mc{R}_{\text{rel}}'(\rho^A)+\mc{R}_{\text{rel}}'(\rho^B)$. Unfortunately, despite having these nice properties, $\mc{R}'_{\text{rel}}$ is usually not convenient to use as a resource measure.  If, for example, the set of free states contains full rank states, then $\mc{R}'_{\text{rel}}(\rho)$ will diverge for all states $\rho$ not of full rank.

\subsection{Geometric Measures}

The relative entropies are not proper metrics in the mathematical sense since they, for example, fail to satisfy the triangle inequality.  One resource measure derived from a true metric is the \textit{trace distance of resource}, defined as
\begin{equation}
\label{Eq:trace-distance-resource}
\mc{R}_{\tr}(\rho)=\inf_{\sigma\in\mc{F}(\mc{H})}D_{\tr}(\rho,\sigma)=\inf_{\sigma\in\mc{F}(\mc{H})}\frac{1}{2}\Vert\rho-\sigma\Vert_1
\end{equation}
for any $\rho\in\mc{S}(\mc{H})$.  Since $\Vert\cdot\Vert_1$ is contractive under CPTP maps, it automatically holds that $\mc{R}_{\tr}(\rho)\geq \mc{R}_{\tr}(\Phi(\rho))$ for any \emph{free} channel $\Phi$.

The trace distance of resource has an appealing operational meaning in terms of state distinguishability.  Namely, for a system prepared in one of two states $\rho_0$ and $\rho_1$ with respective probabilities $p_0$ and $p_1$, the minimum error in guessing the correctly prepared state after measuring the system is given by $\frac{1}{2}-\frac{1}{2}\Vert p_0\rho_0-p_1\rho_1\Vert_1$ \cite{Holevo-1973a, Helstrom-1969a}.  When guessing between the equiprobable state preparation of $\rho$ or some free state in the QRT, the quantity $\frac{1}{2}-\frac{1}{2}\mc{R}_{\tr}(\rho)$ represents the minimum-error probability over all the free states, provided the QRT is convex \cite{Gutoski-2005a}.

Another example of a metric used in quantum information theory is the Bures metric, $D_B(\rho,\sigma)=\sqrt{2}\sqrt{1-F(\rho,\sigma)}$.  Using it, the Bures distance of resource measure can be defined
\begin{equation}
\mc{R}_B(\rho)=\inf_{\sigma\in\mc{F}(\mc{H})}D_{B}(\rho,\sigma).
\end{equation}

Both $\mc{R}_{\tr}$ and $\mc{R}_B$ are called geometric measures because they are built from a                                                                                                                                                true metric.  Historically however, the first type of geometric measure studied was for pure states in entanglement theory, and it involves minimizing $F(\rho,\sigma)^2$ directly \cite{Shimony-1995a, Barnum-2001a}.  This is typically referred to as the geometric measure of entanglement, and we generalize it to a measure on pure states in an arbitrary QRT as
\begin{align}
\mc{R}_G(\ket{\psi})&=\inf_{\sigma\in\mc{F}(\mc{H})}(1-F(\op{\psi}{\psi},\sigma)^2)=\frac{1}{4}\mc{R}_B(\ket{\psi})^4.
\label{Eq:Geometric_Measure}
\end{align}
If $\mc{F}(\mc{H})$ is a convex set whose extreme points are pure states, then this infimum is always attained by a pure state.  Entanglement theory is one such QRT where this is the case, and thus in entanglement theory $\mc{R}_G(\ket{\psi})$ is essentially given by the largest overlap that $\ket{\psi}$ has with a product state.  Such a quantity has wide applications in quantum information theory, such as quantifying performance in LOCC state discrimination \cite{Markham-2007a} and quantum algorithms \cite{Biham-2002a, Gross-2009a} 

There are two ways to extend $\mc{R}_G$ defined in Eq.~\eqref{Eq:Geometric_Measure} to be a resource measure for mixed states as well \cite{Chen-2014a}.  The first and most obvious does not change the functional form at all, and one simply defines
\begin{equation}
\mc{R}_G(\rho)=\inf_{\sigma\in\mc{F}(\mc{H})}(1-F(\rho,\sigma)^2)
\end{equation}
for $\rho\in\mc{S}(\mc{H})$.  As an interesting observation, the $\alpha=1/2$ case of $\wt{\mc{R}}_\alpha(\rho)$ in Eq.~\eqref{Eq:Sandwiched_Renyi_Resource} reduces to a quantity quite similar to $\mc{R}_G(\rho)$.  One sees from Eq.~\eqref{Eq:Sandwiched_Renyi} that
\begin{equation}
\wt{D}_{1/2}(\rho\Vert\sigma)=-2\log\tr\sqrt{\sqrt{\sigma}\rho\sqrt{\sigma}}=-\log F(\rho,\sigma)^2.
\end{equation}
Minimizing both sides over all free states $\sigma$ yields
\begin{equation}
\mc{R}_G(\rho)=1-2^{-\wt{\mc{R}}_{1/2}(\rho)}.
\end{equation}
Monotonicity of $\wt{R}_{1/2}$ then implies $F(\Phi(\rho),\Phi(\sigma))\geq F(\rho,\sigma)$ for any CPTP map $\Phi$.  

The second approach to obtaining a mixed-state measure from Eq.~\eqref{Eq:Geometric_Measure} uses a standard technique in entanglement theory known as convex-roof extension \cite{Uhlmann-2010a, Wei-2003a}.  For any arbitrary mixed state, its extended geometric measure is defined by
\begin{equation}
\label{Eq:convex-roof}
\mc{R}_G'(\rho)=\inf_{\{\ket{\phi_i},p_i\}}\sum_i p_i\mc{R}_G(\ket{\phi_i}),
\end{equation}
where the infimum is taken over all pure-state ensembles such that $\rho=\sum_ip_i\op{\psi_i}{\psi_i}$.  Remarkably, if the QRT has the property that its free states are the convex hull of some set of pure states, then the two proposed geometric measures of resource coincide \cite{Streltsov-2010a}; i.e.,
\begin{equation}
\mc{R}_G(\rho)=\mc{R}_G'(\rho).
\end{equation}
A consequence of this equality is that $\mc{R}_G(\rho)$ necessarily satisfies strong monotonicity since $\mc{R}_G'$ is built from a pure-state function $\mc{R}_G$ that satisfies strong monotonicity under pure-state transformations \cite{Vidal-2000a}.

Even though $\mc{R}_G$ has a relatively convenient mathematical form, it lacks an operational interpretation like the minimum error guessing probability associated with $\mc{R}_{\tr}$.  However, $\mc{R}_G$ is still useful in deriving bounds for the latter since from Eq.~\eqref{Eq:trace-fidelity-relation} we immediately have
\begin{equation}
1-\sqrt{1-\mc{R}_G(\rho)}\leq \mc{R}_{\tr}(\rho)\leq \sqrt{\mc{R}_G(\rho)}.
\end{equation}

\subsection{Witness-based measures}

\label{Sect:Witness}

The next family of measures we consider relies on the idea of resource \textit{witnessing}.  In a general QRT, a \textit{witness} for the particular resource is a quantum observable $W\in\mc{B}(\mc{H})$ such that 
\begin{equation}
\label{Eq:Witness_Definition}
\begin{cases} \exists \sigma\not\in \mc{F}(\mc{H}):\tr[W\sigma]< 0\\
\forall \rho\in\mc{F}(\mc{H}):\tr[W\rho]\geq 0.
\end{cases}
\end{equation}
If $W$ satisfies these two conditions, then it is said to ``witness'' the resource content of $\sigma$.  For convex closed QRTs, the separating hyperplane theorem assures that every resource state possesses at least one observable that witnesses it \cite{Barvinok-2002a}.  Consequently, in such QRTs witnesses themselves can be used to fully characterize the free states.

The theory of witnesses is very appealing from an experimental perspective.  If multiple copies of a quantum system are prepared in the same unknown state, the presence of a quantum resource can be detected whenever a resource witness yields a negative average measurement value.  In entanglement theory, the study of entanglement witnesses is a mature research area \cite{Chruscinski-2014a}, with a number of experimental applications already developed \cite{Guhne-2009a}.  In the QRT of Bell nonlocality, the so-called Bell operators serve as resource witnesses, and the experimental implementation of these witnesses in a ``loophole'' free manner has been a long quest that only recently became completed \cite{Hensen-2015a, Giustina-2015a, Shalm-2015a}.

Beyond distinguishing resource states from free ones, witnesses can also be used for the construction of resource measures.  This approach was originally taken by \textcite{Brandao-2005a} in the context of entanglement, but it has recently been expanded by \textcite{Regula-2018a} to encompass general convex resource theories.  While we encourage the reader to consult the latter for a detailed mathematical development of the subject, here we just review the basic framework and describe the more well-known applications.  

For a collection of hermitian operators $\mc{C}(\mc{H})\subset \text{Herm}(\mc{H})$, we let $\mc{C}^*(\mc{H})\subset \text{Herm}(\mc{H})$ denote its dual cone, i.e., 
\[\mc{C}^*(\mc{H}):=\{X:\tr[XY]\geq 0, \;\forall Y\in \mc{C}(\mc{H})\}.\]  For a given QRT $(\mc{F},\mc{O})$ and any Hilbert space $\mc{H}$, let $\mc{C}(\mc{H})$ be a collection of hermitian operators that is closed under the adjoint of every free CPTP map; more precisely, for spaces $\mc{H}$ and $\mc{H}'$ it holds that $\Phi^*(X)\in\mc{C}(\mc{H})$ whenever $X\in\mc{C}(\mc{H}')$ and $\Phi\in\mc{O}(\mc{H}\to \mc{H}')$.  Then for $\rho\in\mc{S}(\mc{H})$ we define the function
\begin{equation}
\label{Eq:Witness_measure}
\mc{R}_W^\mc{C}(\rho)=\sup\{-\tr[X\rho]:X\in\mc{F}^*(\mc{H})\cap\mc{C}(\mc{H})\}.
\end{equation}
By definition, $\mc{R}_W^\mc{C}(\rho)=0$ for all $\rho\in\mc{F}(\mc{H})$, and monotonicity is easy to verify.  First note that $\mc{F}^*(\mc{H})$ itself is closed under the adjoint of every free map, which follows from the observation that if $0\leq\tr[X\rho]$ for all free states $\rho$, then under any RNG map $\Phi$, we likewise have $0\leq\tr[X\Phi(\rho)]=\tr[\Phi^*(X)\rho]$.  Monotonicity of $\mc{R}^\mc{C}_W$ is then a consequence of the inequality
\[-\tr[X\Phi(\rho)]=-\tr[\Phi^*(X)\rho]\leq \mc{R}_W^\mc{C}(\rho),\]
since $\Phi^*(X)\in\mc{F}^*(\mc{H})\cap\mc{C}(\mc{H})$.  In any QRT with tensor product structure, one can see that $\mc{F}^*(\mc{H})$ is also closed under processing of classical flags; i.e., $\sum_{i} X_i\otimes\op{i}{i}^X\in\mc{F}^*(\mc{H}\otimes\mc{H}^X)$ if and only if $X_i\in\mc{F}^*(\mc{H})$ for all $i$.  If the set $\mc{C}(\mc{H})$ also has this property, then the constructed resource measure $\mc{R}_W^\mc{C}(\rho)$ is convex linear on QC states.  Consequently, the monotonicty 
\begin{equation}
\mc{R}^C_W(\rho)\geq\sum_i p_i\mc{R}^C_W(\rho_i)
\end{equation} 
holds for any free transformation $\rho\to\sum_i p_i \rho_i\otimes\op{i}{i}$.

As an example, for any real numbers $m<n$, the set $\mc{C}=\{X:-m\mbb{I}\leq X\leq n\mbb{I}\}$ is closed under the adjoint of every CPTP map \cite{Brandao-2005a}.  Indeed, if $n\mbb{I}\geq X$, then $0\leq \Phi^*(n\mbb{I}-X)=n\mbb{I}-\Phi^*(X)$, where the inequality follows from $\Phi$ being CP and the equality follows from $\Phi$ being trace preserving, which implies $\Phi^*$ is unital.  A similar argument holds for the operator $X+m\mbb{I}$, and therefore we see that $\mc{R}^C_{W}$ is a resource measure for any choice of $m$ and $n$ \cite{Regula-2017a}.  

In many cases of interest, the chosen set $\mc{C}$ is a cone in the space of hermitian matrices.  That is, $\mc{C}$ is such that $\sum_ic_i X_i\in\mc{C}$ whenever $X_i\in \mc{C}$ and $c_i\geq 0$.  From its definition, $\mc{F}^*$ is also a cone, and thus the value $\mc{R}^{\mc{C}}_W(\rho)$ represents a conic optimization problem \cite{Boyd-2004a}.  Every conic optimization problem has a dual representation obtained by the introduction of Lagrange multipliers.  If $\mc{C}$ is convex with $\mc{F}^*\cap\mc{C}$ having a nonempty interior, then strong duality holds, and we are assured that the dual-optimal solution is equal to $\mc{R}^{\mc{C}}_{W}(\rho)$.  In what follows, we consider three resource measures built using the dual formulation of $\mc{R}^{\mc{C}}_W$ under different choices of $\mc{C}$.  

\subsubsection{Trace Distance Measures}

Building from the example in the previous section, consider the specific choice of $m=n=1$ so that $\mc{C}=\{X:-\mbb{I}\leq X\leq\mbb{I}\}$.  The dual optimization problem of Eq.~\eqref{Eq:Witness_measure} is given by
\[\inf_{\substack{R,S\geq 0 \\ \omega\in\mc{F}_+(\mc{H})}}\{\tr[R+S]:\rho-\omega=R-S\},\]
where $\mc{F}_+(\mc{H})=\mc{F}^{**}(\mc{H}):=\{\lambda\sigma:\sigma\in\mc{F}(\mc{H}),\lambda\geq 0\}$.  It is well-known that this minimization problem is equivalent to minimizing the trace distance between $\rho$ and the set $\mc{F}_+(\mc{H})$.  Since strong duality holds for all non-trivial QRTs in this case, we therefore have
\begin{equation}
\label{Eq:modified-trace-distance}
\mc{R}^+_{\tr}(\rho):=\inf_{\omega\in\mc{F}_+(\mc{H})}\Vert\rho-\omega\Vert_1.
\end{equation}
We will refer to this as the \textit{modified trace distance of resource}, as it has been given such a name in the study of quantum coherence \cite{Yu-2016a}.

One advantage of considering the modified trace distance of resource rather than the standard $\mc{R}_{\tr}$ of Eq.~\eqref{Eq:trace-distance-resource} is that the former satisfies strong monotonicity while the latter does not.  To see where the distinction arises, consider a multi-outcome free operation $\Phi(\cdot)=\sum_i\Phi_i(\cdot)\otimes\op{i}{i}^X$.  For an input state $\rho\in\mB(\mc{H}^S)$, its post-measurement trace distance of resource is given by
\begin{align}
\mc{R}_{\tr}(\Phi(\rho))=\inf_{\omega\in\mc{F}(\mc{H}\otimes\mc{H}^X)}\left|\left|\sum_{i}\mc{E}_i(\rho)\otimes\op{i}{i}^X-\omega^{SX}\right|\right|_1,\notag
\end{align}
where $\omega^{SX}$ is a QC state of the form $\omega^{SX}=\sum_i q_i\omega_i\otimes\op{i}{i}^X$ for quantum states $\omega_i$.  Then writing $\rho_i=\mc{E}_i(\rho)/p_i$ and $p_i=\tr[\mc{E}_i(\rho)]$, we have
\begin{align}
\label{Eq:tr-no-dp}
\mc{R}_{\tr}(\Phi(\rho))=\sum_ip_i \inf_{\substack{q_i\omega_i\in\mc{F}_+(\mc{H})\\\sum_iq_i=1}}\left|\left|\rho_i-\frac{q_i}{p_i}\omega_i\right|\right|_1.
\end{align}
In general the terms in the sum on the RHS will differ from $\mc{R}_{\tr}(\rho_i)$.  On the other hand, if the normalization condition $\sum_iq_i=1$ is removed from the infimum in Eq.~\eqref{Eq:tr-no-dp}, then the RHS becomes $\sum_i p_i\mc{R}_{\tr}^+(\rho_i)$.  Hence, we can conclude that the modified trace distance demonstrates convex linearity on QC states and therefore also strong monotonicity,
\begin{equation}
\label{Eq:strong-monotone}
\mc{R}^+_{\tr}(\rho)\geq \sum_i p_i \mc{R}^+_{\tr}(\rho_i).
\end{equation}
While strong monotonicity of $\mc{R}_{\tr}$ is known to hold in coherence and entanglement theories for special cases \cite{Rana-2016a, Eisert-2003a}, explicit counterexamples can be found \cite{Yu-2016a}.

\subsubsection{Robustness Measures}

\label{Sect:Robustness}

One appealing way to quantify the resource content of a quantum state is by its resilience to hold resource when mixed with some other state.  Let $\mc{T}(\mc{H})$ be any set of quantum states, and for a given state $\rho$ consider mixtures of the form $\omega=\lambda\rho+(1-\lambda)\sigma$ with $\sigma\in\mc{T}(\mc{H})$.  If $\mc{F}(\mc{H})\cap\mc{T}(\mc{H})\not=\emptyset$, there always exists a $\lambda$ such that $\omega$ is a free state for some choice of $\sigma\in\mc{T}(\mc{H})$.  The smallest such $\lambda$ having this property quantifies a resource robustness of $\rho$ against mixtures from $\mc{T}(\mc{H})$.

For many interesting choices of $\mc{T}(\mc{H})$, this notion of robustness can be captured using the framework of resource witnesses.   Specifically, suppose that the set
\[\mc{C}(\mc{H})=\mbb{I}-\mc{T}^*=\{X:\mbb{I}-X\in\mc{T}^*(\mc{H})\}\]
is closed under the adjoint action of every free map.  The dual of Eq.~\eqref{Eq:Witness_measure} is readily found to be
\[\inf_{\omega\in\mc{F}_+(\mc{H})}\{\tr[\omega]-1:\omega-\rho\in\mc{T}_+(\mc{H})\},\]
where $\mc{T}_+(\mc{H})$ is the conic hull of $\mc{T}(\mc{H})$.  If there exists an $\omega\in\mc{T}_+(\mc{H})$ so that $\rho+\omega$ lies in the interior of $\mc{F}_+(\mc{H})$, then strong duality holds.  In this case, we can rewrite the previous equation in a more standard form, which represents the \textit{resource robustness of $\rho$ against $\mc{T}$},
\begin{equation}
\mc{R}^{\mc{T}}_{\text{rob}}(\rho):=\inf_{\gamma\in\mc{T}(\mc{H})}\left\{s:\frac{\rho+s\gamma}{1+s}\in\mc{F}(\mc{H})\right\}.
\end{equation}

Different robustness measures are obtained for different choices of $\mc{T}$.  For example, if $\mc{T}=\mc{F}$, then one obtains the resource \textit{absolute} robustness \cite{Vidal-1999a}, which quantifies the minimum fraction that a given state $\rho$ must be mixed with a free state so that their convex combination is free.  An analogous measure has been studied in the QRT of ``magic states'' \cite{Howard-2017a}.  The other extreme involves taking $\mc{T}$ to be the set of all density matrices.  This is called the resource \textit{global} robustness \cite{Harrow-2003a}, and it is written as
\begin{equation}
\label{Eq:Global-robustness}
\mc{R}_{\text{rob}}(\rho)=\inf_{\gamma\in\mc{S}(\mc{H})}\left\{s:\frac{\rho+s\gamma}{1+s}\in\mc{F}(\mc{H})\right\}.
\end{equation}
Beyond its study in entanglement theory, the global robustness has recently been investigated in the QRTs of coherence and asymmetry \cite{Napoli-2016a, Piani-2016a}.  In addition, the global robustness plays an important role in the study of general asymptotic resource reversibility since it is used to define the class of asymptotically-RNG transformations \cite{Brandao-2008a, Brandao-2015a} (see Section \ref{Sect:Smooth_Entropies}).  A thoughtful comparison between Eq.~\eqref{Eq:Global-robustness} and Eq.~\eqref{Eq:R-max} shows that
\begin{equation}
\mc{R}_{\max}(\rho)=\log(1+\mc{R}_{\text{rob}}(\rho))
\end{equation}
\cite{Datta-2009b}.  The RHS is sometimes referred to as the \textit{log-robustness}.  

A third type of robustness commonly considered is the \textit{resource random robustness of $\rho$}, and it is defined by taking $\mc{T}(\mc{H})=\{\frac{1}{d}\mbb{I}\}$, where $d=\dim[\mc{H}]$ \cite{Vidal-1999a}.  Assuming the maximally mixed state is free, this quantity essentially measures how much ``white noise'' must be mixed with $\rho$ before it becomes a free state.  Alternatively, it can be characterized as the minimal depolarizing parameter $\lambda$ that removes all resource content from $\rho$ when its sent through the completely depolarizing channel $\Phi_\lambda(\rho)=(1-\lambda)\rho+\lambda\frac{1}{d}\mbb{I}$.  However, unlike the absolute and global robustness measures, the resource random robustness is not a monotone in general \cite{Harrow-2003a}.  For the choice of $\mc{T}(\mc{H})$ taken here, it holds that $\mc{C}(\mc{H})=\{X:\tr[X]\leq d\}$.  Consequently, it follows that the resource random robustness is a monotone in any QRT for which every free map has an adjoint that is trace-non-increasing on the set of hermitian operators.  For example, the random robustness is a monotone if the free CPTP maps are unital, a fact which also directly follows from the definition of random robustness.

\subsubsection{Resource-Rank Measures}

\label{Sect:Rank}

We introduce one final family of witness-based measures that extends the general construction described above.  Suppose that $\mc{F}(\mc{H})$ possesses a collection of rank-one free states $\{\op{f_i}{f_i}\}_i$ such that the vectors $\{\ket{f_i}\}_i$ form a basis for $\mc{H}$.  Then one can define the resource rank of an arbitrary pure state $\ket{\phi}$ as
\begin{equation}
\mc{R}_{\text{rk}}(\ket{\phi})=\inf\{r\;:\; \ket{\phi}=\sum_{i=1}^r c_i\ket{f_i},\;\;\text{the $\ket{f_i}$ are free}\}.
\end{equation}
The rank is then extended to mixed states using the prescription of Eq.~\eqref{Eq:convex-roof},
\begin{equation}
\mc{R}_{\text{rk}}(\rho)=\inf_{\{\ket{\phi_i},p_i\}}\sum_i p_i\mc{R}_{\text{rk}}(\ket{\phi_i}),
\end{equation}
where the infimum is taken over all pure-state decompositions of $\rho$.  The most well-known example of $\mc{R}_{\text{rk}}$ is in entanglement theory where it is referred to as the Schmidt rank or Schmidt number of a bipartite state \cite{Terhal-2000a}.  In multipartite entanglement theory, the Schmidt rank generalizes to the so-called tensor rank \cite{Chitambar-2008a}, and it can be used certify the presence of genuine multipartite entanglement \cite{Eisert-2001a}.  Outside of entanglement, the coherence rank has been used to quantify and experimentally measure multilevel coherence \cite{Ringbauer-2018a}.  Additionally, in the resource theory of ``magic states,'' the stabilizer rank of magic states has been related to their classical simulation costs \cite{Bravyi-2016a}.

If we let $\mc{S}_k(\mc{H})$ denote the set of states on space $\mc{H}$ with resource rank no greater than $k$, then $\mc{S}_k(\mc{H})$ forms a convex set.  Thus, there exist observables $W$ which, analogous to Eq.~\eqref{Eq:Witness_Definition}, witness a resource rank greater than $k$ \cite{Sanpera-2001a, Shahandeh-2014a, Regula-2018a}.  That is, $\tr[W\sigma]<0$ for some $\sigma\not\in \mc{S}_k(\mc{H})$ yet $\tr[W\rho]\geq 0$ for all $\rho\in\mc{S}_k(\mc{H})$.  Note that these witnesses depend just on the structure of the free states in the resource theory.  To promote $\mc{R}_{\text{rk}}(\rho)$ to a genuine resource measure, it must be shown to be a monotone under the free operations.  In the resource theories of entanglement and magic states, the resource rank is a monotone under LOCC and stabilizer operations, respectively.  For coherence, the coherence rank has been shown to be a monotone under IO \cite{Baumgratz-2014a} but not DIO \cite{Yue-2018a}.  Operationally, the resource ranks have been shown to quantify one-shot resource costs in entanglement \cite{Buscemi-2011a} and coherence theories \cite{Zhao-2018a}.

\section{General Techniques, Mathematical Tools, and Results}
\label{Sect:tools}

\subsection{Majorization theory}

We saw earlier that the set of free operations induces a preorder. Specifically, we write $\rho\toO\sigma$ if there exists a free operation $\Phi\in\mO$ such that $\sigma=\Phi(\rho)$. In general, determining if $\rho\toO\sigma$ can be a difficult task.  Moreover, even if the solution is computationally feasible, it is still possible that the preorder $\toO$ has no simple elegant characterization. Remarkably, for \emph{pure} bipartite entanglement the pre-order $\toO$ has a simple characterization known as majorization. 

Majorization theory is a topic in matrix analysis with several textbooks written on the subject; see e.g.~\cite{Marshall-2011a} and~\cite{Bhatia-2000a}.
It has many applications in different areas of science, from mathematics to economy and statistics, and more recently quantum physics. Nielsen's discovery \cite{Nielsen-1999a} of the role that majorization plays in entanglement theory has been extended to other resource theories using generalizations of majorization.  These include thermo-majorization~\cite{Horodecki-2013b}, conditional majorization~\cite{Gour2015}, relative majorization and sub-relative majorization~\cite{Renes-2016a}, matrix majorization~\cite{Dahl-1999}, and quantum majorization~\cite{Gour2018}. Before introducing the mathematical definitions, we start with a simple example to gain some intuition behind the definition of majorization.

\subsubsection{Majorization in Gambling}

Consider the following three gambling games. In each game a player is given the choice between one of two biased dice.  The first has probability vector $\p=(1/16,1/2,1/8,1/16,1/4,0)^T$ for the distribution of symbols (outcomes) $x\in\{1,2,3,4,5,6\}$, while the second is described by probability vector $\q=(1/12,1/3,1/3,1/4,0,0)^T$.  In the first game, the player rolls her die, and she wins if she correctly guesses its outcome.  In the second game she can guess two outcomes, and in the third she can guess three outcomes.  Which die should the player choose in each of these games to maximize her winning probability? 

Clearly, in the first game the answer is the first die since if she bets on outcome 2 she will have a $1/2$ chance to win the game, whereas with the second die the highest probability of correctly guessing is $1/3$.  
In the second game, the player should also choose the first die since if she gambles on the outcomes 2 and 5 she will have a probability $1/2+1/4=3/4$ to win, whereas with the second die the maximum probability that she can win is $1/3+1/3=2/3<3/4$. However, for the third game, the second die has a higher probability to win since $1/3+1/3+1/4=11/12$, which is greater than $1/2+1/4+1/8=7/8$. Therefore, the best die to choose in this scenario depends on the game.

In general, if the player can guess $k$ symbols, she should choose a die with probability vector $\p=(p_1,...,p_6)^T$ over a die with probability vector $\q=(q_1,...,q_6)^T$ if and only if the following condition holds:
\be\label{majo}
\sum_{x=1}^{k}p_x^{\da}\geq \sum_{x=1}^{k}q_x^{\da}\;,
\ee
where the symbol $\da$ stands for the rearrangement of the elements of $\p$ in a non-increasing order; i.e., $p_1^\da\geq p_2^\da\geq...\geq p_6^\da$.
If the above relation holds for all $k=1,...,5$ then the player should choose the $\p$-die over the $\q$-die to maximize her chances of winning every such game.  When Eq.~\eqref{majo} is satisfied for all $k$, we say that the probability vector
$\p$ \textit{majorizes} the probability vector $\q$, and we write $\q\prec\p$ or $\p\succ\q$. This definition extends to arbitrary probability distributions $\p,\q\in\mbb{R}^n_{\geq 0}$; that is, $\p\succ\q$ if and only if Eq.~\eqref{majo} holds for all $k=1,...,n$ with equality for $k=n$. 

Majorization is a preorder. 
Moreover, if both $\p\prec\q$ and $\q\prec\p$ then $\p$ and $\q$ are related by a permutation matrix. Therefore, up to permutations, the majorization relation $\prec$ can be viewed as a partial order.
It is simple to check that any $d$-dimensional probability vector $\p$ satisfies
$$
(1/d,...,1/d)^T\prec\p\prec(1,0,...,0)^T
$$
This relationship lends itself to the interpretation that majorization measures how spread-out a probability distribution is over its possible events.
This idea will be further clarified in what follows.  


Consider again the $\p$-die and suppose that the player can permute (relabel) its symbols before guessing an outcome. Such a relabeling corresponds to a permutation of the probability vector $\p$. Denoting this permutation by $\pi$, the relabeled die has probability distribution $\pi\p$. Clearly, such relabeling will not change her probability of winning any of the games described above.  Additionally, suppose she flips an unbiased coin, and if it lands ``heads'' she does nothing, while she relabels the die according to the permutation $\pi$ if it lands ``tails''.  If she \emph{forgets} the outcome of the coin flipping, the probability of the die from her perspective becomes:
$$
\q=\frac{1}{2}\p+\frac{1}{2}\pi\p\;.
$$
Since ``forgetting" information clearly cannot increase the winning probability, we conclude that any die with corresponding probability $\q$ has a smaller winning probability than a die with corresponding probability $\p$. This process of relabeling and forgetting is called \emph{random relabeling}; see e.g.~\cite{Friedland-2013}. The relation between $\p$ and $\q$ can be expressed as $\q=D\p$, where $D=\frac{1}{2}I+\frac{1}{2}\pi$
is a convex combination of the identity matrix and the permutation matrix $\pi$. More generally, we conclude that if $\q=D\p$, where $D$ is \emph{any} convex combination of permutation matrices, then the $\q$-die has a smaller winning probability than then the $\p$-die.

Birkhoff's theorem from matrix analysis states that a matrix $D$ can be expressed as a convex combination of permutation matrices if and only if it is
a doubly stochastic matrix (i.e., a matrix whose entries are non-negative and each row and column sums to one). Moreover, a fundamental theorem of majorization states that $\q\prec\p$ if and only if $\q=D\p$ for some doubly stochastic matrix $D$ \cite{Bhatia-2000a}. 

\subsubsection{Majorization in Entanglement and Coherence theories}

\label{Sect:Majorization-Entanglement}

In entanglement theory, any pure bipartite state is equivalent up to a local unitary operation on a state of the form (known as the Schmidt form):
\be\label{schmidt}
|\psi\ra^{AB}=\sum_{x=1}^{d}\sqrt{p_x}|x\ra^A|x\ra^B\;,
\ee
where $p_x\geq 0$ and $\sum_{x=1}^{d}p_x=1$. Since local unitary operations are reversible, all bipartite states with the Schmidt probability vector $\p=(p_1,...,p_d)^T$ possess the same entanglement \cite{Vidal-2000a}. Nielsen's majorization theorem~\cite{Nielsen-1999a} states that a bipartite pure state with a corresponding Schmidt vector $\p$ can be converted by LOCC to another pure bipartite state $\q$ if and only if $\p\prec \q$.

One can use the definition of majorization in~\eqref{majo} to define $d$-entanglement monotones for the state in~\eqref{schmidt}:
$$
E_k(|\psi\ra)\eqdef\sum_{x=k}^{d}p_{x}^{\da}\quad k\in\{1,...,d\}.
$$
With this definition, Nielsen's majorization theorem can be expressed as:
$$
|\psi\ra\xrightarrow{LOCC}|\phi\ra\iff E_k(|\psi\ra)\geq E_k(|\phi\ra)\quad\forall k
$$
Therefore, the functions $E_k$, known as Vidal's monotones~\cite{Vidal-2000a} (also called Ky-Fan norms), quantify the entanglement of bipartite pure states. They form a complete set in the sense that if $E_k(|\psi\ra)\geq E_k(|\phi\ra)$ for all $k$, then any other entanglement monotone (or measure), $E$, must satisfy $E(|\psi\ra)\geq E(|\phi\ra)$. The Vidal monotones play an important role also in non-deterministic LOCC transformations~\cite{Plenio1999}, and in particular they provide the maximum probability with which it is possible to convert $|\psi\ra$ to $|\phi\ra$ by LOCC~\cite{Vidal1999}:
$$
P^{(\max)}_{\ket{\psi}}(\ket{\phi})=\min_{k}\frac{E_k(|\psi\ra)}{E_k(|\phi\ra)}
$$
where the minimum is over all $k\in\{1,...,d\}$.

The majorization criterion for entanglement transformation implies the existence of entanglement catalysis.  To see this, consider for example~\cite{Jonathan-1999b} the two bipartite entangled states with probability vectors
$$
\p=(2/5,2/5,1/10,1/10),^T\quad\q=(1/2,1/4,1/4,0)^T\;.
$$
Note that $\q\not\prec\p$ since $2/5<1/2$ but also $\p\not\prec\q$ since $2/5+2/5>1/2+1/4$. We say in this case that the two probability distributions are incomparable. Consider another entangled state with Schmidt vector $\rr=(3/5,2/5)^T$. It is straightforward
to check that 
$$
\p\otimes\rr\prec\q\otimes\rr\;,
$$
even though $\p\not\prec\q$. 
That is, the pure bipartite state with Schmidt vector $\rr$ acts as a catalyst in a very similar way as it happens in chemical reactions.

The example above demonstrates that catalyst-assisted LOCC (CLOCC) transformations are more powerful than LOCC alone. Following the notation of Section \ref{Sect:Catalyst}, we write $|\psi\ra\xrightarrow{cLOCC}|\phi\ra$ if there exists a finite dimensional catalyst state 
$|\chi\ra$ such that 
\be\label{chi}
|\psi\ra\otimes|\chi\ra \xrightarrow{LOCC}|\phi\ra\otimes|\chi\ra\;.
\ee
An important question then follows: 
given two bipartite states $|\psi\ra$ and $|\phi\ra$, under what conditions does there exist a catalyst such that $|\psi^{AB}\ra\xrightarrow{cLOCC}|\phi^{AB}\ra$?  Note that if there exists a catalyst $|\chi^{A'B'} \ra$ such that~\eqref{chi} holds, then any measure of entanglement that is additive under tensor product satisfies
$$
E(\psi\otimes\chi)=E(\psi)+E(\chi)\geq E(\phi\otimes\chi)=E(\phi)+E(\chi)\;.
$$
Hence, 
$$
|\psi\ra\xrightarrow{cLOCC}|\phi\ra\Rightarrow E(\psi)\geq E(\phi)\;
$$
for any additive measure of entanglement $E$. Remarkably, the converse is also true! In two independent works by \textcite{Turgut} and \textcite{Klimesh}, it was shown that 
\begin{equation}
\label{Eq:catalysis-cond}
|\psi^{AB}\ra\xrightarrow{cLOCC}|\phi^{AB}\ra\quad\iff \quad E_\alpha(\psi)\geq E_\alpha(\phi)\quad\forall\alpha\in\mbb{R}
\end{equation}
where $E_\alpha(\psi)$ is the $\alpha$-Renyi entropy of entanglement extended for all real $\alpha$ and defined by:
$$
E_\alpha(|\psi\ra^{AB})\eqdef\text{sign}(\alpha)S_\alpha(\rho)=\frac{\text{sign}(\alpha)}{1-\alpha}\log\tr\left[\rho^{\alpha}\right],
$$
where $\rho=\tr_B|\psi\ra\la\psi|^{AB}$ is the reduced
density matrix of $|\psi\ra$. 

Nearly all of these majorization results in entanglement theory can be translated into analogous statements in the QRT of quantum coherence.  For a pure state $\ket{\psi}=\sum_{i=1}\sqrt{p_i}e^{i\phi_i}\ket{i}$ in a $d$-dimensional system with incoherent basis $\{\ket{i}\}_{i=1}^d$, the probability amplitudes $\mbf{p}=(p_1,\cdots,p_d)^T$ play the role of the Schmidt coefficients.  Under the classes of strictly incoherent operations (SIO) and the more general incoherent operations (IO), a transformation $\ket{\psi}\to\ket{\phi}$ is possible if and only if $\mbf{p}\prec\mbf{q}$, where $\mbf{p}$ and $\mbf{q}$ are the probability amplitudes of $\ket{\psi}$ and $\ket{\phi}$  respectively \cite{Winter-2016b, Zhu-2017a}.  The phenomenon of catalytic coherence convertibility is also possible, with the necessary and sufficient conditions being given by Eq.~\eqref{Eq:catalysis-cond} and the obvious replacement $E_\alpha(\psi)\to C_\alpha(\psi):=\text{sign}(\alpha)\frac{1}{1-\alpha}\log\sum_{i=1}^dp_i^\alpha$ \cite{Bu-2016a}.

\subsubsection{Majorization and Statistical Comparisons}

Consider two random variables with probability distributions $\p$ and $\q$, respectively.  For concreteness imagine, as before, that these distributions represent two biased dice.  However, the game is now different.  The player is given one of the two die, and her goal is to determine if it is the $\p$-die or the $\q$-die. Clearly if she can roll the die many times, then by the law of large numbers, she can infer its underlying probability distribution and successfully guess which one she was given.  However, if she can only roll a finite number of times, there is a chance that she will make an error in her identification guess.  Intuitively, the smaller this error, the more distinguishable are $\p$ and $\q$.  In fact, one could quantify the distinguishability between $\p$ and $\q$ as the optimal probability that the player correctly guesses which die she holds after a fixed number of rolls.  This is the problem of hypothesis testing, and its quantum version is discussed in Section \ref{Sect:Smooth_Entropies}.  Other distance measures between $\p$ and $\q$ could also be chosen to quantify the distinguishability of the two distributions.

In statistical comparisons, we are interested in adopting a resource-theoretic perspective and measuring distinguishability in an operational way.  Specifically, we say that a pair of distributions $(\p,\q)$ is more distinguishable than another pair of distributions $(\p',\q')$ if there exists a column stochastic matrix $M$ such that
$$
\p'=M\p\quad\text{and}\quad\q'=M\q\;.
$$
To make the connection with resource theories, interpret $(\p,\q)$ and $(\p',\q')$ as two states and $M$ as a free operation.  The rationale for defining distinguishability in this way is as follows: if Alice is tampering with her die and changing its probability distribution (e.g., replacing the symbols of her dice at random) then this alone cannot improve her ability to distinguish between the original two distributions.  More generally, the matrix $M$ represents a classical channel converting, respectively, the input distributions $\p$ and $\q$ to the output distributions $\p'$ and $\q'$. In this case we write
\begin{equation}
\label{Eq:relative_majorization}
(\p',\q')\prec_{r}(\p,\q)\;,
\end{equation}
and we say the $(\p,\q)$ \textit{relatively majorizes} $(\p',\q')$. Note that relative majorization is a generalization of majorization. Indeed, suppose that all vectors involved are of the same dimension and suppose $\q=\q'=\e\eqdef \tfrac{1}{d}(1,...,1)^T$. In this case we have
$$
(\p',\q')\prec_{r}(\p,\q)\quad\iff\quad\p'\prec\p\;,
$$
since the condition $\q'=M\q$ is equivalent to $\e=M\e$, which implies that the column stochastic matrix $M$ is in fact doubly stochastic.

Originally, the preorder defined above was called $\mbf{d}$-majorization in the case where $\q=\q'=\mathbf{d}$ is some fixed vector $\mathbf{d}$~\cite{Veinott1971}.
  Here we follow the terminology of \textcite{Renes-2016a}, who identified Eq.~\eqref{Eq:relative_majorization} as relative majorization since all relative entropy functions, such as all the quantum R\'enyi divergences, behave monotonically under this preorder. In this sense, relative majorization is the fined-grained version of the relative entropy.

\begin{figure}[h]
    \includegraphics[width=0.3\textwidth]{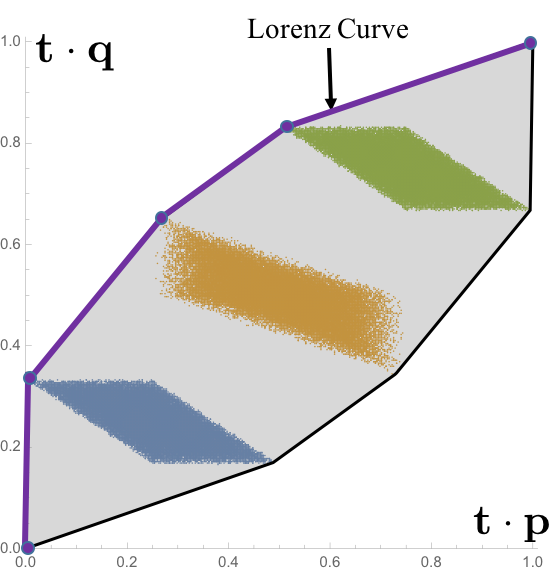}
  \caption{\linespread{1}\selectfont{\small Testing Region of a pair $(\p,\q)$.}}
  \label{testing}
\end{figure}

Relative majorization has a very simple characterization in terms of \emph{testing regions} \cite{Renes-2016a}, which can represent Neyman-Pearson hypothesis testing~\cite{Cover2006}.
The testing region of a pair of $d$-dimensional probability vectors $\p$ and $\q$ (see Fig.~\ref{testing}), is a region on the plane that is defined by:
$$
\mathcal{T}(\p,\q)\eqdef\left\{(\bt\cdot\p,\bt\cdot\q)\in\mbb{R}^2\;\Big|\;0\leq\bt\leq\e\;\;;\;\;\bt\in\mbb{R}^d\right\}
$$ 
where the inequalities $0\leq\bt\leq\e$ are entrywise; i.e.,  the components of $\bt$ are between $0$ and $1$. Such a vector $\bt$ can be viewed as a ``test" since $\bt$ and $\e-\bt$ correspond to a binary-outcome experiment.

Any testing region contains both the origin $(0,0)$ and the point $(1,1)$. Moreover,  for any test vector $\bt$, the vector $\e-\bt$ is also a test. Hence, if a point $(x,y)$ is in the testing region of the pair $(\p,\q)$ so is the point $(1-x,1-y)$. This in turn implies that a testing region is completely specified by its upper (or lower) boundary. The upper boundary is known in the literature (see e.g.~\cite{Gour-2015a}) as the Lorenz curve of the pair $(\p,\q)$ (see Fig.~\ref{testing}).

\begin{figure}[h]
    \includegraphics[width=0.3\textwidth]{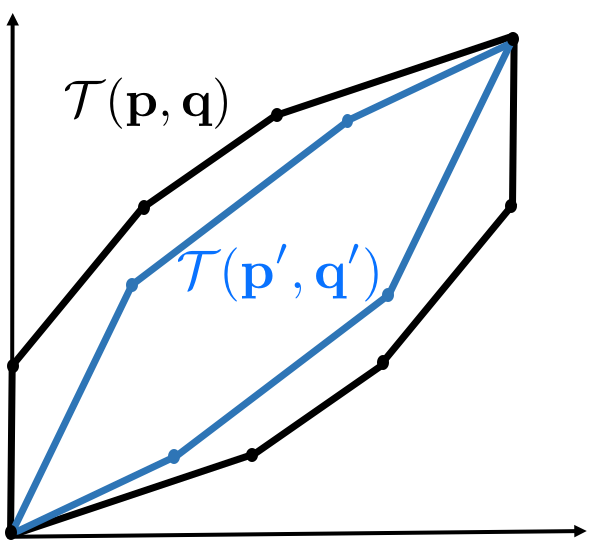}
  \caption{\linespread{1}\selectfont{\small Inclusion of two testing regions.}}
  \label{inclusion}
\end{figure}

It can be shown that a pair of probability distributions $(\p,\q)$ relatively majorizes another pair $(\p',\q')$ if and only if the testing region of $(\p',\q')$ is inside the testing region of $(\p,\q)$ (see Fig.~\ref{inclusion}). That is,
$$
(\p',\q')\prec_{r}(\p,\q)\quad\iff\quad \mathcal{T}(\p',\q')\subset\mathcal{T}(\p,\q)\;.
$$
This remarkable result was already proven by Blackwell in 1953~\cite{blackwell1953}.  However, Blackwell's proof was not direct, and a more direct proof was given later by \textcite{Ruch1980}.  A different proof can also be found in~\cite{Dahl-1999}, where the testing regions are viewed as zonotopes. Since testing regions are specified by their Lorenz curves, the result above can be expressed as follows: $(\p',\q')\prec_{r}(\p,\q)$ if and only if the Lorenz curve of $(\p',\q')$ is never above the Lorenz curve of $(\p,\q)$.

\subsubsection{Majorization in Quantum Thermodynamics}

\label{Sect:Majorization-Thermo}

As discussed in Section \ref{Sect:Thermodnamics}, the set of free operations in the resource theory of athermality are thermal operations.  A strictly larger class of operations are those that preserve the Gibbs state: $
\mathcal{E}(\gamma_H)=\gamma_H$.  In the quasiclassical case, where states are diagonal in the energy eigenbasis, the conversion of a state $\rho$ to $\sigma$ is possible by thermal operations if and only if it is possible by Gibbs-preserving operations~\cite{Janzig-2000a,Horodecki-2013b,Korzekwa}.   To phrase this as a majorization condition, let $\p$, $\q$, and $\mathbf{g}$ be the vectors consisting of the diagonal components of $\rho$, $\sigma$, and $\gamma_H$, respectively. Then in the quasiclassical setting, $\rho$ can be converted to $\sigma$ by thermal operations if and only if there exists a column stochastic matrix $M$ such that:
$$
\q=M\p\quad\text{and}\quad\mathbf{g}=M\mathbf{g}.
$$ 
The above equation is precisely the condition 
$$
(\q,\g)\prec_r (\p,\g).
$$
When the condition above holds we say that $\p$ \textit{thermomajorizes} $\q$. The characterization of relative majorization in terms of testing regions immediately applies to thermomajorization. In particular, $\p$ thermomajorizes $\q$ if and only if the Lorenz curve of $\p$ (with respect to the Gibbs state $\g$) is never below the Lorenz curve of $\q$ \cite{Horodecki-2013b}. 
 
Thermomajorization (which is equivalent to $\mbf{d}$-majorization of \cite{Veinott1971}) has several alternative characterizations. One of them can be expressed as~\cite{Alberti1980}:  
$$
\left\|\p-t\g\right\|_1\geq \left\|\q-t\g\right\|_1\quad\forall\;t\geq 0\;,
$$
where $\|\mathbf{x}\|_1\eqdef\sum_j|x_j|$ is the $\ell_1$-norm.  \textcite{Alberti1980} demonstrated that this characterization has a quantum analog in the case of qubits.  Specifically, given two pairs of \emph{qubit} density matrices $(\rho_1,\rho_2)$ and $(\sigma_1,\sigma_2)$, it was shown that there exists a quantum channel such that $\sigma_j=\mathcal{E}(\rho_j)$ for $j=1,2$ if and only if 
 $$
\left\|\rho_1-t\rho_2\right\|_1\geq \left\|\sigma_1-t\sigma_2\right\|_1\quad\forall\;t\geq 0\;,
$$
where $\|\cdot\|_1$ is the trace norm. Taking $\rho_2=\sigma_2=\gamma_H$ we obtain necessary and sufficient conditions for the existence of a Gibbs-preserving channel that converts the qubit $\rho_1$ to $\sigma_1$~\cite{BuscemiGour}. However, this result only holds for two-dimensional systems and already in three dimensions there are counter-examples~\cite{Winter2004} (i.e., states that satisfy the above inequality even though they cannot be converted by Gibbs-preserving operations).

In the fully quantum case with states not necessarily diagonal in the energy eigenbasis, deciding whether $\rho$ can be transformed into $\sigma$ is a more complex issue.  While the convertibility equivalence between thermal operations and Gibbs-preserving maps no longer holds, one can additionally require the latter to have time-translation covariance in order to obtain a better approximation of thermal operations \cite{Jennings2015, Gour2018} (see also Section \ref{Sect:Thermodnamics}).  It is currently unknown whether the convertibility power of Gibbs-preserving, time-translation covariant maps is strictly greater than the convertibility power of thermal operations.  This problem is important to the field of quantum thermodynamics since the former is much simpler to characterize mathematically than thermal operations. Particularly, it was demonstrated in \textcite{Gour2018} that the problem of converting one general state to another by a Gibbs-preserving and time-translation covariant operation can be solved efficiently and algorithmically using semi-definite programming (SDP). It was shown by employing  a generalization of relative majorization which is called \textit{quantum majorization}.

Quantum majorization is a preorder among bipartite states in $\mS(\mH^A\otimes\mH^B)$ having the same marginal state on system $A$. We say the $\rho^{AB}$ quantum majorizes $\sigma^{AB}$ with respect to a group $G$, and denote it by $$\sigma^{AB}\prec_{q}^{G}\rho^{AB}\;,$$ if and only if there exists a $G$-covariant channel $\mathcal{E}$ such that
$$
\sigma^{AB}=\id^A\otimes\mathcal{E}(\rho^{AB})\;.
$$
To see how quantum majorization is related to thermodynamics, take both $\rho^{AB}$ and $\sigma^{AB}$ to be the following states:
\begin{align}
\rho^{AB}&=\frac{1}{2}|0\lr 0|\otimes\rho_1+\frac{1}{2}|1\lr 1|\otimes\gamma_{H},\nonumber\\
\sigma^{AB}&=\frac{1}{2}|0\lr 0|\otimes\rho_2+\frac{1}{2}|1\lr 1|\otimes\gamma_{H}.\nonumber
\end{align}
With this choice we see that $\sigma^{AB}\prec_{q}^{G}\rho^{AB}$ if and only if $\rho_1$ can be converted to $\rho_2$ by Gibbs-preserving and time-translation symmetric operations. Hence, since quantum majorization can be determined by an SDP so can the conversion of $\rho_1$ to $\rho_2$. Other variants of majorization and their applications in thermodynamics can be found in \textcite{Egloff2015, Philippe2015}.

\subsection{Convex analysis, Semi-definite programming, and duality theory}

\label{Sect:Convex_analysis}

Convex analysis plays an important role in many areas of science (see e.g.~\cite{Barvinok-2002a,  Boyd-2004a}), so it is not surprising that many of its tools are intensively employed in quantum information. To see its application specifically to resource theories, we will consider here a convex resource theory $\mR$ with a convex set of free states $\mF$, and a convex set of free operations $\mO$. 


Recall we observed in Section \ref{Sect:Witness} that if the set of free states is both convex and closed, then 
$\rho\in\mF(\mH)$ if and only if 
\be\label{firstex}
\min_{W\in\mF^*(\mc{H})}\tr[W\rho]\geq 0\;.
\ee
This problem is known as the weak membership problem~\cite{Gurvits2003} for the convex set $\mF(\mH)$.

In some QRTs the optimization problem above is relatively easy, like the resource theory of coherence, and it can be solved using standard techniques in semi-definite programming. For other QRTs, like entanglement theory, the problem above is computationally hard~\cite{Gurvits2003, Gharibian-2010a} (particularly, the computational time is believed to grow exponentially with the dimension of $\mH$). In such cases,  the condition that $W\in\mF^*$ does not have a simple form. 

One approach (see e.g.~\cite{Perez2004} in entanglement theory) is to allow for an $\epsilon$ error, and then intersect $\mF^*(\mH)$ with an $\epsilon$-net, which is a finite set such that every state is at most a distance $\epsilon$ from some state in the set, as measured by the trace distance.  Using an $\epsilon$-net, one can replace the minimization of $W$ in Eq.~\eqref{firstex} with one having a finite number of constraints on $W$, namely $\tr[W\sigma_j]\geq 0$, where $\sigma_j$ is the $j^{\text{th}}$ element in the $\epsilon$-net. This approach can be very useful in small dimensions~\cite{Perez2004} as it provides a way to determine if a state is free or not by using standard techniques from semi-definite programming. Remarkably, it was shown by~\cite{Brandao2011}, that if one replaces the trace distance with an operationally motivated distance, based on the so-called one-way LOCC norm~\cite{Matthews2009}, it is possible to construct a quasi-polynomial-time algorithm for solving
the weak membership problem for the set of separable bipartite quantum states. 

Convex analysis is also very useful for the study of single-shot state transformations.  For a given QRT, denote the set of all free Choi matrices by
$$
\mC(A\to B)\eqdef\left\{J_{\Phi}^{AB}\;\Big|\;\Phi\in\mO(A\to B)\right\}
$$ 
where $J_{\Phi}^{AB}$ is the Choi matrix of the channel $\Phi$. Since we assume here that $\mO$ is convex we get that also the set $\mC$ is convex.  Now consider the problem of deciding whether one quantum state $\rho\in\mS(A)$ can be converted into another $\sigma\in\mS(B)$ by free operations. That is, we want to know if there exists $\Phi\in\mO$ such that $\sigma=\Phi(\rho)$. In the Choi picture, the question becomes if there exists a state $J^{AB}\in\mC$ such that
$$
\sigma=\tr_A\left[J^{AB}(\rho^T\otimes \mbb{I}^B)\right]\;.
$$
By multiplying both sides of the equation above by some matrix $X\in\mB(\mH^B)$ and taking the trace, one can express the equation above in the form:
$$
\tr\left[J^{AB}\left(\rho^T\otimes X-\frac{\tr[\sigma X]}{d_A}\mbb{I}^{AB}\right)\right]=0\;,
$$
where we used the property that $J^A=\mbb{I}^A$. The equation above has to hold for all $X$, but since it is linear in $X$, we just need to check that it holds for all $X\in\{X_j\}_{j=1}^{d_B^2}$, where $X_j$ are some basis elements of $\mB(B)$.  If we assume again that $\mC$ is convex and closed (so that $\mC^{**}=\mC^{*}$) we conclude that $\rho$ can be converted to $\sigma$ by free operations if and only if
there exists a matrix $J^{AB}$ with marginal $J^A=\mbb{I}^A$ that satisfy the following two conditions:
\begin{align}
& \tr\left[J^{AB}\left(\rho^T\otimes X_j-\frac{\tr[\sigma X_j]}{d_A}\mbb{I}^{AB}\right)\right]=0\quad\forall j=1,...,d_B^2\nonumber\\
& \tr[J^{AB}W]\geq 0\quad\forall W\in\mC^*\;.\nonumber
\end{align}
The problem above, similar to the weak membership problem~\eqref{firstex}, is a feasibility problem in (conic) linear programming. In some QRTs (e.g. affine QRTs~\cite{Gour-2016a}, QRTs of asymmetry and thermodynamics~\cite{Gour2018}), the condition that $W\in\mC^*$ can be expressed as an SDP problem, and in this case determining whether $\rho\toO\sigma$ can be solved efficiently and algorithmically by SDP. In other QRTs, the problem can be much harder.

\subsection{Smooth Entropies and the Generalized Stein's Lemma}

\label{Sect:Smooth_Entropies}

In the imperfect one-shot scenario, the goal is to transform an initial state $\rho$ into some target state $\sigma$ within an $\epsilon$-error.  A fundamental technique in the study of this problem involves the ``smoothing'' of some function over a small subset of density matrices. The general use of $\epsilon$-smoothing in quantum information theory was pioneered by \textcite{Renner-2005a}, originally for application in quantum key distribution (QKD) (see \textcite{Cachin-1997a} for classical origins).  Since then it has been used prominently in quantum hypothesis testing and other studies of one-shot quantum Shannon theory \cite{Renner-2005b, Konig-2009a, Tomamichel-2010a, Renes-2011a, Tomamichel-2012a, Wang-2012a, Datta-2013a, Datta-2013b, Datta-2013c, Tomamichel-2013a, Matthews-2014a,  Radhakrishnan-2016a, Wang-2017a}. Here we review two smooth entropic quantities and describe their application in terms of single-shot resource formation and distillation.  This will set the stage for the asymptotic reversibility result based on quantum hypothesis testing and the Generalized Stein Lemma.

We begin with the quantity $\mc{R}_{\max}$, which is defined in Eq.~\eqref{Eq:R-max} as
\begin{equation}
\mc{R}_{\max}(\rho)=\inf_{\sigma\in\mc{F}(\mc{H})}\!\{\lambda \;|\;\rho\leq 2^\lambda\sigma\}.
\end{equation}
As noted in Section \ref{Sect:Robustness}, this is equivalent to the log-robustness of resource.  For $\epsilon\in(0,1]$, a smoothed version of $\mc{R}_{\max}$ is given by 
\begin{equation}
\mc{R}^\epsilon_{\max}(\rho)=\inf_{\hat{\rho}^{AB}\in B_\epsilon(\rho^{AB})}\mc{R}_{\max}(\hat{\rho})
\end{equation}
where $B_\epsilon(\rho)=\{\sigma:F(\rho,\sigma)\geq 1-\epsilon\}$.  Essentially, $\mc{R}^{\epsilon}_{\max}(\rho)$ finds the smallest value of $\mc{R}_{\max}(\rho)$ within an $\epsilon$-ball centered at $\rho$.  In the QRT of entanglement, $\mc{R}^\epsilon_{\max}$ quantifies the one-shot catalytic entanglement cost under $\delta$-resource generating operations \cite{Brandao-2011a}.  
 In the QRT of coherence, the relaxations of catalytic convertibility and $\delta$-resource generating operations can be dropped, as $\mc{R}^\epsilon_{\max}$ provides the one-shot (non-catalytic) coherence cost of a given state using (strictly) resource non-generating operations \cite{Zhu-2017a}.

The second smooth entropic quantity we discuss is based on the problem of quantum hypothesis testing.  In quantum hypothesis testing, the goal is to distinguish one state $\rho$, called the null hypothesis, from another $\sigma$, called the alternative hypothesis \cite{Hiai-1991a, Ogawa-2000a}.  Typically one attempts to minimize the identification error when the system is in state $\sigma$, given some threshold in the identification error when the system is in state $\rho$.  In more detail, one considers a two-outcome POVM $\{M,\mbb{I}-M\}$ with associated error probabilities
\begin{align}
\alpha(M)&=\tr[(\mbb{I}-M)\rho]\notag\\
\beta(M)&=\tr[M\sigma].\notag
\end{align}
Then for any $\epsilon>0$, the problem asks to compute the smallest possible value of $\beta(M)$ under the constraint that $\alpha(M)\leq\epsilon$.  This can be phrased as an entropic quantity known as the \textit{hypothesis testing relative entropy}, which is defined as
\begin{align}
D^\epsilon_{\text{H}}(\rho\Vert\sigma)= \sup_{0\leq M\leq\mbb{I}\atop \alpha(M)\leq \epsilon} -\log\beta(M).
\end{align}
This quantity, which involves an ``operator smoothing'' \cite{Buscemi-2010b},  provides the appropriate one-shot quantifier for many other information-theoretic tasks \cite{Wang-2012a, Tomamichel-2013a, Matthews-2014a, Dupuis-2013a}.  Note that $D^\epsilon_{\text{H}}(\rho\Vert\sigma)$ is expressed as a semi-definite optimization, and it can therefore be efficiently computed.  

For a QRT with a compact convex set of free states, we can introduce the resource measure
\begin{align}
\mc{R}^\epsilon_{\text{H}}(\rho)&=\inf_{\sigma\in\mc{F}(\mc{H})}D^\epsilon_H(\rho\Vert\sigma)\notag\\
&=\sup_{0\leq M\leq\mbb{I}\atop \tr[(\mbb{I}-M)\rho]\leq\epsilon}\inf_{\sigma\in\mc{F}(\mc{H})}-\log\tr[M\sigma],
\end{align}
where the minimax theorem has been applied to switch the order of extrema.  For entanglement theory, this quantity corresponds to the one-shot distillable entanglement under maximal operations, i.e., the largest $R$ such that $\rho^{AB}\toOmax_\epsilon\phi^+_{2^R}$ \cite{Brandao-2011a}.  For coherence, an analogous result holds in terms of distillable coherence provided the allowable set of $\sigma$ is enlarged slightly \cite{Regula-2017a}.  The quantity $\mc{R}^\epsilon_{\text{H}}(\rho)$ also has application in quantum thermodynamics.  Note that in thermodynamics, the Gibbs state $\gamma_H$ is the unique free state for a given thermodynamic system, and so we have the reduction 
\begin{equation}
\mc{R}^\epsilon_{\text{H}}(\rho)=\sup_{0\leq M\leq\mbb{I}\atop \tr[(\mbb{I}-M)\rho]\leq\epsilon}-\log\tr[M\gamma_H].
\end{equation}
\textcite{Halpern-2016a} have shown this to, roughly speaking, quantify both the one-shot extractable work of a thermodynamical state, as well as the one-shot work cost of forming it.

A more traditional version of hypothesis testing is in the asymptotic setting where the two hypotheses are presented in many-copy form, $\rho^{\otimes n}$ and $\sigma^{\otimes n}$.  For any $\epsilon\in(0,1)$, the asymptotic rate of $D^\epsilon_{\text{H}}$ is given precisely by the relative entropy:
\begin{equation}
\lim_{n\to\infty}\frac{1}{n}D^\epsilon_{\text{H}}(\rho^{\otimes n}\Vert\sigma^{\otimes n})=S(\rho\Vert\sigma).
\end{equation}
This is known as the quantum Stein's Lemma, and its proof was given by \textcite{Hiai-1991a} and \textcite{Ogawa-2000a}.  This result is quite appealing since any task quantified by $D^\epsilon_{\text{H}}$ in the single-shot level can then be quantified by the relative entropy in the many-copy setting. 

To apply the results and techniques of quantum hypothesis testing to quantum resource theories, one needs to generalize the problem.  One such scheme involves a scenario in which either the null hypothesis $\rho$ or a set $S$ of alternative hypotheses is possible.  The goal then is to distinguish $\rho$ from the states belonging to $S$.  In the context of quantum resource theories, it is typical to let $S=\mc{F}(\mc{H})$ be the set of free states.  \textcite{Brandao-2010b} have proven a generalization of the quantum Stein's Lemma that holds for most well-structured sets of free states.  Specifically, suppose that $\mc{F}(\mc{H})$ has the properties of being
    \setlist{nolistsep}
    \begin{itemize}[noitemsep]
        \item[1.] Closed and convex,
        \item[2.] Closed under tensor products,
        \item[3.] Closed under the partial trace,
        \item[4.] Closed under permutation of spatially separated subsystems.
    \end{itemize}
Note that the free states in any QRT having tensor product structure will satisfy these conditions.  The generalized quantum Stein's Lemma ensures that
\begin{equation}
\label{Eq:quantum_stein}
\lim_{n\to\infty}\frac{1}{n}\mc{R}_{\min}^{\epsilon}(\rho^{\otimes n})=\mc{R}^\infty_{\text{rel}}(\rho)
\end{equation}
for any $\epsilon\in(0,1)$ \cite{Brandao-2011a}, and furthermore, \textcite{Brandao-2010b} were able to also show that
\begin{equation}
\label{Eq:R_max-limit}
\lim_{\epsilon\to 0}\lim_{n\to\infty}\frac{1}{n}\mc{R}^{\epsilon}_{\max}(\rho^{\otimes n})=\mc{R}^\infty_{\text{rel}}(\rho).
\end{equation}
In subsequent work, \textcite{Brandao-2015a} explicitly connected Eqs.~\eqref{Eq:quantum_stein} and \eqref{Eq:R_max-limit} to the problem of asymptotic resource convertibility under \textit{asymptotically resource non-generating (RNG)} operations.  This involves transformations of the form $\rho^{\otimes n}\toOn_\epsilon\sigma^{\otimes \lfloor nR'\rfloor}$, where $\mc{O}_n$ is the class of $\epsilon_n$-resource generating operations (see Section \ref{Sect:Other_Classes_Operations}) such that $\lim_{n\to\infty}\epsilon_n=0$.  If the free states in a QRT satisfy the four properties listed above, then the result of \cite{Brandao-2015a} says that \textit{any} two states $\rho$ and $\sigma$ are reversible under asymptotically-RNG transformations, with a rate given by the ratio of the regularized relative entropies of resource.  That is,   
\begin{equation}
\label{Eq:asymptotic_rate}
R(\rho\to\sigma)=\frac{\mc{R}_{\text{rel}}^\infty(\rho)}{\mc{R}_{\text{rel}}^\infty(\sigma)},
\end{equation}
provided that $\mc{R}_{\text{rel}}^\infty(\rho),\mc{R}_{\text{rel}}^\infty(\sigma)\in (0,\infty)$.  Note, the restriction that $\mc{R}^\infty_{\text{rel}}$ be nonzero and finite is more than just a mathematical detail.  Physically relevant QRTs, such as the QRT of asymmetry, have free states which satisfy the four necessary properties, and yet $\mc{R}^\infty_{\text{rel}}(\rho)=0$ for all resource states \cite{Gour-2009a}. In this case, the results of \cite{Brandao-2015a} cannot be directly applied.  Nevertheless, when $\mc{R}^\infty_{\text{rel}}$ is nonzero and finite, Eq.~\eqref{Eq:asymptotic_rate} along with the arguments of \textcite{Horodecki-2002c, Gour-2009a} imply that
\begin{equation}
\frac{\mc{R}_{\text{rel}}^\infty(\rho)}{\mc{R}_{\text{rel}}^\infty(\sigma)}=\frac{f^\infty(\rho)}{f^\infty(\sigma)},
\end{equation}
where $f^\infty$ is the regularized versions of any asymptotically continuous function for which $f^\infty(\rho),f^\infty(\sigma)\in(0,\infty)$.   This says that the regularized version of all asymptotically continuous resource measures are equivalent up to an overall proportionality factor.  Thus, the regularized relative entropy of resource can be interpreted as the unique measure of resource for the task of asymptotic convertibility.

\section{Outlook}

A common theme in physics is the unification of theories and models that at first glance may seem completely unrelated. 
Most notable in this regard is the successful unification of the three non-gravitational forces in nature. 
Such an amalgamation not only leads to new discoveries, but it also has the potential to profoundly change the way we perceive the world around us. 
With the advent of quantum information science, many seemingly unrelated properties of physical systems,  
such as entanglement, asymmetry, and athermality, have now become recognized as resources. This recognition is profound as it allows them to be unified under the same roof of quantum resource theories. 
Entanglement, athermality, and asymmetry, are no longer regarded as just interesting physical properties of a quantum system, but they now emerge as resources that can be utilized and manipulated to execute a variety of remarkable tasks, such as quantum teleportation.  

This review article began with a precise definition of quantum resource theories and then considered general structural features of different QRTs.  As discussed in Section ~\ref{Sect:processes}, all QRTs in quantum information theory can be viewed as a resource theory of processes.  For example, in Section \ref{Sect:Steering} we discussed how the phenomenon of quantum steering can be cast as a resource theory of incompatible (semicausal) multi-sources, just one specific type of quantum process.  More work is needed in the future to better understand unifications like this for other QRTs.  

Section \ref{Sect:Specific_QRTs} has provided a summary of specific QRTs that reflect recent and ongoing developments in the field.  We hope this choice of examples sparks the reader's interest on new topics or motivates the construction of novel QRTs.  Unfortunately, since the subject of QRTs spans a large range of topics, we could not cover all resource theories previously studied in the literature.  Notable omissions include the resource theory of knowledge~\cite{Rio2015,Kraemer2016}, imaginarity~\cite{Hickey2018}, superposition~\cite{Plenio2017}, and others.   Moreover, this review did not discuss the characterization of resource theories as symmetric monoidal categories~\cite{Coecke-2016,Fritz-2015}.  This category theory approach to QRTs can be useful when considering other models beyond quantum physics, such as the framework of generalized probabilistic theories, or when incorporating resource theories in other fields of science. 

\noindent\textit{Acknowledgments:---}  We would like to thank David Jennings, Matteo Lostaglio, and Nicole Yunger Halpern for helpful discussion and insight on quantum thermodynamics, as well as Gerardo Adesso, Eneet Kaur, and Xin Wang for constructive feedback.  We are gracious to Earl Campbell for clarifying certain aspects of magic state resource theories.  Finally, we thank Mark Wilde for carefully reading an earlier draft of this manuscript and providing a number of helpful comments.  E.C. is supported by the National Science Foundation (NSF) Early CAREER Award No. 1352326.  G.G. research is supported by NSERC.

\bibliography{QRTbib}

\end{document}